# Secure and Trustworthy Computing 2.0

# Vision Statement

**Prepared for the Secure and Trustworthy Computing Program,
National Science Foundation**

**Date: July 27, 2023**

# Table of Contents

# 1    EXECUTIVE SUMMARY

The Secure and Trustworthy Computing (SaTC) program within the National Science Foundation (NSF) program serves as the primary instrument for creating novel fundamental science in security and privacy in the United States with broad impacts that influence the world. The program funds research in a vast array of research topics that span technology, theory, policy, law and society.

Once a decade, SaTC revisits its mandate and undertakes the task of envisioning a future for cyber-security research. This comprehensive vision considers the needs and opportunities for research conducted on behalf of the United States, contributing to the nation's advancements in security and privacy. This document serves as the culmination of that effort, providing valuable insights for the future of this critical field.

We find ourselves at a unique moment in time, witnessing the rapid advancement of technology that is poised to reshape our society and daily lives. However, along with these transformative capabilities come significant threats from adversaries. Our reliance on online and interconnected systems has never been greater, as the public increasingly depends on technology for personal and professional activities. As such, it is imperative to address the associated risks and vulnerabilities.

Recent advancements, such as the remarkable progress in artificial intelligence (AI) and particularly in large language models, underscore the need for heightened attention to security and privacy. The uncertainty surrounding the security and safety of these new capabilities, especially in the complex political and social landscape, emphasizes the urgency for robust research in these areas. By addressing these imperatives, we aim to safeguard the well-being of the people in the United States and around the world.

The critical need for advancements in the science of security and privacy is evident. The interconnectedness of systems, the pervasive use of technology, and the potential risks posed by emerging threats necessitate a deep understanding of these domains. Moreover, the benefits resulting from this research should be accessible to all individuals, regardless of their communities, resources, or technical expertise. The inclusivity of the research outcomes is paramount for a secure and trustworthy computing ecosystem.

The subsequent sections of this document offer a comprehensive view of general themes and specific areas of focus for future research. While we have made extensive efforts to be exhaustive, it is practically impossible to identify all potential opportunities and needs. Nevertheless, this view provides a representative snapshot of the research areas, acknowledging the existence of many others. It is important to consider this perspective as a moment in time, with its recommendations and insights warranting frequent revisitation.

The detailed technical content presented in the following sections are the result of over a year's worth of efforts, drawing from the collective input and contributions of numerous individuals. We express our heartfelt appreciation to all those who participated in the in-person workshops, engaged in informal conversations, and contributed through online communications. Their valuable insights and collaboration have been instrumental in shaping the contents of this report.

This document is complete as of July 2023. We encouraged readers to provide feedback or additions. In the Spring of 2023, SaTC accepted feedback via the NSF REI process.

We extend our gratitude to all participants, the National Science Foundation, and the dedicated staff and students across various organizations who have supported us throughout this journey. We hope that the discussions presented herein prove useful, stimulating further thought and leading to new research initiatives that serve the betterment of united states and society as a whole.

Patrick McDaniel (University Wisconsin, Madison)
Farinaz Koushanfar (University of California, San Diego)
(July 2023)

## 1.1 OVERVIEW

In the Fall of 2022, several members of NSF Secure and Trustworthy program approached Patrick McDaniel and Farinaz Koushanfar to create a vision of NSF-funded cybersecurity research for the next 10 years. Over the next 10 months they collaborated to collect information informally, through online forms, and through a workshop held in the Spring of 2023 (see next section). This input was organized and synthesized to highlight challenges, opportunities and needs for new science and outreach within the SaTC community. This document is the result of that effort.

## 1.2 History of Secure and Trustworthy Computing

The National Science Foundation's (NSF) Secure and Trustworthy Computing (SaTC) program was established in 2007 to address the growing concern over attacks on industry, government, and the public. The SaTC program focuses on advancing the science of secure computing, with the goal of ensuring that computer systems and networks are resilient against cyber threats and can be trusted to protect sensitive information and critical infrastructure. The program supports research that spans a wide range of areas, including (among many) cryptography, software security, hardware security, network security, human factors and usability, and socio-technical aspects of security and privacy.

The SaTC program has been instrumental in advancing the state of the art in secure and trustworthy computing. For example, it has supported research on the development of new cryptographic techniques, such as homomorphic encryption and post-quantum cryptography, which have the potential to significantly enhance the security of data and communications. The program has also funded research on the development of secure software and hardware, including the design of secure programming languages and operating systems, and the development of hardware-based security mechanisms, such as trusted execution environments and secure enclaves.

In addition to supporting research, the SaTC program places a strong emphasis on education and workforce development. The program supports the development of cybersecurity curricula and training programs at all levels, from K-12 to graduate school, and provides funding for cybersecurity education and workforce development initiatives. The program also funds research on the socio-technical aspects of security and privacy, which seeks to understand the human factors that contribute to security breaches and to develop strategies for improving security awareness.

Overall, the SaTC program has made significant contributions to the field of secure and trustworthy computing. Its investments in innovative research, education, and workforce development have helped to build a stronger and more secure digital ecosystem, and its impact is likely to continue to be felt for generations to come.

## 1.3    The SaTC 2.0 Workshop

The Secure and Trustworthy Cyberspace (SaTC) Program of the National Science Foundation held a two-day workshop on March 8-9, 2023, at the University of Texas, Dallas Davidson-Gundy Alumni Center in Richardson to gather input and insights from industry, academic, and government partners about the key research challenges and opportunities for the cybersecurity and privacy community in the next decade.

The event was attended by a total of 87 individuals, with the majority being from the academic community. Out of the total attendees, 66 were from academic institutions, including universities and research centers. 15 attendees were from the National Science Foundation. The presence of these NSF representatives indicates that the event was focused centrally on scientific research and education. Finally, there were 6 attendees from the industry sector. This suggests that there was some interest from companies or organizations that may be interested in collaborating with academic researchers or utilizing research outcomes in their work. Overall, the event was successful in bringing together a range of stakeholders from academia, government, and industry.

The workshop was designed to assess the key research challenges and opportunities facing the cybersecurity and privacy community over the next decade. Participants included a wide range of stakeholders, including industry leaders, academics, and government officials. This diversity of perspectives allowed for a comprehensive exploration of the most pressing issues facing the field of cybersecurity and privacy.

During the workshop, participants were organized into small groups to discuss various topics related to cybersecurity and privacy. These groups worked together to generate a set of textual observations and recommendations, which were collected by the workshop organizers. The artifacts produced during the workshop were one of the key resources used to create this document. Indeed, several of the passages in this document appear from the breakout notes unaltered, while many others appear in only slightly edited form.

The breakout sessions for the workshop were created through a collaborative effort involving members of the national science foundation, industry, government, and academia. Initially, the list of potential topics was developed by studying past efforts at organizing the field of cybersecurity. The organizers then used this exercise, combined with their personal judgment and expertise, to identify 38 topics that they felt were relevant and important to the field. During the workshop, a brainstorming session was held, resulting in the selection of an additional 5 topics to be included in the breakout sessions. By incorporating a diverse range of perspectives and expertise, the workshop organizers were able to identify a forward-looking list of topics relevant to SaTC.

## 2 EMERGING THEMES IN SECURITY AND PRIVACY

The scope of discussion and idea generation activities were breathtaking in their scope, diversity and intellectual depth. At the same time, several broad themes emerged that spanned many areas and discussions. Explored below, these themes signal subtle (and not so subtle) shifts in technology and thought that will require new science, policy, and outreach.

### 2.1 Rethinking Hardware and Systems for Security and Privacy

The ongoing battle between security practitioners and adversaries has perpetuated a relentless cycle of attacks and countermeasures. A significant reason for this persistent state is the lack of strong security guarantees provided by underlying software and hardware architectures upon which secure systems can be built. The vulnerabilities inherent in these components hinder the development of truly secure and privacy-preserving systems.

The absence of scientifically rigorous guarantees from hardware (e.g., CPUs, GPUs) and system components (e.g., clouds, operating systems, VMs) severely limits the extent to which security and privacy-preserving systems can be constructed. It is imperative to redefine and reconceptualize the nature of security guarantees at the physical, architectural, and system design levels. A fundamental shift is needed to ensure that these guarantees are more robust and effective in mitigating threats.

However, it is important to acknowledge this is a unique moment in time to realize new systems. The evolution of cyber-security technology, recent advancements in hardware and system architectures—coupled with significant investments by both government and industry—present an opportunity to reimagine the foundations of hardware and systems. These advances enable a new science of system security, starting from hardware-based guarantees, e.g., memory protection, hardware-assisted control flow integrity, and isolation primitives, that enable secure system designs that will provide new capabilities in current and emerging systems architectures, e.g., virtualized environments, serverless computing, and cloud architectures.

The field of research emerging from this theme will bring together scientists and practitioners at the intersection of hardware architecture, operating systems, programming languages, formal methods, and cyber-security. Collaborative efforts will aim to develop new models, designs, and methods that enhance existing computing architectures. They will also focus on creating new operating systems and software designs that leverage these guarantees to establish secure foundations. Additionally, strategic approaches for transitioning these technologies to the public and private sector will be identified.

Success in this field will yield groundbreaking technologies that lay secure foundations for future systems. By bridging the gap between hardware and system design, researchers will pave the way for more robust, secure, and privacy-preserving computing environments. The collaboration of experts from diverse domains will contribute to advancements in computing architectures, enabling the development of secure systems that can withstand the ever-evolving landscape of cyber threats. Ultimately, this research theme holds the potential to revolutionize the security landscape, creating a future where systems are built on secure foundations.

## 2.2   The Needs and Opportunities of Robust Generative Models

Research across a wide range of scientific endeavors—particularly in machine learning, AI, and data science—over the past decade has led to significant new capabilities. The most disruptive recent technology with immense power and vulnerabilities yet to unfold are formed by generative AI models (also dubbed as "foundation models"). These models are constructed by learning a significant portion of the digital data generated by humanity (which in itself has raised notable concerns over the data ownership, trustworthiness, and privacy of the distributed clients). However, the community's understanding of the security issues and susceptibility to adversarial attacks or unwanted behavior for large foundation models is at present limited. Here, the super-sophisticated gigantic models bring upon new issues that are far beyond the better understood vulnerabilities of the conventional deep learning models.

Society and the technical community are at a point of uncertainty surrounding the new capabilities of these models. Skeptics of technology have pointed out potential negative consequences of these models, with accompanying adverse impacts on education, markets, professional careers, human-interactions, and democratization. At the same time, others cite the immense potential of this disruptive technology in the betterment of human lives. Regardless of one's perspective, it is clear that recent progress in this technology is irreversible. If revolutionary machinery can be developed in a trustworthy way, it will be a wonderful assistive technology with paradigm-shifting impacts on life. In short, these new technologies have the potential to change life on this planet.

Therefore, new research is needed to ensure safe and responsible adaptation of generative models. The scientific community must mount an interdisciplinary, multi-pronged approach to ensure trustworthiness, safety, security, and privacy of the foundation models. On the one hand, issues that have been identified earlier for AI-based models take on new forms and challenges due to the size, use-case, and capabilities of the foundation models. Efforts over the past decade have sought to address adversarial examples (samples), data and model poisoning (a.k.a., backdoor/Trojan), learning and retraining in malicious environments, data integrity, data confidentiality, lack of generalizability, output bias, and output integrity/trustworthiness.

Research on foundational models encompasses several critical areas spanning many different academic disciplines, several exemplars of which are highlighted here: (a) IP generation and protection are crucial in addressing the rising number of plagiarism cases involving automated content generation across various media types. Existing detectors often fall short, requiring advancements to detect and protect against such misuse while safeguarding the IP of data owners and model generators. (b) Impersonation attacks by fake avatars pose a significant threat, with deepfakes being utilized for scams and frauds. Developing effective methods to detect deepfakes and countering attacks on deepfake detectors are pressing research questions in this domain. (c) Bias in future generated data introduces concerns regarding biased content generation and unfairness. This includes the potential for biased data insertion to influence model responses and transfer biases across models and generated data, creating new vulnerabilities. (d) Magnified privacy concerns arise as personal statements, essays, and legal documents are processed in platforms like ChatGPT4, presenting challenges in unlearning and ensuring data privacy. (e) The fast adoption of foundational models by businesses without adequate consideration of limitations and inherent randomness brings forth risks. Understanding and addressing these risks is essential to prevent overreliance and mitigate potential damages. (f) Explainable security is another important aspect, as current models lack transparency in their generated responses. Research into providing probabilities and parallel surrogate models for security can address this challenge, despite the immense size of contemporary foundational models. (g) The lag in policy development poses a significant concern, as the rapid development and adoption of foundational models outpace the establishment of appropriate regulations. Urgent investigation into policies capable of addressing potential damages is necessary to keep pace with technological advancements.

Foundational models open exciting new avenues to securing computing, communication and cyber physical systems. For example, automated methods based on foundation models can be used for finding software or hardware bugs, static/dynamic program analysis, composing and compiling safer programs, robust co-design of hardware/software systems, and even identifying the security threats of AI based systems. The vast number of potential applications call for more funding to materialize the potential of the emerging generative AI technologies on safety and security of our modern world.

## 2.3 Security in the Era of Indistinguishable Synthetic Artifacts.

The recent advancements in artificial intelligence, specifically in the realm of generating synthetic speech, images, videos, and text, have reached a point where these artifacts are virtually indistinguishable from real human creations. While these capabilities hold great potential for benefiting society, such as aiding the disabled and supporting under-served communities, they also pose significant risks and challenges to cybersecurity. Adversaries can exploit these technologies to carry out novel and highly harmful attacks, leveraging virtual humans with realistic appearances, speech patterns, emotions, interests, and life stories.

Such adversaries represent new risks for several reasons, including: (a) *Persistence and trust*: Virtual humans can maintain long-lasting relationships, building trust over time, which can later be exploited for malicious purposes. Their persistence enables them to establish a deep level of trust with individuals or communities, making it difficult to detect their malicious intent. (b) A*bility to target attack*: Adversaries can tailor virtual humans to target specific victims, leveraging shared interests, biases, and backgrounds. By customizing their interactions, these virtual humans can manipulate individuals more effectively, enhancing their ability to deceive and exploit trust. (c) *Scalability*: The ability to scale these automated attacks is a significant concern. As technology advances, it is plausible that a substantial portion of social media interactions could be generated by virtual humans, posing challenges in distinguishing between genuine and synthetic entities.

Broadly speaking, the emergence of indistinguishable synthetic artifacts generated by AI presents both opportunities and challenges for society. While these technologies can offer significant benefits, they also introduce new avenues for adversaries to carry out sophisticated attacks. To address these threats, cybersecurity research must prioritize the detection and identification of virtual humans, develop preventive measures, establish legal frameworks, and educate users. By fostering collaboration between academia, industry, policymakers, and the public, we can work towards ensuring the responsible and secure integration of these technologies into our society.

Research in this area should, at least, include: (a) *Detection and dentification*: Cybersecurity research needs to develop robust methods for identifying virtual humans and differentiating them from real users. Techniques such as machine learning, natural language processing, and behavioral analysis can be explored to detect anomalies and patterns associated with synthetic artifacts. (b) *Prevention and mitigation*: Efforts should focus on developing preventive measures to minimize the harm caused by virtual humans. This includes exploring methods to detect and counteract manipulation tactics employed by these entities. Technical solutions like AI-based defenses, reputation systems, and anomaly detection algorithms could play a crucial role in safeguarding against such attacks. (c) *Societal norms and legal frameworks:* Collaboration with policymakers and legal scholars is essential to establish appropriate societal norms and legal frameworks to protect the public from the potential risks associated with synthetic artifacts. This includes defining regulations, guidelines, and ethical standards to ensure responsible usage and mitigate the potential for abuse. (d) *User education and awareness*: Promoting user education and awareness is vital to equip individuals with the knowledge and skills to identify and protect themselves from synthetic artifact-based attacks. Empowering users through cybersecurity education, promoting critical thinking, and creating communities of interest focused on cybersecurity can contribute to a safer online environment.

## 2.3 Responsible Cyber-Security: Privacy, Accountability, and Societal Impacts

In the ever-evolving landscape of cybersecurity, it is crucial to recognize and address the inherent tensions between privacy, utility, accountability, and societal impacts. As online services and communication increasingly shape systems, infrastructure, and society, the security and privacy community must adapt and develop techniques to mitigate the effects of technology on the physical world and its inhabitants. This theme encompasses various research areas, including combating misinformation, propaganda, and hate speech on social media, establishing provenance tracking for digital artifacts, redesigning the internet for proper attribution and filtering, addressing social needs, and minimizing the environmental impact of technology.

Misinformation, propaganda, and hate speech have profound effects on society. They can manipulate public opinion, sow division, and erode trust in institutions. By spreading false narratives, these harmful elements distort reality, hinder informed decision-making, and potentially incite violence. Addressing this issue requires innovative cybersecurity research to develop effective strategies for detecting, mitigating, and countering the spread of misinformation, propaganda, and hate speech on social media platforms.

Another critical area of exploration is the development of techniques to track the provenance of digital artifacts, such as images, videos, and communications. With the rise of synthetic content and deepfakes, it becomes increasingly challenging to distinguish reality from fabricated or manipulated information. By establishing reliable provenance tracking mechanisms, we can safeguard users from abuse and bolster defenses against the proliferation of misinformation. This research is vital for maintaining trust in digital content and ensuring that individuals can make informed decisions based on authentic information.

The structure of the internet itself may require redesign to accommodate the attribution and accurate filtering of network traffic. As we envision a future with billions or even trillions of interconnected devices, managing network environments efficiently and securely becomes paramount. By developing new techniques, such as advanced attribution and filtering mechanisms, we can enhance cybersecurity, protect against network-based threats, and enable the effective management of vast networks, devices, and data flows.

Beyond technical considerations, security and privacy research must also address the social needs associated with technology. Equitable access to technology should be a priority, along with protecting vulnerable communities from abuse and discrimination. Cybersecurity initiatives must actively work towards fostering inclusivity, fairness, and social justice, ensuring that the benefits of technology are accessible to all and that vulnerable populations are not disproportionately harmed.

Moreover, it is crucial to recognize the environmental impact of technology and incorporate sustainability into cybersecurity research. Computational systems must be equipped with the capability to accurately measure, manage, and report their environmental footprint, including energy consumption and material impacts. By promoting environmentally conscious design and minimizing the negative consequences, such as the energy-intensive nature of blockchain mining, the security and privacy community can contribute to a more sustainable technological future.

In summary, work in relation to this theme should emphasize the need to navigate the tensions between privacy, utility, accountability, and societal impacts. By addressing issues related to misinformation, propaganda, and hate speech; establishing provenance tracking mechanisms; redesigning the internet for enhanced attribution and filtering; considering social needs; and promoting environmental sustainability, cybersecurity research can contribute to a safer, fairer, and more sustainable digital landscape.

# 3    RESEARCH OPPORTUNITIES

## 3.1    Systems Security

In the ever-evolving landscape of computing, ensuring the security of systems has become a critical imperative. This section delves into the multifaceted realm of systems security, encompassing a range of interconnected domains including, for example, security policy, distributed systems security, hardware security, and web security. By examining the challenges and opportunities within each area, we aim to unravel the complexities associated with protecting the foundation of computing systems. From establishing robust security policies to defending against sophisticated attacks, securing distributed systems, fortifying hardware components, and safeguarding web applications, this section explores the multidimensional aspects of systems security and offers insights into the cutting-edge research and practices driving its advancement.

### 3.1.1    Security Policy

**Area Description**

This topic area, Policy Based Security, encompasses all aspects of policy control and governance of secure systems.  As secure systems are pervasive in modern life, policy-based security plays a vital (if often hidden) role in everyday society.  There are several broad sub-areas which should be foci for policy-based security over the next several years:

*Supply chain security*: Policy issues in supply chain security management: Modern applications are a composition of many third-party libraries. It is important to understand what permissions a specific library/module should have in the context of the application where it is used. For example, how should security policies and formal methods research apply to Software Bill of Materials (SBOM) components. This may drive research at the intersection between policy-based security and cryptography.

*IoT/Distributed coordinated system*: Distributed coordinated systems (e.g., IoT-Cloud) are associated with different entities with various semantics, capabilities, and rules to coordinate access control policies ensuring the security and privacy for all entities in the system. This requires research to study the consensus foundation among different entities in the distributed coordinated systems (devices, organizations, geo-political borders), how to design efficient formal method analysis satisfying the security and privacy requirements of different entities.

*Security policy verification*: Once secure policies are designed, they must be verified to ensure that security postures are properly enforced. Applying and improving both formal methods and falsification solutions can occur in the verification of security policies.

*Auditing/proof of security policy enforcement*: Security policies must be implemented in software (or even hardware) for them to take effect on real-world systems. Even the most secure set of security policies may be vulnerable or by passable if their actual implementations are vulnerable. Therefore, one research area in policy-based security is the semi-automated and automated auditing of the implementation of security policy enforcement. A related research area is providing users with semi-automated or automated proofs that certain policies are being enforced to gain trust from the users (and the general public).

*Usability*: Although security and privacy policies in systems and software are provided to users, they are obfuscated with law jargon and hard to understand to the layman. There is a need to make security policy language more accessible to the general public. Possible remedies can include standardization of policies, determining granularity of user control (so they are not insensitive to options - push a button to make pop-up go away vs not able to work), and including readable security stickers on devices, as part of software systems, applications etc.

Another aspect is organizational and corporate policies on data, access control, and monitoring and how that relates to employees' personal devices, activities and right to privacy.

**Technical Efforts**

**Composability.** An important feature of policy languages is *composability*. Some desirable features for policy languages include compositionality, explainability, and clarity about the assumptions made by underlying formal models. Modern systems may have many stakeholders with competing priorities and security policies; for instance, on the web the user, website, browser manufacturer, and many third parties (advertisers, analytics services, etc.) are all participants. Compositionality may also arise when technology is available internationally, or when collaboration happens within an organization. Having composable policy languages allows all stakeholders to combine their own policies in meaningful ways but requires policy conflicts to be resolved and communicated. Conflict resolution is an area where more research can be done, and especially how conflicts are communicated and the potential risks of the new policy (if any).

**Architectural and policy co-design.** A second technical effort involves the opportunity for *architectural and policy co-design*. These fields are often separate and regard the other field as immutable. Prior experience shows that breaking down these silos can result in real steps forward. For example, security policies applied to networking struggled for years to tame the challenges of middle boxes, routers, and end systems, each with separately configurable mechanisms that need policy control. However, the architectural change of SDN (Software Defined Networking) suggested replacing these many mechanisms with a central manager. The result was a wealth of new, centralized policy approaches that were effective at taking control of the distributed elements that make up a network, and eventually widespread deployment and a new multi-billion-dollar industry in SDN. Our question: is there an opportunity for similar architecture rethinking in other areas that could produce similar outcomes? As one potential direction, today's web-based permissions and cookie management feel burdensome to users and somewhat ineffective at protecting privacy–is there a different architecture that would improve this situation?

**Usability.** The third technical effort is *designing for usability*, which applies to multiple domains including policy language, formal methods, and permissions. Policy languages, as they stand, often lacked clarity and conciseness in both the requested permissions as well as the resultant data collected and how it is used. (For example, GDPR-motivated cookie consent in browsers and web pages are both too intrusive and not as effective as preferred.) Further, the frequency of requests to the user (e.g., every time they visit a new web page), risks desensitizing users to the impact of their agreement to data collection and usage. Efforts in this space should be catered towards creating policy language that is more understandable and informative to facilitate policy-writing, even without technical expertise, while also being less burdening to the end user. The same recommendations apply to other types of policies, like application permissions on the Android platform. Additionally, efforts should be directed towards ensuring policies are informative about how the requested information will be used. Finally, in the case of violations to the agreed upon policy, stakeholders should be informed, in clear language, what the incurred risks of such violations would be.

**Formal methods.** There is continued potential to apply *formal methods* to prove properties of policy composition and enforcement. Such methods have the promise of providing guarantees about outcomes in implementations or compositions. However, the long-term challenge remains in understanding the assumptions of formal methods, how closely they match the real system, and how well they are understood by users or policy writers. Potential applications of formal methods include policy verification, conflict resolution (for composition), and assurance that data is used the way it is intended. Efforts to make these formal methods usable to non-experts would also better allow subject matter experts to verify policies over their own systems.

**Technical and Societal Impacts**

The technical impacts for each of the areas identified under include:

*Composability:* Scientific success in improving the composability of security policies would likely enable better composition of other technical components (since their security policies would be composable). Improved composability might also lead to be better integration of security system and the policy-oriented aspects of civil society.

At best, the impact of *architectural and policy co-design* can be a dramatic simplification of the problem, and resulting in lower cost, greater functionality, or better performance. In SDN (Software Defined Networking), this simplification has resulted in widespread adoption of the technology by multiple companies and the creation of a new multi-billion-dollar industry.

Usability, as it applies to policy language and permissions, will see substantial impact on end users and how they interact with online systems. The lack of clarity surrounding information collection, usage, and risks, as well as the frequency of these unclear requests, has left users desensitized to information agreements. Success in efforts addressing these issues would substantially lessen cognitive burden on end users and enable better mechanisms for informed consent.

Finally, the impact of improvements to *formal methods* are the ability to have certainty in the outcome of our systems. An even greater impact is the potential for formal methods to be usable to non-experts, allowing much broader use.

**Cross-Discipline and Outreach Opportunities**

The area of policy-based security connects to many other disciplines, both within and outside of computer science. For example, policy-based security overlaps strongly with CS research on usability and usable security as security policies are not functional if they are not usable and understandable by their intended communities. Policy-based security also overlaps with other policy-oriented sectors of civil society, such as public policy, law (domestic and international), law enforcement, and national security. There is also an overlap with psychology (ties into usability) and medicine (which provides many examples of "break the glass" use cases where in an emergency a policy override is necessary to administer life-saving aid). The resulting security improvements that policy-based security leads will benefit multiple industries, including smart buildings, automobiles, etc.

### 3.1.2   Storage Security

**Area Description**

Storage security can be defined along two rough dimensions: consumer and enterprise. In both, security questions surrounding data confidentiality (e.g., encryption), access control, durability/recovery in the event of compromise, and more, and these issues extend to considering local data versus data stored off-premises, particularly in cloud environments.

Efforts in storage, in particular storage security, have led to mature systems that address many confidentialities, integrity, and availability (in the short, medium, and long term) needs. For this reason, much of the future work in this area may be driven by contextual needs (in clouds, service-oriented systems, within massive scale data farms). Thus, most of the future work in storage security will be conducted in concert with other fields described throughout this document. Some areas of interest that are more general are below.

**Technical Efforts**

**Auditability.**   A major research problem to study in this area is the auditability of stored data and the auditability of the security of the stored data. In short, auditability is about ensuring -- in an economic manner -- the storage hardware and providers provide the promised service level agreement. Prominent examples include ensuring the integrity of data that is stored in cloud data storage, ensuring the data is properly encrypted in hardware-based storage (which is obviously not the case with many SSDs that are equipped with hardware-accelerated data encryption), and ensuring the availability of data backups.

The auditability of secure storage is closely related to recent progress in the field of cryptography, especially homographic encryption (which will enable auditability of data stored in the cloud without (a) revealing the full decrypted content to the cloud provider, and (b) transferring all the stored data from the cloud storage provider to the user who wants to audit the data) and secure.

multiparty computation (which will enable auditability of data without accessing the entirety of the stored data).

**Support distributed databases / object storage.** Challenges in implementing and supporting distributed databases also exist. The goal is to enable better accessibility, however, storing data at multiple (e.g., geographic) locations also opens a wider threat surface. Mitigation for such attacks can be on the storage security side but blends heavily with network security.

**Storage in emerging system architectures.** More remotely, issues pertaining to private storage (oblivious RAM), secure computing, and secure in-memory computing could have connections to storage security, although main challenges must be addressed in other areas.

**New storage mediums.** Although this may involve more electrical / material science / mechanical engineering, new storage mediums with unique physical properties will provide new building blocks for novel secure storage mechanisms, e.g., quantum storage and DNA-based storage. For example, a high information-density system that has a physical write-once, read many (or read once) properties could provide higher and more robust security assurances.

**Consumer cloud storage.** Consumer cloud storage systems often perform client-side encryption. This makes applications fundamentally distributed across user devices, and therefore requires secure storage of data that is replicated and synchronized. There will be emerging problems here in terms of compression plus encryption, traffic analysis type issues, logical vulnerabilities due to manipulation by adversarial platforms, and more.

In another vein, there are concerns that current cloud infrastructures lack guarantees for individuals maintaining access to their data (a key availability need). Lockouts by abusers or other adversaries that gain access to an account and prevent the legitimate owner of the data from regaining access to their data is an acute problem, and one for which mixed policy plus technical mechanisms are needed to help resolve. (e.g., legislation requiring companies to have processes in place for regaining access to accounts.)

**Distributed access control.** Another area of effort that is needed is to develop privileged access control mechanisms for consumer cloud storage systems like enterprise storage systems. For example, some data is more valuable and access to such data should be more privileged, i.e., not by the same password/access to the account. The research question is how to automatically classify user data as privileged (e.g., via a privacy-preserving ML-based model) and control its access via more secure but easy-to-use/transparent manner (e.g., muti-factor, additional biometric, etc.)

Object-level security, where fine-grained encryption is applied at the object rather than the storage medium level (e.g., individual ML model / file vs. the whole hard drive or filesystem), in conjunction with hardware accelerators for said types of objects is an interesting direction for future research. Combined with appropriate hardware and access control / key management mechanisms, this type of storage security has the potential to protect against several classes of attacks that manage to operate within the security perimeter (e.g., through exploiting a software flaw), as well as protect intellectual property (e.g., custom ML models) in the context of outsourced and/or cloud computation.

**Power aware storage security.** Advances such as processing in memory and processing in storage can allow for more efficient and power-efficient processing. These mechanisms can also allow for security mediation of data closer to where it is being stored, but these mechanisms could potentially be vectors for attack as well such as tampering and data leakage.

**Non-destructive writes.** Enterprises are unlikely to be able to store all data in a nondestructive way, i.e., so that future updates are appended to an update log, instead of overwriting the data being updated. However, nondestructive writes are likely the most general protection for ransomware and denial-of-service attacks. As such, automatic mechanisms to identify critical data that should be written nondestructively (with other data being written conventionally) would be highly beneficial.

### Technical and Societal Impacts

Successful research in this space will enable better privacy protections for individuals and increased assurance for enterprises and large organizations. Ransomware attacks are one of the most common and destructive malware variants and assuring protections against these attacks that can be based on secure storage mechanisms can mitigate their harmful effects.

Auditability of secure storage helps guarantee and improve users' trust in storage hardware and cloud storage service providers. It will also help alleviate emerging cyber threats, such as ransomware attacks and insider threats to critical data assets.

### Cross-Discipline and Outreach Opportunities

The field of storage security presents various cross-disciplinary and outreach opportunities that can be explored to advance research, education, and practice. Some of these opportunities include:

*Industry partnerships*: There is a growing demand for storage security experts in various industries, such as healthcare, finance, and e-commerce. Collaboration between academia and industry can facilitate knowledge transfer and industry-relevant research.

*Public awareness and outreach*: As data storage becomes ubiquitous, it is essential to raise public awareness of the risks associated with storage security. Outreach activities such as public talks, workshops, and cybersecurity awareness campaigns can help educate the public about storage security risks, best practices, and emerging trends.

*Policy and regulation*: Storage security policies and regulations are crucial in ensuring data privacy and security. Cross-disciplinary collaboration between storage security experts and policymakers can lead to the development of effective policies and regulations that balance security needs and privacy concerns.

### 3.1.3   Distributed System Security

**Area Description**

Computing systems are becoming more and more distributed as well as more complex—and such trends are likely to accelerate in the future. The determination of what is or is not a distributed system has become itself murky. One aspect that may help is scoping the problem on an algorithmic level: those algorithms that specifically rely on distributed systems. Another aspect to scope the problem can be on the technical architecture, e.g., ad-hoc, service-oriented, peer-to-peer or client-server architectures. However, even within these well-scoped sub-problems, distributed systems will emerge as complex behavior of the interactions between systems and are named emergent distributed systems. Understanding and securing these systems will be challenging, as the security of one node or system will affect the security of the whole system.

**Technical Efforts**

**Emerging distributed systems security.**  Distributed systems have heterogeneous and emerging computer architectures, and they have different security requirements, attack surfaces, and security primitives. For instance, GPUs/FPGAs in a single computer can be considered as a distributed system. They have different hardware architectures, software and hardware side-channels, and root-of-trust. These configurations and uses require unique security and privacy analysis and guarantees.

**Denial of service detection and mitigation.** One of the most interesting challenges in distributed systems will be detecting and mitigating distributed denial of systems attacks. As users become more networked, the stability and availability of such systems will be critical. Studying novel attacks and designing systems that are resistant to DoS will become important. Note that there has been a sharp increase in the number of DoS attacks against distributed systems over time.

**Zero-trust architectures.** An active area of research is to support distributed authentication and attestation, without presuming trust between nodes in the system. Zero-trust architecture (ZTA) has recently been talked about a lot. Next generation distributed systems may adopt the "do not trust, always verify" principle when it comes to authentication and attestation.

Moreover, Zero-trust architectures are becoming the default authentication approach in distributed systems. It is non-trivial to properly authenticate users and devices in a non-home environment (e.g., the SaTC2.0 workshop :-)), so that they can work together in a secure and trusted fashion in specific semantic contexts of the distributed system/environment. The problem becomes even more challenging, when there exist cyber-physical systems with cyber components, physical components, and human users.

**Security and privacy in emerging distributed systems architectures.** Distributed systems are diverse and heterogeneous using a variety of layers, programming models, and interfaces between the underlying components. This provides a large attack surface. An attack or failure in one or more components may have a severe impact on the whole system. Research needs to determine concepts and mechanisms that provide proactive isolation rather than known static system isolation that are not able to confine increasingly complex IT systems.

**Observability.** Distributed system security of the future must leverage better observability. This becomes a design challenge and research issue – how to make complex distributed systems more observable. This makes security analysis possible.

**Trust.** Extending trusted computing notions to larger units of distributed infrastructure is an important challenge. How does it work for complex applications and with multiple interacting parties? How do device building blocks play into a larger notion of robust trust.

**Auditing.** Tomorrow's distributed systems should also feature auditing features – an issue that is often omitted in research. How can the system prove security or privacy to operators and auditors. This is a distributed systems design point that needs attention at every level – hardware, systems software layers, data, protocols, and more.

**Data provenance for distributed systems.** With so many entities that are collecting and analyzing data, Data provenance is a big issue. Particularly with a heterogeneous network with a diverse set of devices that integrates physical and cyber components. How to ensure trustworthy provenance across the network in a way that becomes amenable for higher level functions (e.g., attribution)? Further how such provenance is represented and verified in complex, heterogeneous systems is an open question.

**Performance.** A key question when considering solutions that attempt to address security properties of distributed systems is in performance. This is particularly stark when attempting to implement a "zero-trust" architecture. Many of our modern performance improvements, and the direct result of being able to leverage trust in a component, and when this trust evaporates, so too does performance. For example, the Spectre and Meltdown hardware mitigations reduced performance of all systems that implemented them 10%--15% (NOTE: these numbers are from memory, look them up whoever is reading and using this text). Therefore, in a complex distributed system, consideration must be given to performance and how to achieve a system that is usable by end-users while achieving our security goals (such as zero-trust).

**Interactions between media performance and security.** Many distributed systems can be categorized by the *communication latency* inherent in the system. For instance, client-server architectures are bound by the client's internet connection (which is typically the bottleneck). As society reaches for the stars, a key question will be how the nature of the security of distributed systems changes as the communication latency increases. How will modern distributed systems scale with a space-level communication latency? What new security challenges will arise due to orders-of-magnitude increases in communication latency of distributed systems? In addition, satellites often have very infrequent communication windows, which can significantly increase the latency for a request and a response. An example that was given was the technical challenge of how to revoke a PKI certificate for a satellite that has very infrequent and short communication windows.

In addition, space offers interesting geo-political challenges compared to other aspects of computing. Particularly for satellites in Low Earth Orbit, there are many such satellites, controlled by multiple entities. In addition, it is unclear what the legal requirements are for communication between satellites as well as communication back to Earth.

**Attribution.** In distributed systems, *attack attribution* (i.e., figuring out where the attackers are) will continue to remain an interesting and important research problem. For example, can research create automated tools that analyze complex attacks, and that can tell us who (e.g., which groups, which countries, etc.) are launching them?

**AI support for distributed system security.** One key direction to many of the technical challenges may reside in effective use of AI. An "AI power tool set" that can help security operators to make sense of how the system is performing under adversarial scrutiny and attack. AI tools can help stitch distinct data domains that humans do not do well naturally, identify patterns in data that are hard to decipher by human or manually driven tools (e.g., elastic search) and the leverage the powerful combination of cyber-human reasoning.

**Detecting and mitigating adversarial campaigns.** A large-scale distributed system (e.g., a social network) may potentially become a favorite "host" of large-scale, low-and-slow cyber-attacks/campaigns, due to its connectivity and (relatively) uniform software/hardware. It is important to proactively identify potential threats and unconventional attack scenarios (e.g., stealthy APTs) that a subject distributed system might (inadvertently) help foster.

### Technical and Societal Impacts

Distributed systems security will continue to have an important impact on legacy systems. Many systems out there were built many years ago and are increasingly being connected to the Internet. If these systems can be secured, the overall security of the Internet will increase.

Distributed systems security research also has a direct impact on topics such as the underground economy. Studying how criminals build their systems and earn illicit profits is of interest from a research point of view, and important. Distributed systems security is impacting not only system builders, but also a diverse set of people working in different domains (e.g., law enforcement).

**Cross-Discipline and Outreach Opportunities**

Existing research shows one out of every five enterprise users falls victim to spear phishing. In addition, cyber criminals can conduct even more sophisticated attacks after collecting and correlating victims' sharing and information on multiple social networks. Therefore, the end user is the key, and educating the end users is important for all stakeholders.

An example of a combination of legal and regulatory problems is evident in IoT devices. A recent new example showed that law enforcement subpoenaed Ring for access to a neighbor's Ring camera. These recordings are stored on Ring's servers, due to their technical architecture, which is a distributed system. The data provided to law enforcement by Ring, included recordings from *inside* the neighbor's house—which was not material to the case and violated the privacy of the neighbor. This example demonstrates how technical solutions (the architecture of Ring's distributed system) combined with legal and law enforcement to cause a surprising situation for the end-user. The end-user did not expect that law enforcement would be able to access their internal home Ring camera recordings for an (to them) unrelated investigation.

A key question in distributed systems is who owns the data and who is responsible for the protection of that data. Is it the user that owns their data and is the only one that can access it? As demonstrted from modern systems, this is not the case: the application owner typically can access the data in addition to the users (as well as others).

With the increasing deployment of small satellite systems, or CubeSats, in Low Earth Orbits (LEO), there is a need to develop a reliable and secure computing model for digital infrastructure in space. The space context presents new considerations for network communications and maintenance models, as communication among space segments and between space and ground segments would have different latencies and communication throughputs. In addition, legal and regulatory implications surrounding digital infrastructure deployed in space are not yet fully settled, requiring new considerations. Moreover, there is a need to propose new threat modeling and attack vectors for computing models in space. Addressing these emerging challenges requires interdisciplinary efforts encompassing researchers from various disciplines including aerospace, material science, nuclear physics, and electrical engineering researchers. Finally, there is a need to renew education and training materials as it is essential to educate end-users with up-to-date security contexts.

### 3.1.4   Operating Systems Security

#### Area Description

Operating systems (OS) provide the foundations of computing across devices, from embedded system and vehicles to cloud computing environments, Software applications and environments also provide OS-like functionality and require similar protections, such as web browsers and smart contract interpreters. Assuring the security of operating systems is a field of research that dates back 60 years and while foundational concepts of secure systems remain relevant, looking forward to how new technologies such as new hardware architectures and hardware-based protection mechanisms will be used, the continued increase of Internet-of-Things devices, and the increasing role of AI in OS configuration and potentially as a major component of the OS, demonstrate that OS security is an area that will require significant research efforts over the next decade and beyond.

OS security should consider many different OS types based on hardware platform and contexts. Some of these include lightweight IoT OSes, mobile device OSes, client computing OSes, richly provisioned cloud server OSes. Some hardware platforms are used in special contexts like robotics (ROS) or autonomous driving. What security features are needed for each?

OS security research should address not only attack prevention but resiliency. Complex OSes will unavoidably have vulnerabilities over time but has enough been done to think through containment strategies and the broader management of attacks across large user and device populations.

OS security research includes the topic of virtualization which, like OSes, looks at the system software layer between hardware and application software. In this case, their system management layer lies between hardware and multiple operating systems and shares many of the same research issues. Containerization within an operating system is likewise in scope.

#### Technical Efforts

**Robustness and pervasive vulnerabilities**. Modern operating systems are incredibly complex systems that have evolved over a substantial part of the life of computing. Due to this nature, there are a plethora of security vulnerabilities that are found every year in modern operating systems. Therefore, instead of playing "whack-a-mole" and fixing each of these individual vulnerabilities, the security of the OS must be considered. There must be the assumption that there are and will be vulnerabilities found in the OS and research directions that explore how to operate in such an environment, which can be thought of as robustness. Research should attempt to either mitigate the damage done by security vulnerabilities (by either eliminating exploitation or making exploitation very difficult) or redesign the OS in such a way that a vulnerability discovered in one component does not impact the security of the whole system. These approaches should consider both legacy OSes as well as new greenfield designs.

**Formal verification of mainstream OSs.** Despite the recent successes in formal verification technology in OS verification, it may be reasonable to conclude that it may not be possible to do push-button verification of mainstream OSs. The interesting questions are therefore about whether meaningful assurances about component subsystems of OSs can be provided, and what that may entail. For instance, providing immutable provenance / audit information would be of high interest to attack detection, attribution and response, and any efforts towards assurances that the audit system will be resilient will be welcome. Particularly of interest are techniques that go beyond the realm of traditional formal verification (e.g. the use of guardrails around subsystems that cannot be verified easily). In short, more "practical formal methods" in bringing the technologies to mainstream OSs would be welcome areas.

**Trusted Computing Base.** The discussion of Trusted Computing Bases for OSs has evolved as with various developments over the last two decades on virtualization and hypervisors. Currently, a stage of evolution exists where large-scale containerization as a way to scale applications to cloud driven deployment, and questions of how an application could escape the container to impact the OS (and other containers) have resurfaced. The broader questions in this context are about what assurances can be provided about resiliency to attacks on containerization -- possibly involving formal assurance techniques as mentioned above, or by way of runtime techniques or some combination thereof.

**Attack surface of new OSes**. As applications and hardware evolve, new OSes will be designed such as domain-specialized OS, database-based OS, serverless OS, and distributed OS. Thus, one critical research direction is to study the attack surface of these new OSes, especially new attacks to them.

**Security-oriented re-design of OS**. Existing OSes are mainly designed to optimize performance, not security. This group thinks it is critical to re-think the design principles of OS from the perspective of security. For instance, how to redesign OS to prevent vulnerability exploits, be resilient to module failures, and make formal verification easier and more scalable.

**Software security techniques to OS/hardware**. A key aspect that materially and measurably improves the security of modern operating systems is when hardware features are created that help the OS to provide stronger security guarantees. Hardware support here is important due to performance improvements. However, the hardware features that are needed are usually first demonstrated in software (often all the way at the application level), then implemented in the OS, and finally in hardware. An example of a success story here is control-flow integrity, which was first introduced in 2005 (TODO: approximate time from brief search) as a software implementation with a ~50% overhead in some cases. 15 years later (2020), Intel introduced control-flow integrity into their hardware, and still, as of 2023, Linux has not widely deployed support for this hardware feature. A goal of SaTC should be to create a pipeline to foster research into security primitives that can be translated into hardware. By creating teams consisting of researchers that are experts in software security, OS security, compilers, and hardware, perhaps the timeframe from idea into practice can be shortened, thus improving innovation in this space.

**Protection against side channels.** With the advancement of hardware design, there can be new vulnerabilities coming from side channels that may affect OS. How to systematically prevent side channel attacks is worth investigating. Previous solutions mainly lie in detecting and coping with each individual side channel attack and are thus ad hoc. With the support of machine learning technologies, it is important to derive systematic ways to detect and protect against side channels.

**Protecting small OS on mobile devices**. There is an explosion of mobile edge devices (e.g., smartphones, wearables, and AR/VR headsets). The OS running on these mobile devices is small confined by the resource-constrained hardware support. These small-form OS pose unique security vulnerabilities. One interesting research direction for the future is to understand the security vulnerabilities and provide light-weight security solutions to these small OS.

**Use of Trusted Execution Environments.** An important area of OS security research is trusted execution environments (TEEs). OSes of the future could be modified to both use TEEs for improved security of the OS itself, and to better secure applications and services being managed the device. But how does this or could this work on both fronts? Really, OSes could be completely restructured to feature TEE protections and the extensive use of attestation proofs throughout. Note that there are an increasingly large number of TEE technologies offered by hardware vendors. Each of these technologies needs exploration in the context of OS security research.

**Virtualization**. Virtualization enables the multiplexing of hardware resources across multiple OSes. With this, however, comes many security issues as hardware resources may or may not be designed to support multiple users. For example, a GPU may not support isolation between co-located tenants or may leak data through caching mechanisms. The broader picture of OS security following the evolution of future hardware should include considering security from a virtualization point of view.

**Protection against side channels.** How might OS techniques help with the problem of side channels like timing attacks, cache snooping attacks, and PCIe snooping and tampering. It could be that some types of hardware attacks are only preventable through software mitigations. But what are the techniques, and can they be offered as OS services or features? What hardware platforms could be helped by these techniques?

**Quantum Security.** A quantum computer, in principle, needs to support multi-tenant, multi-user operation and utilization. As resources of a quantum computer (e.g., circuits) are shared among multiple computations, there is a need for OS-like scheduling, supervision, and isolating computations from each other. Faulty or incomplete isolation creates an opportunity for attacks and misuse. There is a need to develop OS-like capabilities that will provide security and isolation among computations in a multi-tenant use of a quantum computer.

**Formal verification of security properties.** Machine-checked formal methods potentially offer highest attainable assurance of software security properties but are currently difficult to apply to OS code that ranges from high-level, application-facing services to lowest-level, hardware-facing machine code. Future research must address this gap by discovering new low-level (e.g., "bottom-up") methodologies for verifying mission-critical low-level code not expressible in high-level source languages, and connecting it to compiler-based (e.g., "top-down") formal methods evidence of assurance of source codes that incorporate it. Such research should prioritize scalability and automation technologies that facilitate code-proof co-development, so that proofs of OS security can be feasibly maintained alongside rapidly evolving OS implementations.

**Memory safety.** Memory-safe languages, such as Rust, may play a more important role in OS and OS module development. It is, however, unclear whether such languages are efficient and expressive enough to develop all OS modules. The cost of using such languages does not only lie in the run-time performance but also during the development stage.

**IoT operating system security.** IoT and embedded systems are connected to the Internet now and face the same cyberattacks our desktops, mobiles, and servers face. They, however, lack the security mechanisms that have been deployed on desktop, mobile, and server systems. For instance, the software systems on IoT devices largely lack privilege separate, stack canaries, address space layout randomization, etc. Implementing such mitigations on IoT also face unique challenges and opportunities because the hardware and ISA layers of IoT systems provide different security primitives than desktops, mobiles, and servers. Security solutions on such systems also need to take memory footprint and power consumption into consideration. Another research direction is how to retrofit security into an OS designed for a device with limited capabilities (e.g., no internet) that starts being used for something more. This may involve research at the compiler-layer that can automatically partition and secure legacy code.

**Operating system support for zero-trust systems.** One area of development and research should be to consider implementing the Zero Trust Architecture (or its principles) within the OS. Some attempts were made in the past to modularize the OS into mutually distrustful components that established trust among themselves only based on strong security attestations, for a given purpose. Given the attention to and work in the Zero Trust design, it would be worthwhile to revisit this approach and apply it within the OS.

### Technical and Societal Impacts

OS is a fundamental component of any computing system. A handful of OSes are widely used across the universe of smartphones and cloud computing, for example. Breaches in security impact the user community in far-reaching ways. The trustworthiness of operating systems enables using them in devices that people rely on for health and their lives.

OS security could impact hardware design specifications. Insightful re-designs of OS security could create new standards in peripheral interfaces, communication protocols, drivers, and more. Moreover, OS security services could make applications more secure (or less secure) by handling core security functions in robust ways (memory isolation, data encryption, secure storage, etc.)

If the scientific community fails to make progress in OS bug/patch analysis, test, and verification, the Linux community will not be able to address their known security risks timely. If successful, the Linux community will be able to secure their system in a more efficient and effective fashion.

If the scalable formal methods and language-based security assurances are not built and applied to the broadly defined OSes (such as browsers and smart contract interpreter), the OS security has to rely on the high-cost manual auditing which provides weak security guarantees.

**Cross-Discipline and Outreach Opportunities**

Increasingly there are areas outside of traditional OS development where OS system design is occurring. For example, web browsers evidence functionality very similar to traditional OSes, and IoT devices represent classes of devices that traditionally did not have computing and networking capabilities but now do. Best practices of OS security are required in these areas and other emerging areas such as AR/VR devices to ensure that these devices are secure.

An enduring challenge when developing trustworthy and secure operating system primitives is the risk that they can be used in ways that reduce user autonomy and control. Assuring a balance of autonomy and empowerment with computing devices while preserving security and safety guarantees is a critical ongoing research area, one that involves multi-disciplinary research into how devices are used and whether current OS design can be disempowering to populations. Outreach that helps to communicate to a broad spectrum of users what security guarantees they can rely on from their OS and their devices, and importantly what is not provided, can also provide increased autonomy and control for users.

Developing security for OSes has historically been an issue in tension with other demands such as power consumption. For example, non-interference can be gained by maximizing computation and bandwidth so that high and low subjects cannot be distinguished but these are extremely inefficient and in an era of climate change, developing trade-offs that assure secure computing without excess resource consumption will be critical for sustainability.

Opportunities exist within this space for collaboration with law scholars. Operating systems are appearing in systems that have critical responsibilities for human safety, such as autonomous cars, commercial aircraft, and medical devices. When they fail to meet expectations, users and our legal system demand greater accountability than for less critical computing systems, such as laptops and mobile phones. The ability to formally guarantee performance (such as in real-time operating systems) links to legal guarantees for that performance, and thus to liabilities when expectations are not met.

Opportunities also exist for collaboration with human-computer interaction research to effectively communicate permissions on devices, especially mobile and IoT devices. While permissions are implemented in the operating system, it is the responsibility of the user interface to expose permission functionality accurately and meaningfully to the user, so that they can make decisions that align with their privacy preferences. Looking beyond the status quo, sensors will become increasingly pervasive on computing devices and communicating permissions to users will become a greater challenge on devices that have limited use interfaces, such as smart speakers or Bluetooth beacons.

### 3.1.5 Hardware Security

**Area Description**

Securing the lifecycle of hardware systems, beginning with design and extending through deployment and use, is essential to the future of computing systems. At the beginning of the lifecycle, primary concerns are the correct specification of intended functionality; verification of the design to ensure that functionality (and ideally nothing more); integrity of the design and manufacturing process; protection of intellectual property; and verification of the result. In deployment, primary concerns range from properly informing users of the security and privacy features of the device; the discovery of unintended functionality (e.g., side channels) that might leak secret data (intellectual property, user data, etc.) or admit other forms of abuse; and modes of use (e.g., wearables) and the consequences thereof in particular applications.

**Technical Efforts**

**Specialized hardware for crypto.** Crypto techniques such as homomorphic encryption, secure multi-party computations are computation demanding. Although researchers have made great progresses in optimizing the designs of algorithms, there is still a gap due to performance overhead. Hardware acceleration for crypto is an emerging area that shows promising breakthroughs in recent years to further improve the performance. There have been some explorations in designing domain-specific hardware for crypto techniques such as homomorphic encryption, which shows significant speed up. Methodologies include optimal memory hierarchies, computation units designed for special crypto operations (e.g., fast-executing, and pipelined arithmetic units). There are still many research questions about how to optimize the co-design of hardware and crypto, and how to adapt the design for special domains such as machine learning, and how to protect the security of the hardware accelerators.

**Hardware assisted security.** Hardware plays a big role in providing security in computation. Computing is facing problems in providing speed and security at the same time. The increasing demand for machine learning models on resource-limited hardware makes the problem worse. For example, one interesting direction is to explore increasing the memory footprint of TEEs to provide secure executions while keeping the same hardware footprint. Another interesting direction is to provide light-weight authentication on implantable devices, critical to real-time and secure hardware computation.

More broadly, hardware-based primitives and abstractions can offer more efficient and tamper-resistant security measures than software-based solutions. In the past, a variety of hardware-based primitives have been developed, such as physically unclonable functions (PUFs) and enclaves. With emerging applications like virtual and augmented reality, wearables, and artificial intelligence, there is a growing demand for new hardware security primitives that can support these technologies. Therefore, the invention of new hardware security primitives is an ongoing necessity to keep pace with evolving technological advancements.

**Side-channel identification and mitigation.** Hardware-based security solutions have been found to be vulnerable to side-channel attacks. For example, trusted execution environments (TEEs) have been shown to be susceptible to various side-channel attacks, including those arising from speculative execution, cache timing, and even privileged software (known as controlled side channels). It is important to continue researching, identifying, and defending against these various side channels for hardware-based security solutions. Such effort is crucial for ensuring the reliability and trustworthiness of hardware-based security measures.

**Hardware security in low-resource environments.** Edge computing and mobile sensing are booming in recent years. They provide a broad range of mobile applications, which require high security and privacy. Purely relying on software-based security is not efficient. Developing small footprint security mechanisms at the hardware level is desirable for resource-constrained edge computing environments. One interesting problem is to explore the available resources such as CPU and GPU together for the use of new security mechanisms.

**Hardware support for next generation applications.** Many emerging applications need re-thinking of hardware security. On the one hand, the researchers would like to reuse the existing security solutions as much as possible. On the other hand, these emerging applications have new requirements, advancements and limitations that need to rethink how to provide hardware security in an efficient way while meeting the needs of these applications. For example, the zero-permission motion sensors on AR/VR headsets can leak user's privacy information including gender information, user identity and even some speech information when wearing the headsets.

**Power-efficient security.** Hardware security primitives require additional computing and networking – causing an increase in power usage (e.g., Keys, Crypto, Authentication protocols etc.) . In emerging domains this is sub-optimal (e.g., Wearables, IoT etc.). As additional solutions, and vulnerabilities are developed, this issue will generate more concern. Further research needs to be done in constructing solutions that pay attention to the power draw – across threats as well as defenses. More generally, developing solutions that can scale up when power is available (in domain, during operating phase etc.) and scale down when it is not would be relevant across all hardware security domains. This scalability may come at the expense of reduced security posture but needs to provide a pareto surface of 'efficiency'.

**Supply chain security and assurance.** Today's hardware development & manufacturing value chains span many countries and actors. Assurance of design across the full value chain is critical towards hardware security – across the full lifecycle. Delivering an untampered product is critical to hardware security. A key need is to provide an accepted framework for sharing security practices (requirements, tests etc.) across the ecosystem and value chain – many companies with differing practices and many different functions. During design, the design tools have to be assured, but also need to provide the mechanisms for inserting assurance primitives in the design (and the associated tests) across various levels of abstraction, e.g., tamper resistance, observability, side channel resilience. During manufacturing, assurance of manufactured design – from chip set through system solution – requires additional, newer tools to provide 100% coverage efficiently.

Although the supply chain is 'less relevant' in operational phases, the ability to test replacement parts and spares in-field is a critical component of such holistic supply chain security management. This work will need to include participants from all parts of a lifecycle, but also participants in organizational design.

**Intellectual property protection.** Another key aspect of managing the value chain for Hardware security is providing the ability to manage intellectual property across many providers (eg: cores) permitting for inclusion of such IP into design/solutions without permitting reverse engineering. This approach needs to provide for such anti-reverse engineering, while also permitting the ability to prevent &find tamper/malicious activity.

## Technical and Societal Impacts

Addressing hardware security at all points in its lifecycle is simply essential, as hardware forms the cornerstone for all computing. In many cases, hardware cannot be easily replaced or modified once deployed, even if unintended functionality is discovered that might be highly damaging. Significant progress in hardware security will support trustworthiness in essential domains in which computing is important.

## Cross-Discipline and Outreach Opportunities

Several disciplines, aside from core hardware research and engineering, need to be involved in the hardware security lifecycle research. The formal methods community will continue to be essential in aiding the verification of hardware designs and devices. Systems security researchers will be critical in developing techniques to discover unintended functionalities of designs and to facilitate the prevention or mitigation of these "surprises". Usability researchers have an important role to play in ensuring that consumers of devices are properly informed of components' uses and potential pitfalls. The emergence of new types of devices (e.g., quantum components, innovative wearables) and devices having considerably lessened power footprints will, of course, require innovations leveraging knowledge in their domains.

### 3.1.6 Software Security

### Area Description

This technical area encompasses the lifecycle of software creation and execution, and the security aspects thereof. Subtopics include security features of programming languages (and proving these properties); methods for translating software to achieve an artifact with provable properties; assessing the trustworthiness of software for particular purposes, based on automated analysis or provenance; translating natural-language requirements into code specifications; resolving conflicting software requirements; languages and software for novel domains (e.g., gaming) or hardware (e.g., quantum computers) and the security properties thereof; and educating the next generation of software developers on the security features of the code they produce.

Software security is by its nature extremely broad. It touches on (i) how software is developed, (ii) verification of software at compile time, (iii) vulnerabilities at runtime, (iv) patching of software, (v) side-channels, (vi) how software development tools are used by developers, and many more topics. Moreover, as the community strives to see this research deploying in the real world, it is necessary to touch on many of these topics simultaneously.

Moreover, as the boundary between software and hardware gets increasingly blurry, blended/hybrid software-hardware architectures and systems become more common. While there are separate suites of techniques for software and hardware verification (and two very separate research communities dedicated to each), there appears to be no principled/coherent means of verifying blended systems and architectures.

### Technical Efforts

**Certification of Hardware/Software Co-Designs.** While it is possible to certify the reliability and performance of hardware and software packages as independent tools, it has become increasingly important to certify these packages in unison with each other. While certified software or hardware may work independently, if one of these fails when combined with the other, it sacrifices the security guarantees that were previously present during independent analysis.

**Developer-Oriented Software Verification.** There are decades of excellent research from the Programming Language (PL) community on verifying the correctness of software. Unfortunately, there has been limited adoption of this technology by the wider developer community. This lack.

of adoption significantly stymies secure software development. There needs to be more funded research aimed at jumpstarting efforts to bring software verification to the broader developer community. It would involve, PL, security, and HCI researchers working together to establish (i) how formication can and should be introduced to developers, (ii) how would developers use these tools when working with large software systems, (iii) and how effective are these tools at scale.

**Automatic Software Writing/Debugging.** The advances in AI/ML can be used for automatic application-specific, hardware-aware, secure AI-generated software and debugging tools. Establishing measures to safeguard such systems against poisoned code injection attacks (attackers release insecure code so that the LLMs will learn it) using adversarial ML. Automotive software writing/debugging for teaching programming and software could be one of the applications.

It is also critical to understand the quality of the code output by these systems, i.e., is it usable in software systems or does it introduce new vulnerabilities. Future work must evaluate the quality of the output and recalibrate methods to address its limitations.

**Side channel recognition and resistance.** With the growth of various types of hardware side-channels targeting cryptographic secrets, it becomes important to develop software (especially, security-relevant portions thereof) to be side-channel-resistant. Hiding software inside TPMs or TEEs is not effective since even these "secure" environments are themselves subject to side-channels. Thus, there need to be ways to either write or instrument code that is impervious to (e.g., all timing) side-channels.

**Redundant Software and triple security.** There are opportunities to establish measures of safety for large-scale software by providing redundant and yet error-diverse component design and implementation. The study of theory and engineering of redundant software design and implementation can be an important area of technical effort for delivering truly fault-tolerant and self-resilient software systems in the presence of software security threats.

**Human-in-the-loop secure software verification and software assurance**. For code components used in mission critical software systems, it is important to perform software assurance testing and verification. Such auto-testing and auto-verification should be accompanied by human-expert involvements in critical and vulnerable component interactions to provide diverse channels of software security guards.

**The impact of Input-and-output data on software security**. Software security is not just about the computational components of the piece of software (system) but also about the capability of the software system to mitigate adversarial inputs on the verified software as well as mitigate the malicious transformations of software output data.

**Secure patching.** Software will all need to be patched at some point in its lifecycle. Identifying ways this can be done more securely will be needed. Research should be encouraged into building mechanisms for securely updating software in a reliable method that can be rolled back (to ensure availability). Also, how is it possible to help users know what software they are running? In modern systems with auto-updating, users may struggle (or lack the ability) to verify that the software they are running is indeed what they think they are. While auto updates have had positive impacts, they also open up users to attack and there needs to be more reasoning about that.

**Metamorphic programs.** Distributed systems remain vulnerable to attacks that leverage software vulnerabilities. A buffer overflow on one machine will likely look on all devices. This is very problematic in systems such as blockchain that rely on an assurance that only a small number of nodes will be compromised at a time. There is research into metamorphic programs—programs that all accomplish the same functionality, but do so using different machine code, data structures, etc. to provide implementations that have different vulnerabilities. Ideally, such programs could make it more difficult for an attacker to find a single attack that brings down the system. Research into automating the creation of such metamorphic binaries would be helpful for blockchain and other distributed systems.

**Evolution of programming languages.** Future language will support, for example, safety, privacy, and information flow, properties. Programming languages need to be extended to support different (and broader) security properties such as privacy and compliance and consider all these properties holistically. To this end, property specifications need to be developed, as well as automatic methods for generation from higher level specifications (combine LLM generative models with formal verification tools). This can help solve issues with insufficient, imprecise, or ambiguous specifications.

**Secure and privacy-respecting software development.** Developers play an important role in terms of producing secure and privacy-respecting software. Research is needed to understand, characterize, and model developer behaviors, including challenges and pitfalls in incorporating.

security and privacy requirements and considerations into software development. Furthermore, there is a need to better support developers – through improved software engineering processes and tool support to write secure and privacy-respecting code and help them avoid mistakes when integrating security code or privacy-enhancing technologies (e.g., misconfiguration of crypto parameters, copying outdated/insecure code, mismatch between chosen PETs and desired privacy guarantees, etc.). Solving these challenges will necessitate deeper integration of software engineering practice and processes with security engineering and privacy engineering, as well as advancing security engineering and privacy engineering as software engineering disciplines.

**Usability of secure programming tools and languages.** There are numerous research issues associated with existing tools for secure programming and programming languages. For example, considering secure / memory safe languages - are they expressive enough, for example to develop code for kernels? Do they require significantly more expertise to program in? How are performance challenges solved?

**Hardware dependencies and reconfigurable hardware.** Reconfigurable hardware / agile hardware can change how programs are written. Software proofs can be carried out at the hardware level, but these can themselves lead to denial-of-service attacks. There are also issues with memory sanitization. It is important to explore the means software can protect sensitive data in code, in memory, across the entire data flow, how do programmers write programs that bridge the hardware/software divide and protect data in flow, data at rest, data in use.

**Software provenance and software supply chain.** Supply chain problem - especially with cyberphysical supply chain. If legacy code is embedded inside applications, how is attribution written? Vulnerability analysis at the supply chain level? Who is accountable? Blockchain based systems can be used to integrate supply chain into smart contracts, where based on the root cause of the vulnerability, remediation can be redirected to the appropriate owner. This work integrates well with designing accountable software systems programs, and there are clear overlaps with SaTC.

**Roots of trust for disaggregated systems.** Disaggregated systems are systems where storage, compute, hardware may be at different locations - what if a TPM is not available (i.e., in a centralized sense)? Can this be carried out with automated testing, automated integration, distributed (cloud) applications, web services / software composition, orchestration, where the dependencies are not clear. New debugging tools need to be developed.

**Translation of natural language requirements to software requirements and code.** Software is subject to an increasing set of legal and regulatory requirements regarding security, privacy, and data protection. Yet, there is often a mismatch and gap between an organization's claimed policies and values and their actual implementation in software. A challenge is the effective translation of legal requirements to code specifications and code in order to facilitate compliance. Furthermore, this necessitates new approaches for resolving conflicting regulatory requirements on software, such as from different jurisdictions (e.g., US federal vs. state privacy laws, GDPR and other countries' regulation with extraterritorial applicability). Furthermore, internal data management approaches and integration with programming languages are needed to ease tracking and monitoring of compliance-related aspects (e.g., data origin, jurisdictions, legal bases for processing, etc.).

**Formal verification that a language meets its advertised goals.** Formal verification of software has been useful but faces several challenges: (1) it is a huge undertaking to formally verify a relatively small piece of code, e.g., a memory allocator. How to make formal verification scalable is of great importance. (2) software will be constantly updated but most formal verification must restart after software update. How to incrementally conduct formal verification is also important.

**Domain-specific languages.** Domain specific languages provide both an opportunity and a challenge. They may directly increase the use of computational approaches to solve domain specific problems since they have specialized properties tailored to the application, but they also tend to introduce vulnerabilities. For example, web assembly language instead of javascript are now used for web software. However, these can lead to new vulnerabilities. Similarly, IOT languages can open more vulnerabilities. The key problem is that domain experts may be unaware of security vulnerabilities or issues, and we need development that is mutually carried out.

**Security of AI/ML software.** Programming frameworks should support mechanisms and development of policies for security of data and models as well as inference: data-specific security, privacy, and compliance requirements for ML datasets to protect from data poisoning, backdoor attacks; Adversarial attacks.

**Network protocols.** Given constantly updated RFCs and extension, how developers could easily understand the protocols and make software consistent/compliant with the protocols. A potential direction is tools to translate high-level languages to code specifications, or check consistency with protocol traces.

**Missing and imprecise security specification challenge.** One keystone for secure software development is the availability of complete and precise security specifications. However, emerging platforms (IoT, smart, mobile) often lack these specifications. Developers may then rely on their own understanding of a given platform-specific functionality to implement security measures (e.g., access control). Considering codebase size and complexity, this task is inevitably error prone. It is challenging to demystify the security specifications, which can lead to developers making mistakes and various security issues such as overprivileged software and insufficiently protected software components. Research efforts should be dedicated to the automatic generation of security specifications from the platform's codebase (e.g., through analyzing, modeling and extracting security properties from the platforms via program analysis and formal reasoning).

**Economic models:** Decentralized applications rely on economic models such as DeFi token mint, burn as well as on consensus methods to provide a coherent outcome of the decentralized computation and to incentivize the peers to be part of this network. Development of the smart contracts or software as it evolves requires game theoretic algorithms, consensus protocols as well as blockchain-based requirements (re-entrant behavior, transactional semantics, cryptographic capabilities) together to be used in the development and entire DevOpS/ProdOps environment.

**Reliable software decommission and data deletion.** How to ensure that data is provably deleted in response to data deletion requests or de-identification requirements across distributed systems and backup solutions. Such work would explore automatic code generation with provable properties, autonomous code (self-protecting and self-defending, self-modifying systems), and software recycling and sustainability (is the software code ready to be recycled or entirely removed from the ecosystem such as insecure cryptographic libraries).

## Technical and Societal Impacts

The impact of improving the security of software cannot be overstated. Software powers our lives. When software is insecure the effects range from minor annoyances up to and including death. In a more positive phrasing, the more trustworthy our software systems become, the more likely they are to be adopted into widespread usage. This will allow new technology to flourish, improving the lives of citizens and improving national defense.

Success in these areas will mitigate security vulnerabilities in software and system, facing new programming languages, applications, or hardware. Software and system security protects systems from compromise or attacks. For end users, it prevents privacy information leakage, For companies or organizations, it ensures the software would not be abused or compromised by attackers, which avoids financial loss or societal impacts. The success in the areas will also Improve software and system development. The tools can allow developers to better understand specifications or protocols, to develop secure software or systems. The results from the research can benefit education for secure software development and analysis, to train the next-generation workforce.

Traditionally, this research in software security has (mostly) been confined to researchers. By helping work to broaden its base and make it applicable to software engineers, the impact that it has on society will significantly increase. Moreover, it is expected that as it is used in more real-world deployments, new issues with formal identification will be identified, spurring innovation within this area.

**Cross-Discipline and Outreach Opportunities**

Software security is ultimately about ensuring that software behaves as expected. This includes behavior that is consistent with the developers' expectations/objectives/intent. This also includes ensuring that the software behaves in a manner that is consistent with users' expectations. This in turn requires modeling people's expectations such as capturing design requirements from developers or capturing software behavior expectations from users. With more environments where pretty much anyone can become a developer, new interfaces will need to be developed that allow people to develop software using natural language specifications ("a la ChatGPT"). This also includes developing functionality that will help developers more effectively specify the behavior and properties of the software they are developing and modify code when they modify their requirements or find that their software does not conform with their specifications. This type of functionality will have to include support to ensure that software meets various security properties. It will have to also extend to the development and refinement of AI/ML-based software.

*Legal*: Software security is not a purely technical domain and instead intersects with public policy and legal issues. This includes exploring new regulatory requirements, which may include requiring that software come with various security guarantees. This may also include the exploration of certification programs designed to verify security guarantees provided with software. Such guarantees would ensure that when software is composed, it is easier to ensure that it continues to meet various types of security guarantees.

*Business:* As new regulatory requirements are explored; it is important to understand their business and economic impact. The benefits of new regulations would have to outweigh the economic impact of providing security guarantees and complying with new certification requirements. Historically, industry has pushed against such requirements. Given how pervasive and complex software has become and the way in which software is composed, requiring that software comes with guarantees seems unavoidable, yet research will be needed to determine how far one can realistically go.

### 3.1.7  Web Security

**Area Description**

Web security includes a broad set of topics such as various web servers, browsers, protocols, and applications running on top of the server and browser. Web security differs from traditional software security because it directly interfaces with humans. The future of web interaction will change and evolve with the emerging technologies, such as AI-based content creators, AI assistants, and AR/VR technologies. This will change the way users "surf" the web as well as the way web content is generated. New interaction mediums will create new types of security and privacy challenges that SaTC should focus on. Future of web security research can be broadly divided into three directions: (a) how to secure the (future) web technologies such as browsers and servers, (b) how to make user interactions with the web secure and safe, (c) how to help users gain trust on the web.

**Technical Efforts**

**Browser security.** Modern browsers have become complex operating systems that can concurrently run a large variety of web applications. However, browsers are not like traditional operating systems and are characterized by a unique set of security problems. Some of these problems arise from the fact that web applications consist of code that is dynamically loaded from potentially untrusted servers, and that most modern applications are built by composing untrusted code originating from a variety of sources. There has also been a recent push from industry to make web applications look and feel like native applications, which requires an ever more intimate interaction with the underlying device's software/hardware. This trend will likely continue. For instance, mobile applications can already be built as a hybrid of native and web code, with the latter running on a rendering engine. There will be a need to better define security policies that the browser/web-rendering system will need to enforce to protect the security of users' devices and the privacy of users.

**Web application security.** As native applications move more and more to the web, new security challenges will continue to arise. Web applications are composed of client-side code and remote code, making it challenging to apply traditional software analysis approaches to assess the security of web applications (e.g., in terms of both vulnerability detection for benign apps and malicious behaviors). Furthermore, web applications do not follow the traditional software update cycle, in which applications are explicitly updated by installing new code on a device. Instead, web applications' code is loaded dynamically from a remote source and can change at any time. There will need to be significant research dedicated to developing new methods for a continuous security evaluation of future web applications.

**Human aspects to web security.** The introduction of new ways to access web content (e.g., via voice-activated devices like Alexa) and new engines for synthesizing that content "in the network" (e.g., the integration of generative models in search engines like Bing) pose new challenges to humans in assessing the trustworthiness of the information they retrieve. Our web security indicators (e.g., the lock icon) were already poor vehicles for conveying trustworthiness of content, and even these are jeopardized by new modalities for retrieving content. Moreover, the in-network generation of content poses challenges to determining the provenance of information.

**Human authentication technologies and the web.** Though not unique to the web, challenges for human authentication to services have been exacerbated by it. Not only do passwords continue to pervade this ecosystem, but mechanisms to improve security (e.g., multi-factor authentication) and/or usability (e.g., password managers) have been slow to gain adoption and come with their own challenges. The prevalence of passwords, in turn, gives rise to account takeovers via credential stuffing, themselves seeded by password database compromises. Research in more effective, and more accessible, means to authenticate users continues to be needed.

**Measurement on the Web.** One of the unique factors of the web is the diversity of entities that interact in a shared environment. Servers, browsers, users, companies, advertisers, etc. all collide on the web. Unlike other computer systems with more engineering control (e.g., an individual operating system), web technologies have, by design, enabled the growth of the web beyond any one entity's control. As such, there may need to be a shift in mindset when studying the web to incorporate empirical, observational research of the web. For example, natural ecosystems are studied to seek to understand the innumerable relationships and interactions that exceed our current knowledge. Likewise, the web must be measured in order to discern the different actors (benign, malicious, in-between) and how they interact with each other to understand new/emergent security threats, quantify harms/risks, inform defenses, and prioritize security efforts.

**Securing New Web Interaction Modes.** New modes of accessing and securing content over the Internet are emerging (voice-based, XR/AR) and the security and privacy challenges posed by these emerging modes must be addressed. Given the new ways and devices users interact with the Web, how would the Web evolve over time? What unique security and privacy issues these additional interfaces will introduce to the Web? The attack surface is different with the addition of these interfaces. For example, with voice-based interaction, there is a possibility of injection attacks to the audio input device. It is useful to think about a broader threat model that takes these new attack vectors into consideration. Users will likely have multimodal interfaces to interact with the Web which brings additional security and privacy challenges. Further, web content may be dynamically generated or modified before consumption which brings its own additional challenges to Web security.

**Trust and safety on the web.** The web has enabled unprecedented communication between a large, diverse set of entities. This has led to wonderful new capabilities (video conferencing, research collaboration), but at the same time, given adversaries unprecedented access to victims. In order to defend against these threats, web entities must be authenticated - this allows permissioned communication, and more importantly, rejection of unknown or malicious entities. If everyone on the internet "looks" the same, steps towards security cannot happen. This is a fundamental starting point for web and internet security, and it will become more and more.

important over time as more and more entities join the web. Research in this area will look at both user/client and server-side authentication. Based on authentication, a higher level of trust and safety can be built, for example, PKI, provenance of data, code, and information. By identifying who is being communicating with, then authorization/access control can be performed, and support auditing of data (e.g., code, websites, news, health information, etc.).

**Securing User Interaction.** Users are evolving the way they interact with the Web/Browsers via AI plugins thanks to advancement in HCIs and ML. This emerging AI-assisted interaction creates new adversarial models and attack surfaces that require attention to enhance the security and reliability of the Web. For example, AI-assisted plugins (e.g., BingGPT) can provide better and more insightful information to the user experience. However, these tools also create vulnerabilities for the attackers to populate misinformation via compromised AI models that target specific user groups. Future research should pay attention to techniques that can counter these new attack surfaces. Also, since web interaction is getting more automated, user data privacy is also a critical factor to be considered, given that more user information is implicitly collected to refine these AI-assisted recommendation/ information retrieval models.

**Tracking and Privacy on the Web.** Privacy consciousness and tools are limiting the usefulness of third-party cookies forcing content generators and publishers to rely on first-party cookies to monetize their content. This unintended consequence is creating a perverse incentive leading to a worse privacy situation. Research needs to determine approaches for sustainability of the web in terms of enabling monetization of content while preserving the privacy of the users. Efforts like Google's Topics framework may provide a path to weighing the tradeoffs between privacy and utility, but more formal analysis of these tools is needed.

**Anonymity and Accountability on the Web.** There is a need for technologies that can balance the tension between anonymity and accountability in web-based interaction models. Without progress on such technologies social issues such as bullying, child abuse in cyberspace will be hard to detect and prevent.

## Technical and Societal Impacts

Achieving security and privacy on the web and browsers is critical as "web browsing" is a key channel that attackers try to exploit to gain foothold into systems/networks — phishing, typosquatting etc. Web-based tracking technologies are also at the heart of data collection by companies and third parties and significantly pose privacy risks to end users; user tracking is also a predominant way to monetize many of the web-based services/content.

Failing to address privacy issues will either result in significant harm to end users or loss of economic activity surrounding web content creation and consumption. Failing to address security issues will result in a vulnerable cyberspace that acts as a remote vector into public and private infrastructure.

**Cross-Discipline and Outreach Opportunities**

There is a need for a new economic model for monetization of content by providers, publishers, and online marketing while preserving privacy on the web. The web privacy policies would require cross-disciplinary efforts between economic, cybersecurity, ethics, and policy experts.

There are social issues that arise from the ubiquitous use of the web, such as online bullying. Study of such issues require cross-disciplinary efforts between cybersecurity, social sciences, and education – and increasingly AI. The web also provides an important platform that supports the online economy (e.g., e-commerce, advertising) hence the financial and economic risks of the web-based economy are closely associated with the technical risks and vulnerabilities of the web system described above.

## 3.2    Network Security

Network security research studies the methods and techniques used to secure computer networks from unauthorized access and attacks. This field encompasses a wide range of topics, including next-generation systems, cryptography, network architectures, intrusion detection and prevention, firewalls, and network isolation primitives. Researchers explore innovative methods to detect and prevent attacks on computer networks, as well as to develop tools and techniques to mitigate the impact of successful attacks. The field of network security research is constantly evolving, as new threats and attack methods emerge, and as new technologies are developed to defend against them. Moreover, recent moves toward new network architectures and models of computer will require addressing rapidly advancing security and privacy needs.

### 3.2.1    IDS, Detection and Prevention

**Area Description**

Intrusion Detection and Prevention today focuses primarily on the real-time detection and blocking of attacks to a victim network, using network-based and/or host-based telemetry. Considering traditional network-based intrusion detection, this topic has been explored in a vast number of academic publications and many industry solutions have been developed and deployed in the real world. In addition, the transition to end-to-end-encrypted network communications represents a very significant challenge for NIDS, in some scenarios. This area of research should therefore evolve to focus more generically on *Attack Detection, Mitigation and Response*. On the attack detection front, anomaly detection and behavior-based approaches will need increasingly more research efforts. Also, the automatic generation of attacks (e.g., using ML) is in scope. In addition, this area should focus on the prioritization and explainability of alerts generated by attack detection systems, to help human security analysts to better assess and respond to potential intrusions.

One aspect that is of particular interest is the *Response* to attacks. Alert explainability and attribution may help reduce the time needed for an attack to be noticed by an analysis, thus reducing response time. At the same time, alert explainability may also enable the development of autonomous response systems that can mitigate an attack in real-time and prevent future attacks.

**Technical Efforts**

Intrusion detection research has focused on developing systems and algorithms for identifying the fact, origins, and parameters of cyber-attacks and attackers. There is an arms race that requires constant vigilance and investment to prevent new waves of attacks from yielding catastrophic consequences on the public and private sectors. Exemplars of future work in this area include:

**IDS secure and privacy-preserving data cultivation and management.** Some of the challenging problems of traditional IDS are the availability of data to share, reliable labeled data with no negative privacy implication, scale of data, etc. Synthetic data could be interesting in this context to allow for designing realistic attack detection but also for sharing the data.

**Addressing AI-generated attacks.** AI-generated attacks are already a concern for intrusion detection, but their frequency and potency are likely to increase with advances in AI. Generative AI models may pose a greater risk for security: by training them upon large volumes of attack data, it could be possible to automate the development of new attacks. Defending against these attacks will require an investment in AI to detect this approach, like detecting telltale signs of text or images generated by AI models.

**Managing alert overload.** Typically, alerts generated by commercial attack detection systems (i.e., IDS as they are traditionally known), are too many to analyze by human operators. Some systems might generate thousands of alerts each day. Clearly, research is needed to reduce the number of alerts that are generated, and to allow the human operator to focus on the meaningful and important ones, potentially based on the values of the assets that are being monitored. In this space, AI methods and algorithms might come in handy in automating alert reduction. Further, the identification of features that would allow the automated analysis of alerts to rate them based on their importance would also be interesting.

**Optimizing human/IDS interaction.** Currently IDS systems are not fully automated. Outputs serve first level alerts to human analysts. Quite often, IDS system outputs have a high false rate: a high level of false positives cause analysts to be overworked, and efforts wasted. Even worse, it makes them less sensitive to true alerts. On the other hand, false negatives will give analysts a false sense of security, leading to missed attacks. Research is critically needed to make the IDS/human analyst team more efficient and more productive while reducing the cognitive load on analysts.

**Human + IDS teaming.** Similar (but distinct from) the previous area is human/computation teaming. If an IDS generates an alert, but a human never sees it, was the attack detected? IDS systems, by their very nature, require humans to take actions on the alerts. Therefore, future research should consider the Human and IDS system team and should study what is the optimal configuration of this system. The goal would be to spur research to reduce the time that it takes for humans to respond to an alert, improve the accuracy of a human responding to an alert, and to improve the ability of the system to detect and remediate attacks.

**Alert interpretability and explainability.** A key challenge in current IDS systems is that the systems generate an alert (i.e., that a likely attack was detected), however the human must perform significant effort in understanding the impact and threat of the alert. The field of AI has developed the subfield of explainable AI, in which the outputs of AI classifiers are explained to humans. Perhaps this idea can be brought to IDS: research should investigate if alerts of IDSs be explainable to the SOC analysts that must act on the alerts.

**Behavior-based detection.** For modern attack detection, mitigation, and response systems, one very promising research direction is anomaly detection that takes behavioral aspects of the network into consideration. Signature-based detection systems are quite outdated these days. Looking at the system as a whole and trying to identify benign behavior and differentiating it from malicious behavior is a promising approach going forward. For this type of research, AI might become useful in creating behavioral models, and in the creation of systems that focus on "hunting" threats rather than simply detecting them. That is, the system should be able to identify threats *before* they become attacks and take a more proactive approach rather than a reactive one.

**IDS in untrustworthy environments.** IDS mechanisms can be applied to less trustworthy components, e.g., in the context of supply-chain protection concerning attacks like SolarWinds. A key aspect of defense is not to rely on any specific system and to assume that, eventually, system will be compromised. Therefore, when thinking about how to prevent attacks on a network, in addition to monitoring the network, research should consider end-point security. One avenue to achieve this might be a zero-trust environment, where no component implicitly trusts other components. While this approach cannot completely prevent compromise, it can enhance more traditional IDS systems.

**Host-based IDS.** Host-based IDS (HIDS) should also be considered to complement NIDS. More comprehensive and dynamic approaches beyond traditional remote attestation on the static integrity of code should be explored. Novel approaches should be able to detect the integrity of all sorts of information flow in the system, including control-flow integrity, data-flow integrity, etc. Implementing information flow violation detection may be a challenging task for IoT and embedded systems due to the power consumption budget and limited memory and computing resources. Novel mechanisms that take advantage of new hardware features, such as confidential computing mode, hardware debuggers, hardware tracers, hardware memory tagging modules, etc., may help make such solutions more efficient and practical.

**Prioritizing attack evidence.** Both because of the use of network encryption and the subtlety of sophisticated attacks, signature-based detection is unlikely to be particularly useful in preventing significant cyber-attacks. Consequently, anomaly/behavioral-based detection will play a larger role in attack detection. To handle both the volume of alerts and the problem of false positives, new techniques for prioritizing and focusing investigation on the most-likely-to-be-significant security events are needed. Providing additional context to generate sufficient confidence that an alert/event is actionable is important. Such context may include activity and associated attributes from across an enterprise network, correlated alerts, etc.

**Containment, Mitigation, and Response.** based on empirical evidence and considering the state of practice, the expectation is that systems and networks will continue to experience security incidents. As a result, defenders need to be able to contain a detected attack, respond to eliminate the attacker's presence and any latent backdoors, and mitigate the possibility of the same attack being successful (whether against the previously compromised or other similar systems). This means that techniques, tools, and mechanisms are needed for precise identification of relevant attack parameters and for containing their effects. Conducting effective defensive cyber operations is an open problem, especially when considering the challenges of system/network complexity, scale, and potential impact to legitimate system operations by defensive actions. The application of autonomous agents to act and respond to contain an attack and prevent it from happening again while providing certain safety guarantees is an important problem as well.

**Measuring Success.** reducing the time to discovery of attacks (e.g., for APTs) and reduction of exfiltration of data. More generally, better metrics on attack detection and mitigation are a critical area of research.

**IDS for decentralized network protocols.** Decentralized networks, such as those used to coordinate blockchains, also face the threat of intrusion. However, traditional intrusion detection systems are difficult to implement on such networks due to several reasons. First, some decentralized networks are operated in a decentralized manner, making it challenging to impose traffic limits or block suspicious access. Second, decentralized networks lack centralized data collection capabilities, which makes data-driven intrusion analysis challenging. Third, even if an intrusion is detected, decentralized protocols impose many restrictions that make it difficult to respond effectively. Therefore, addressing these issues may require an overhaul of existing protocols.

**AI-assisted IDS** (amount of training data, adversarial ML attacks, data poisoning). Current IDSs rely on AI-assisted tools to make informed decisions. In the future, with the amount of data being generated massively across the network from many nodes, this poses critical challenges to form a robust AI model to effectively detect the intrusion because the adversary can populate numerous malicious data across the network to pollute the AI model. Moreover, AI tools are also fragile and vulnerable to data injection attacks. There will be a need to develop new approaches for AI-assisted IDS such as reinforcement learning.

**AI-assisted intrusion attacks:** While IDS uses AI for intrusion detection, the adversary can also utilize AI tools to perform intrusion more effectively and bypass the AI-defended tool. Further analysis needs to be investigated for such AI vs. AI combat.

**Handling large data volume:** The fast advancement of many sensing modalities in daily life activities will largely increase the amount of data transmitting in a network. Different parts of the network suffer from various vulnerabilities caused by the large amount of data. New research is needed to study the impact and vulnerabilities at different network levels. New mechanisms need to be designed, considering the impact of machine learning techniques as these techniques will be employed at multiple network levels to handle the large volume of data. The security protection and the network transmission performance need to be considered in an integrated manner to provide efficient handling of large data volumes while preserving the data security.

**Security architecture:** The network security architecture continues to be a key aspect of how to achieve better network security, including how intrusion detection can be established. New security architectures, such as Zero Trust Architecture, should be investigated and potentially utilized at the network layer.

**Network virtualization**: Network virtualization (incl. NFV, SDN, etc.) offers several operational benefits to network providers. At the same time, by performing what used to be specialized functions on general-computing HW/SW, NV may enlarge the attack surface of the network layer. An open problem remains as to how to take advantage of NV (including virtualized IDSs) to improve security, without introducing vulnerabilities in general-purpose HW/SW.

**IDS in decentralized networks:** In the centralized system, one can build an effective IDS due to the availability of all data collected at a single point. However, in the decentralized network, it is more challenging to have an effective IDS since the data is distributed. There should be an investigation on decentralized/distributed IDS, where multiple entities can collaborate with each other for effective IDS. The data of each entity can be of privacy, an investigation on privacy-preserving distributed IDS is also desired.

**IDS in constraint network systems:** Specialized network systems have certain constraints. For example, in V2V networks, the entities geographically located close to each other may use special protocols for communication. In the future, there may be new network protocols for specialized systems (e.g., IoT medical network). It is worth it to investigate how to characterize intrusion attacks in such constrained systems and develop corresponding effective IDS mechanisms.

**Securing Specialized cyber-physical system (CPS) Networks:** As more and more of our infrastructure digitizes and is networked many networks with special requirements have emerged. Some past examples include electrical grid control networks. Some emerging examples include – networks in smart manufacturing facilities (i.e., industrial IoT), networks in automated industrial warehouses (e.g., robot networks). These networks impose additional constraints (e.g., timing, legacy) and have different traffic patterns than traditional enterprise networks. Traditional solutions are not suitable and typically require domain specific solutions. Design of efficient intrusion prevention, detection and recovery mechanisms for such networks taking advantage of their unique characteristics and respecting their constraints remains an open problem.

**Detecting Advanced Persistent Threats:** Cyber-attacks have become a part of the arsenal for nations to achieve their geopolitical objectives, putting our networks underlying key services at significant risk. However, detecting such attacks in a timely manner remains notoriously difficult. There is a need for tools and technologies to prevent and better detect such attacks in our networks.

**Dealing with false alarms:** IDS systems generate false alarms that are hard to deal with. Typically, human analysts need to review many potential false alarms. As with any human task, there is a possibility of error which could have significant consequences. Human speed is too slow. Machine learning is heavily used in intrusion detection systems, and there is a need to augment analysts' ability to understand the decisions of intrusion detection systems. These will include explainable AI techniques for intrusion detection and better visualization.

Another aspect of ML-based intrusion detection is the idea that systems could become the attack targets themselves. Research needs to investigate potentially unique attacks targeting the ML-based intrusion detection systems and develop countermeasures.

Additionally, deep learning-based systems may not be the best fit for IDS due to lack of training data. There is a need to investigate other AI approaches such as reinforcement learning.

**Self-adaptive detection systems:** Since attacks keep changing, the network is dealing with a moving target. Deploying detection systems efficiently and effectively in the network becomes critical in the ever-growing network domain. Reinforcement learning and machine learning are very useful tools that can be explored in designing self-adaptive detection systems. It is important to perform intelligence training as attackers come in different forms and have different effects. Building effective reinforcement learning and machine learning models will be key research areas.

**IDS on Encrypted Traffic:** As more and more traffic become encrypted for privacy purposes, there is a need for techniques to perform intrusion detection over encrypted traffic especially when network layer inspection is not sufficient. For example, an open problem is how to detect data exfiltration with encrypted traffic.

**Intrusion Response:** Responding to intrusions and being resilient to them is also a key challenge. Automated response approaches that can tolerate intrusions and respond by either re-configuring or other means remains an open problem.

**Technical and Societal Impacts**

The impact concerns many computer systems, where rapid and effective discovery of (large scale) cyber-attacks, as well as the response to them is critical to the future systems. It is also essential to reduce cognitive overhead for human operators and enhance the accuracy of human/IDS team.

Networks are a key component of our technology fabric that enable communication and interaction between network nodes and make many of the technologies and applications (e.g., the Internet, Internet based applications such as online banking & video calling, smart functions in IoT devices, etc.) users rely on everyday possible. Without securing such networks it is not possible to ensure the security and availability of the applications that users have come to rely on.

With the popularity of special-purpose network protocols (IoTs, smart homes, decentralized networks), if research fails to make progress in this area, there will likely be massive attacks that are likely to happen ubiquitously due to the unexplored vulnerabilities in such protocols.

**Cross-Discipline and Outreach Opportunities**

Numerous Intrusion Detection Systems (IDS) incorporate machine learning and data analytics capabilities. However, to further advance IDS techniques, the ML and security communities should collaborate and make concerted efforts. It is worth noting that the adoption of IDS in practice may be limited by users' budget constraints, and therefore, economic analysis may be necessary to facilitate widespread adoption. Finally, the adoption of IDS requires policy support.

While security solutions exist for prevention in networks, many of them are only recommendations in networking and communications standards and go unimplemented by carriers or vendors. There are opportunities to explore economic incentives and policy factors to improve or mandate implementation of security technologies.

There are opportunities to achieve intrusion detection based on the physical properties of communication. For example, quantum networking offers detection capabilities at the physical layer. There are potentially other such opportunities to investigate, with cross-over and multidisciplinary efforts including electrical engineering, material science, etc.

### 3.2.2    Wireless and Sensor Network Security

**Area Description**

Wireless communication technologies are becoming increasingly pervasive via last-hop links to user devices (e.g., smart phones / Wi-Fi-connected devices), mobile systems (e.g., cars), physically embedded devices (e.g., medical implants), and remote computing (e.g., outer space, oceans). Wireless communication presents many unique security challenges. 1) Because wireless is a default-broadcast communication mechanism, privacy is a major concern: malicious observers can easily/passively access communications. 2) At the same time, authentication is a major challenge to distinguish between benign / malicious active communicators on a wireless network. 3) Wireless modalities are also capable of physically tracking users/devices, which can be used for both beneficial and harmful purposes. Such issues are made more complex (and have additional advantageous properties) when sensors are embedded in such networks.

**Technical Efforts**

**Human sensing**: Related to wireless security is sensing, this focus explores using signals to detect human presence via backscatter or other approaches. Work in this area introduces opportunities for new security mechanisms, but also privacy issues since they may represent new tracking mechanisms. Flipped around, mechanisms for sensing/detecting wireless devices (or non-wireless) devices. In a world where IoT devices are ubiquitous, users may increasingly need mechanisms to enumerate and audit the physical location of devices, for example, hidden camera or microphone detection (highly relevant in settings like Airbnb or domestic violence). This is particularly true for malicious devices; research should examine how to move towards standards and regulatory oversight for manufacturers to integrate explicit protocols for device discovery.

**Wireless tracking:** Research should investigate the issues coming out of wireless tracking technologies, such as tracking devices (AirTags, Tile). Wireless tracking technologies use Bluetooth mesh networks to perform physical tracking of the device and are advertised for use for lost or stolen devices. Beyond their stated purpose, they pose a significant safety risk as they are dual use technologies that can be exploited unknowingly being used to stalk their user. With this consideration, research should be aimed at understanding how to design tracking device protocols to allow safe discovery, i.e., we should avoid a world in which cheap, covert, widely accessible tracing devices allow widespread stalking and other harms. This may require new protocol designs since discoverability itself may be a privacy/tracking problem from the perspective of Bluetooth beacon-based tracking. There is some overlap with the work on COVID-era Bluetooth-based contact discovery.

**New/evolving Application Areas:** The expanded use of wireless technology in specific application areas such as vehicular or medical contexts may be considered a major driver in the development of new standards to ensure security and privacy guarantees. For example, while future vehicular networks are expected to help improve situational awareness and facilitate necessary and helpful applications (e.g., roadside assistance or improve the flow of traffic), proper measures should be employed to avoid privacy invasions (e.g., monitoring or tracking of movement patterns) or integrity violations (e.g., tampering with information). Similarly, while wireless technology has proven beneficial in addressing, managing, and monitoring medical conditions (e.g., pacemaker or insulin pumps) at the same time this introduces new (often unanticipated) attack vectors. Future technologies need to be developed to better address this challenge. For example, in an emergency setting, accessing the data of a pacemaker should be possible while in other non-emergency contexts such a readout should be prevented.

**Offensive techniques in wireless security:** Although there has been a rise in security in over-the-air attacks during the shift from 4G to 5G, there still exists a wide array of undiscovered vulnerabilities in our wireless systems. There is great need for more focus on offensive techniques in wireless security to further discover these vulnerabilities and address them. These offensive techniques allow us to craft defenses and protect wireless security before an actual adversarial entity takes advantage of the vulnerabilities. This includes further analysis of side channels in signaling protocols and applications.

**Analysis of Human Activity Recognition (HAR) via Wi-Fi**: New techniques have been developed to enable systems to classify human activity based on the Wi-Fi backscatter off the human in a room. It is assumed that these Wi-Fi signals are not privacy-compromising, however it is important to analyze these systems for security and privacy risks. As these systems become more prevalent and well-implemented, attacks will rise, either on the software side or network side.

**New wireless modalities:** Several new wireless modalities may emerge in the future that change some of the fundamental assumptions currently held for wireless communications. One example is directional beams / antennae that can reduce the publicly observable/eavesdrop-able region of wireless communications. Consideration must also be put into new communication mechanisms that do not rely on electromagnetic waves. For instance, quantum entanglement over increasingly large distances may enable direct wireless communication that is not observable by passive adversaries/auditors.

**Physical layer security.** It is important to explore novel uses of some physical layer/hardware assistance for better security. For example, in the case of location fraud issues, it could be useful to think of location/distance binding/authentication (absolutely as well as relative location) with physical information. In location/tag tracking, there are also privacy issues. It is also possible to think of new emerging data-based approach (AI) to solve physical layer security issues.

**Sensing and actuation security.** Physical constraint of sensing can bring both security challenges and opportunities. The community needs to investigate new attack-surfaces in this domain. Further, it will be interesting to study the integration of sensing into new security applications, even in non-traditional settings. There are also potential privacy issues to consider, particularly in the new joint sensing communication scenarios.

**New wireless scenarios.** What are new security and privacy issues in emerging CPS/IoT networks such as home/industrial IoT, smart grid, smart building? It is important to systematically investigate new attack surfaces, such as those on analog/sensing interfaces (e.g., radiation, sound, lasers). It will be important to investigate how to perform functional attestation of large scale IoT devices. Security and privacy of critical data generated by new types of sensors are of interest as well. The environmental impact of security/privacy techniques in the context of very large-scale CPS/IoT deployments is another area of research consideration.

**Technical and Societal Impacts**

**Impacts on future generations of wireless and sensor network standards:** Standards and specifications in wireless and sensor networks rapidly evolve. While fixing security issues with the current generation of standards can be difficult (because hardware- and software-based implementations for these networks are usually difficult to alter or update), security issues are usually quickly mitigated when the next-generation standards are formulated. Therefore, findings in attacks to wireless and sensor networks are critical to the security of these networks in the future.

**Impacts on space communication:** Wireless communication is a must to achieve communication in space. Such scenarios include satellite communication and soon, interplanetary communication. Improvements in wireless security, especially in the robustness and reliability of wireless communication will help achieve more secure and reliable space communication. It is important to recognize the fact that the infrastructure for satellite or interplanetary communication is controlled by some major players, which is very different from more traditional wireless communication infrastructures.

**Societal impacts:** The Internet-of-Things is gradually becoming the first entry for users to the Internet and the digital world, and the communication security between IoT devices is increasingly critical to the security and safety of human beings. Examples include wireless networks involving health devices, vehicular networks, and smart-home devices.

Another societal impact arises from defending against or regulating the emergent use cases of wireless technologies, such as low-energy wireless tracking devices (e.g., AirTags), wireless-based location tracking, wireless-based human gesture tracking, and wireless-based indoor location tracking. The abuse of these capabilities will cause privacy infringement for subjects being tracked; research is urgently needed to ensure privacy-protection solutions in these areas.

Additionally, self-organizing wireless networks, especially self-organizing P2P networks using short-range communication techniques (e.g., Bluetooth) have shown great potential during periods of social unease (e.g., riots under oppressive regimes, communication with Internet or mobile network censorship) or natural disasters (e.g., earthquakes, severe fires, etc.). Improving the availability and robustness of self-organizing wireless networks has the potential to save lives and foster better citizen societies in these scenarios.

**Cross-Discipline and Outreach Opportunities**

The opportunities and challenges of wireless and sensor network technologies intersect with several disciplines (beyond computer science and election engineering) and will include techniques from many areas of computer science such as systems, model checking/formal methods, cryptography, and others. Understanding how human interactions play a role in security is an area of research need.

One compelling application domain for wireless technologies is medical device design. It will be valuable to include medical and biomedical researchers, as well as public health experts, in research, development, and evaluation efforts.

How users perceive these technologies will be a critical dimension in adoption, use, and maintenance. Amidst growing distrust in science, technology, government, and other institutions, new systems based on wireless and sensor network technologies may become the focus of rumors, conspiracy theories, distrust, and even efforts to physically disrupt their development. For example, in 2020 activists motivated by conspiracy theories related to Covid-19 attacked the physical infrastructure of 5g networks. It will be valuable to incorporate psychologists, sociologists, experts in science and technology studies, and researchers in other related social science fields into conversations about these technologies.

Another relevant domain is disaster or crisis response. When these technologies become part of critical infrastructure, then their failure can cause and/or complicate the responses to crises. Experts in disaster response, including researchers from crisis informatics and the sociology of disaster, could be valuable in thinking through and designing technology, considering these risks.

### 3.2.3   Future Network Security

**Area Description**

Research in Future Networks should include securing the underlying network infrastructure and architecture such that all communication carried over the network provides foundational guarantees (e.g., confidentiality, integrity, availability). There is a need to consider how identity and authentication plays into new networks. This will be critical in helping to combat network borne, scams and phishing attacks carried out over current networks.

There should also be research on "operating through" networks that are untrustworthy, that is, ensuring that security properties hold for devices on a compromised or hostile network. By simultaneously considering both situations, the impact of research will be much greater, provider end-users the greatest benefits possible regardless of the network they connect to.

Research into satellite communications should be supported. These communications represent a fundamentally new set of capabilities as compared to cellular communication. This research will also include definition of protocols, support for testbed development, research into how key management will be used, how updates are managed, and how end-to-end encryption will be used.

There have already been several research attempts that are trying to redesign the network protocols of today. Efforts that focus on keeping the existing networks but improve the problems of today would be more valuable (i.e., the legacy problem).

From the future networking point of view, an interesting direction is to study if existing networking protocols can be modified or if systems can be built on top of them that would give users more control over how their traffic is being routed. This can be very interesting for many people who would like to, for example, prevent their traffic from going over specific countries or regions. It could also help them to have better anonymization of their traffic.

There will also need to be consideration on how to secure data computation as it becomes less decentralized and split between the cloud, the fog, and local devices.

Finally, there is a need to build a testbed where this research can happen. The bar to participating (in terms of cost and sophistication) in this testbed should be low.

**Technical Efforts**

Future networks and applications open new opportunities and challenges that connect security and privacy research with network science, signal processing, cryptography and many other fields. Exemplars of work in this space include:

**Future 5G/6G networks**. There needs to be research examining how the unique nature of 5G and 6G natures present a novel attack space. For example, these networks often have strong time synchronicity requirements. What happens if those are disrupted? Similarly, these networks are often short distance, high-bandwidth networks. How does this impact potential security issues? How is interference handled? Similarly, research needs to determine how to use these new network infrastructures to do novel things. For example, using them as sensors.

**Identity and Authentication.** Future networks have the potential to provide new concepts of identity. These primitives can potentially be used to better deal with scam/nuisance messaging (in particular, robocalls). While individual users in current networks can generally elect to not answer any phone calls (and selectively return calls when it suits them), businesses (especially SMBEs) have little option but to answer calls. Being able to interact with customers with at least the same identity guarantees provided to websites currently would dramatically impact the overall security posture of SMBEs. Any such considerations would need provisions to operate without necessarily mandating end-to-end authentication for every call - activities by anonymous whistleblowers should be able to maintain plausible deniability in the presence of such mechanisms. As scams become increasingly more sophisticated (e.g., via deepfaked voices and realistic-sounding scripts from Large Language Models such as ChatGPT), the ability to reason about who is on the other end of a transaction will only become more important.

As identification and authentication improves, care must be taken so that networks are not overburden and result in harming privacy goals. Concepts of identity may also be useful in contexts well outside the network, allowing users to potentially make attestations about their location to traditional networked systems (e.g., banking).

**Attributing location.** Future networks need better attribution and tracking in terms of determining where attacks are emerging, so they can be detected and mitigated. From a forensics point of view, being able to determine who is behind an attack, when an attack was launched, and what happened in detail during the attack would be highly useful. However, research on minimizing the impact of tracking is also critical. For instance, in the context of 5G, while it is now much more difficult to obtain long-term identifiers over the air, intermediary networks still eventually receive such information. Because of this, nation-states can arbitrarily track users, making it hard for individuals (and especially USG employees) to "operate through". Specific mechanisms to better protect long term identifiers from potentially hostile nation-states, which also understand the needs for CALEA lawful interception, must be designed. There needs to be some physical layer/hardware assistance to prevent location fraud (e.g., cuckoo attacks), and provide location/distance binding/authentication. Doing this at scale is a major challenge.

**Anonymity and censorship resistance.** Future networks must provide some level of anonymity on the Internet. Research should determine if there are ways to build systems that provide a better "Tor" (i.e., better and more performant anonymous surfing capabilities)? Building these capabilities into the future network would allow more people to use it, and it would also be useful against censorship attempts. This can also include thinking about non-traditional methods of anonymity or censorship resistance. For example, sneaker nets.

**Rearchitecting the Internet for security.** Many efforts such as IPSec, BGPSEC, DNSSEC have seen very poor deployment. There should be research into understanding why this has been the case, including technical, economic, and human factors reasons. Using this information, there should be a theoretical framework that describes how to build secure network technologies if they are to be deployed. Old protocols should also be updated to work within this framework.

**Formal security verification of network protocols.** It may be time to explore formal verification of future network protocol specifications (thousands of pages of natural language text) and implementations (protocol stacks that go into operating systems and drivers); defining new security and privacy properties that should be proved and discovering vulnerabilities, ambiguities, and flaws via verification.

**The security of IPv6.** Research that inspects the socio-technical impacts of IPv6 is needed. For example, do users understand the risk of globally accessible IPs? Do they know how to configure IPv6 networks? What are the implications if they make mistakes? Research should explore what new attacks are available, how traditional attack such as spam or DoS attacks evolve combined with new network infrastructure. What existing defense mechanisms such as firewalls become less effective when facing IPv6 related attacks. Research should explore whether bad configurations will lead to dangerous routing situations.

**Security testbeds.** There is also interest in designing and developing the next-generation testbed. Such a testbed should support fully contained, ethical hacking as well as testing on both clean and adversarial networks. The testbed should be designed such that it does not just support one technology but instead be as diverse as possible in technologies in order to allow for testing and research on interoperability, backwards compatibility, and fallback scenarios. The testbed should also support a wide range of network protocols, with instrumentation and measurement support to assess impacts of attacks and defenses.

**Secure distributed computation.** With future networks, computation has become decentralized, occurring at the edge, the core, the fog, the cloud, and everywhere between. At the same time, systems build on networks are becoming dependent on the security of network communication, e.g., microservices, cloud infrastructures. There should be more research into what this now enables and how networks can support their underlying performance and security goals. Additionlly, rearch investigating how this impacts the security model, whether it is feasible to secure data when it is stored on many systems and the performance impact related to security is warranted.

**Interoperability and backwards compatibility.** A greater focus must be put on the issues of interoperability and backwards compatibility. In the past, many attacks and vulnerabilities in this context were discovered once technologies were implemented. Instead, more efforts should be made at the time of design and specification stages. A focus should be placed on using formal verification methods in the context of enabling interoperation and backwards compatibility (across the full network stack).

**Technical and Societal Impacts**

One key impact could be leveraging future networks to help combat scam calls and phishing attacks, as well as address network borne attacks, e.g., DDoS, network scanning, bots, and worms. These cause immense harm, particularly to vulnerable populations. Research into protecting users based on authentication properties provided into the network would have a strong positive impact on this community. Research could also help provide censorship resistance, allowing people living in repressive regimes to have greater access to information. Not only does this help the public good, but also promotes national security by weakening rogue or repressive governments. Research into surveillance resistance can help protect individual fundamental rights.

Building the capability to "operate through" compromise networks would help protect US citizens against attacks against US infrastructure. It would also allow the DoD to operate in regions where the networking capability cannot be relied upon. By supporting the creation of a testbed for future network, the ability for researchers to conduct research will be greatly enhanced.

**Cross-Discipline and Outreach Opportunities**

Efforts should work with economists and other business analysts to understand the economic and business implications of future networks. This could explore what will be necessary to incentivize the deployment of security technologies. Similarly, there needs to be collaboration with human factors researchers to ensure users understand the security guarantees of these future networks. There is a need to understand how these security guarantees can be used to help prevent scam calls to vulnerable populations. This could also involve collaborations with ML/AI research into creating personal assistants who help protect vulnerable users from scams. Research should also explore the future equivalent of today's ransom attacks in the context of future networks and how can this drive the development of proper security measures.

Other potential areas of cross-discipline research include incorporating formal verification of network specifications and interconnections between different networks (e.g., allowing secure interconnect between 4G and 5G networks). It could also involve hardware co-design to bake security properties into the devices powering the network.

Recent wireless network technology advances largely depend on the new technologies in the underlying hardware technologies. The security of future networks needs joint security across different layers including hardware design and spectrum/waveform analysis, and leverage data analytics to understand system behavior and identify vulnerabilities.

Future networking efforts may include research into satellite communications and other sensitive areas that will require engagement and support from the Department of Defense (DoD) and other agencies. Other governmental agencies (e.g., NASA) might also be natural partners. Working with law enforcement will also prove beneficial.

### 3.2.4   Media Spectrum Security

**Area Description**

Wireless networks at the physical layer are vulnerable to a range of security and privacy threats, including eavesdropping, denial of service attacks, spoofing, and signal interference. It is essential to develop robust, trustworthy, verifiable, and effective security and privacy mechanisms to ensure the security and integrity of wireless communication. This involves implementing measures such as encryption, access control, redundancy, monitoring, and intrusion detection to prevent unauthorized access and protect sensitive data. Additionally, ongoing research and development of security and privacy solutions are necessary to address emerging threats in emerging applications and ensure the continued security of wireless networks. Domains of research interest: wireless networks, IoT devices, CPS infrastructure, network infrastructure generally (wired).

**Technical Efforts**

**Legacy System Security.** One area of research includes what to do with the many, many legacy devices using insecure physical interfaces and network technologies. To ensure long term impacts, future efforts should avoid research investment in band-aids (small improvements, single vulnerabilities), but promote research in fundamental fixes that would have a significant impact across millions of devices.

Legacy network devices are a large problem, with the likelihood of becoming larger over time. Aside from research in new technology domains, the research community needs to investigate potential progress in adding security to legacy devices. Research possibilities include investigating "protocols within protocols" or isolated network domains that have new governing elements, and other considerations.

**Hardware and software verification.** Research is needed in the area of validating and protecting the security of hardware and associated drivers (software). There needs to be more robust notions of attestation, validation, and proof techniques, at both design time and runtime. At runtime, robust notions of trust should be added including "proofs" of integrity and attestation, including how to engineer a new generation of robust devices that is provably trustworthy.

**Physical attack recognition and resilience.** A related area for research may include the physical vulnerability of network infrastructure located outside of national borders, including, but not limited to underwater cables. In the context of Russian threats to escalate its war efforts, the ability to attack undersea cables has been raised.[1] Unlike the more technical challenges around securing vulnerabilities in hardware, these challenges more likely lie in the physical realm of improving security of cables and/or ensuring redundancy for network traffic traversing through potentially vulnerable pathways.

---

[1] https://www.foreignaffairs.com/ukraine/russias-halfway-hell-strategy

**Security management.** The trend is toward more devices installed across an infrastructure. For example, home thermostats and sensing devices, but there is no higher level "orchestration" layer to configure and audit security. Research should examine security management across larger device deployments and creating a more integrated picture.

It is important to recognize that any kind of orchestration should account for various settings or contexts. For example, a home environment for IoT devices is much more clearly defined than if the same devices are used in the context of a smart campus. A smart campus may see more dynamic settings with a bigger variety of user groups where some contexts may also be subject to specific policies or legal regulations (e.g., FERPA). It is also important to understand and account for the range of stakeholder groups that might vary across these contexts: collective orchestration and the interfaces that support this orchestration may vary drastically between the home IoT context (in which the stakeholders may all be individuals in a single family) versus a smart campus (in which the stakeholders may all be part of different organizations with competing preferences).

**Trust.** Trust is a key issue in network devices. There needs to be enough exposure to make trust possible. Users need to perceive and understand trust features. Meanwhile, system notions of trust should make use of robust methods of proving software sources, device integrity, and the identity of the device and its user. Confidential computing (aka, trusted computing) should become more widely used in networking infrastructure.

**Observability.** Security requires observability. Devices should be designed to expose functionality and operation so that security analysis is possible. These observability channels, however, need to be isolated from shared network media to avoid exposure to an adversary. It should be a research challenge to define how network devices provide observability and what the properties are to best support security challenges.

**Defending against DDoS in the context of IoT.** Attackers can leverage insecure IoT devices to mount DDoS. The abundance of connected IoT devices implies a high amplification DDoS factor with a relatively lower cost (The Mirai botnet is a good example). It is imperative to account for this attack vector. A challenge that needs to be addressed here is ensuring that these devices are protected. Unfortunately, the current security state of IoT devices is very weak. The devices are widely known to be under-protected because of a variety of reasons: (1) insecure default security configurations, e.g., default credentials, (2) over privileges, (3) failure to adapt secure updates, and (4) inadequacy of existing security models due to the lower computational power of IoT devices.

**Future hardware architecture.** A key frontier for future investment is the security of new hardware. Network hardware is becoming more complex as domain specific hardware components are added and single systems become ensembles of hardware components. (e.g., modern SmartNICs may include ASIC acceleration elements and general-purpose ARM processes to augment network processing) Understanding the security and privacy consideration of these more complex data path devices is an area of need.

Computational materials in which computing functionality is embedded into everyday materials present both opportunities and challenges. If computing functionality is seamless and hidden, how might mechanisms be designed to provide appropriate notice and consent to stakeholders who use and/or are subject to those technologies? Similarly, how might the new affordances and capabilities of these technologies be leveraged to provide perceptible assurance of privacy and security (e.g., through intentional powering)?

**Security economics.** The economic incentives behind security are a complex and interesting area that needs research. Areas include understanding what it takes to get users to pay for security features, let alone use security features, exploring how technology venders can be incentivized to provide better security features, and how security be monetized for technology providers.

**Anonymity.** Ensuring anonymity in wireless communication is a major concern, particularly for mobile devices that may roam across different networks. Devices at the MAC layer often carry MAC addresses, making users vulnerable to tracking attacks. While randomizing MAC addresses can mitigate this issue, such randomization methods have been shown to be vulnerable to side-channel attacks. Therefore, it is crucial to protect the user's anonymity by implementing robust and reliable measures. In addition, the rise of tracking devices, such as AirTags, presents another challenge in preventing their potential abuse for tracking individuals. Developing effective solutions to address these challenges is an ongoing area of research in wireless security.

**Technical and Societal Impacts**

With the increasing reliance on wireless communication systems (e.g., cellular network, Wi-Fi, and Bluetooth) in everyday life, media/spectrum security has become an essential area of research. Ensuring the security and integrity of wireless networks is critical to protecting user privacy (sensitive data transmitted), preventing cyber-attacks (anti-sniffing, spoofing, etc.), and enabling new applications and services such as massive IoTs, connected and autonomous vehicles (CAVs), and smart cities.

Scientific success in the areas outlined above could lead to significant improvements in physical-layer security for the existing and emerging hardware that power these application areas. For example, a well-understood taxonomy of threats can help inform standards that engineers and manufacturers rely upon when implementing security best practices in hardware. Improving user interfaces for individuals, communities, experts, and data stewards to help monitor data flows and mitigate security threats in the context of consumer-focused hardware, such as IoT technologies, could help with a range of privacy and security challenges from identity protection to physical safety. Providing perceptible assurance for when sensors are active and separating sensor manufacturers from those that provide edge-computing capabilities could engender greater trust that environmentally situated sensors are only active and monitoring as expected. Proactively working on the security and privacy of novel and emergent hardware technologies (e.g., computational materials) should also help get ahead of the new and emergent threats entailed by these technical advances.

**Cross-Discipline and Outreach Opportunities**

With the increasing reliance on wireless communication systems (e.g., cellular network, Wi-Fi, and Bluetooth) in everyday life, media/spectrum security has become an essential area of research. Ensuring the security and integrity of wireless networks is critical to protecting user privacy (sensitive data transmitted), preventing cyber-attacks (anti-sniffing, spoofing, etc.), and enabling new applications and services such as massive IoTs, connected and autonomous vehicles (CAVs), and smart cities.

Scientific success in the areas outlined above could lead to significant improvements in physical-layer security for the existing and emerging hardware that power these application areas. For example, a well-understood taxonomy of threats can help inform standards that engineers and manufacturers rely upon when implementing security best practices in hardware. Improving user interfaces for individuals, communities, experts, and data stewards to help monitor data flows and mitigate security threats in the context of consumer-focused hardware, such as IoT technologies, could help with a range of privacy and security challenges from identity protection to physical safety. Providing perceptible assurance for when sensors are active and separating sensor manufacturers from those that provide edge-computing capabilities could engender greater trust that environmentally situated sensors are only active and monitoring as expected. Proactively working on the security and privacy of novel and emergent hardware technologies (e.g., computational materials) should also help get ahead of the new and emergent threats entailed by these technical advances.

Failure in the areas outlined above could entail several privacy and security threats. First, there is the possibility of a total loss of physical privacy, as ubiquitous sensors may be able to always access and detect individuals, in all contexts. The tampering of physical devices could result in threats to physical safety. For example, if devices can be set on fire or control other aspects of a user's physical environment. Lack of trust in hardware could also result in the stifling of adoption of these technologies, limiting the potential utility these technical advances could engender in society. There are also several sociotechnical ramifications. If users cannot trust hardware, it could lead to chilling effects in behavior and free expression. Privacy and security are fundamental to an empowered and informed population, and thus essential to debate and democracy.

## 3.3    AI Security and Privacy

As artificial intelligence and machine learning continue to advance and permeate various aspects of our lives, ensuring the security and privacy of these technologies is paramount. AI and ML systems handle vast amounts of sensitive data, and any vulnerabilities or breaches can have severe consequences, including unauthorized access, data leaks, and misuse. Additionally, AI-driven applications often make critical decisions that impact individuals' lives, such as healthcare diagnoses, financial transactions, and autonomous vehicle operations. Safeguarding these systems against adversarial attacks, ensuring data protection, and maintaining the integrity and fairness of decision-making processes are crucial for building trust and confidence in AI/ML. By investing in research and development to enhance AI/ML security and privacy, research can create a more resilient and responsible AI ecosystem that benefits society while mitigating risks and preserving individual rights and freedoms.

### 3.3.1   Logic Based AI Security

**Area Description**

Logic-based AI relies on logic and deductive reasoning to represent knowledge. It is applied in domains such as planning, knowledge representation, reasoning, causal inference, and probabilistic graphical models, and is widely used in various applications such as access control, system scheduling, distributed systems, and autonomous systems. Further, logic-based AI can aid in the design and deployment of secure systems such as detecting ongoing attacks.

Logic-based AI security is more challenging since the attacks are more strategic and targeted on the AI planning or knowledge reasoning workflow. Hence, it is critical to create foundational theory and countermeasures for quantifying and improving the security, reliability, and robustness of logic-based AI.

**Technical Efforts**

Logic-based AI may be able to learn higher-order features than other AI methods. This could aid in explainability, interpretability, and more human-aligned models. While logic-based AI seems to facilitate explainability compared to other forms of AI (such as statistical AI), it may be vulnerable to several attacks. Attacks identified include but are not limited to poisoning, evasion/adversarial examples, backdoor attacks, model inversion, membership inference, and policy-based attacks. The full extent of these attacks' likelihood is to be investigated.

There exists a tradeoff between efficiency and robustness of logic-based AI. Defining robustness metrics in the logic-based domain is important for effective defenses of poisoning evasion/adversarial attacks. Clear definitions of attack metrics, equivalent to metrics in statistical AI (e.g., imperceptibility of adversarial inputs) are necessary.

A worthwhile effort could be combining logic-based AI and statistical AI to achieve more robust AI systems. This includes investigating how each of these paradigms could help the other. For example, could approaches developed for robust statistical AI be applied to logic-based AI or could logic-based AI be used to address some of the security challenges of statistical AI? For instance, can logic-based AI be used to secure statistical AI models from adversarial inputs? There also exist scenarios in which logic-based AI can be used to address general security challenges. An example of this is using logic-based AI as an effective firewall to protect users from harmful traffic.

Another possible example domain is the use of logic-based AI in distributed systems. In this sense, logic-based AI can help statistical AI (for example rules-based methods for designing personalized models in a federated learning setting). Additionally, logic-based AI can itself be applied in a distributed setting, which can improve robustness in some respects, or otherwise introduce a larger attack surface in other aspects.

**Technical and Societal Impacts**

Bridging the gap between statistics-based AI and logic-based AI requires the clear definition of distinctions and similarities between the metrics in which performance, robustness, and security (attack/defense) are measured. Success in defining these characteristics enables the establishment of quantifiable security guarantees for a logic-based model. Providing quantifiable security guarantees allows the broader community to employ logic-based AI systems in practice with more confidence. Additionally, discovering translations between logic-based AI and statistics-based AI with respect to metrics, representations, and more could also lead to increased robustness and safety in statistics-based AI.

Efforts aiding in explainability, interpretability, and more human-aligned models, as well as other qualities that have a greater presence in logic-based AI compared to statistics-based AI, will have a large positive impact downstream. Such systems would likely be more robust and inherently safer, but also investigations and quantifications on both fronts would be easier.

Large scale real world intelligent systems typically consist of statistical learning and logic-based AI methods. Another important technical impact of logic-based AI security is the theory and practical methods for such large-scale AI systems level security, which requires formal verification, logic-based AI methods, and human-in-the-loop design to develop effective safeguards.

**Cross-Discipline and Outreach Opportunities**

There are many opportunities for outreach to different disciplines. For example, Bio-inspired AI may help/inspire logic-based AI and logic-based AI security is critical for securing the real-world deployment of Bio-inspired AI in science and engineering domains. Logic based AI security methods and theory can benefit significantly from the inherent ease of explainability of logic-based

AI research, especially logic-based AI security solutions. Similarly, logic-based AI security can also benefit mechanical engineering and other disciplines with heavy use of control theory.

Logic-based AI can benefit from mathematical theory to improve robustness and performance of such systems. For example, formal verification methods in domain specific applications can be adapted in logic-based AI.

Human-AI teaming and Human in the loop of AI are critical to many cross-discipline areas, such as safety in smart health, in smart city, in cyber-manufacturing, and in aviation (GPS based, etc.). Concretely, the medical diagnostic decisions and robotic-based surgical operations are safety-critical and those systems are typically logic-based AI, combined with statistical data driven AI. Safety in such systems can be more challenging than adversarial attacks to AI/ML model training or pre-trained AI/ML models.

### 3.3.2   Reinforcement Learning Security

**Area Description**

Reinforcement learning (RL) concerns sequential decision-making problems. Compared to classic supervised learning, RL can make more complicated decisions and thus be potentially utilized in more sophisticated applications and complex systems.

One promising area is to explore using RL to solve complicated security problems.  RL is good at searching for or learning policies in a giant problem search space by using feedback. Such problems are common in the security domain, e.g., program fuzzing. If RL is utilized for such problems, it can potentially obtain solutions in a more efficient and scalable way. There are some key challenges to achieving this goal; how to design proper environments and agents, including their actions and rewards. This requires extensive domain knowledge. Another problem is how to better combine the perception ability of deep learning with the decision-making capability of RL.

Research must also explore how to ensure RL (and the systems it supports) are secure and trustworthy is also important. This challenge can be tackled using the following dimensions: 1) Environment and agent setups:  e.g., single-agent, multi-agent, competitive, or collaborative, 2) Threat models: RL will bring in new attack surfaces compared to classic supervised and unsupervised machine learning. Based on the new attack surfaces, research can explore and define novel threat models, 3) Security properties: Like supervised/unsupervised deep learning, attention must be given to security properties, such as robustness, privacy, and fairness, 4) Applications: the specific security risks of RL in different applications needs further research.

**Technical Efforts**

**Novel-use cases of RL for security.** RL is better at solving sequential decision-making problems compared to traditional ML. This potentially enables it to solve more complicated applications that require solving a problem via multiple stages. In addition, since RL requires design environments, actions, and rewards for specific applications, it provides a perfect opportunity for embedding expert knowledge. The strength of RL is aligned with continuous improvement based on feedback. The community can explore more novel use cases of RL in security scenarios that will benefit from the iterative learning from RL. Research efforts are needed to build up an open and easy-to-use RL infrastructure or testbed to support the implementation and deployment of such use cases.

**Understand attack surfaces of RL and its application.** RL itself is an active research topic in the ML research community, however RL-based security applications are still limited. The community needs to study the unique attack surface of RL and its applications. The community also needs to be mindful of RL for engineering attacks and malicious models. Researchers need to be mindful about interaction and inference from partial (or potentially poisoned) information. The definition of RL agents needs expert input to design the RL system, to some extent, it makes the attack easier, in the sense that the attack could keep feeding the system malicious feedback that leads the RL agent to fail. The community could study efficient ways of launching such attacks, as well as effective defenses against the attacks.

**Explainable RL.** RL has great potential to support sequential decision making. Embedding RL with explainability will help usability of the RL mode. The explainability of RL includes supporting local/micro explainability at each step of the agent's decision-making point, as well as global/macro explainability at the policy level. There is a need for the research community to study, define and deploy explainability in RL.

**Other trustworthiness concerns about RL.** To make RL more trustworthy in general and a better fit for security applications, research needs to study the security concerns of RL other than adversarial robustness, for example, privacy-preserving, and fairness. More work is required on formalizing these properties and in studying the trade-off between these properties.

**RL data opportunities and challenges.** Security research progress has been inhibited by availability of big datasets, especially anomaly ones. RL requires less labeled data to begin with, better aligned with security research where the ground truth is lacking. RL has potential to help grow security research and gather more data. Of course, generating high-quality data for RL is still a challenge/opportunity. There are also challenges aligned with interactive exploration.

**Adversarial learning against RL.** Adversarial learning against reinforcement learning poses greater challenges than adversarial learning against supervised learning techniques. The primary obstacles lie in manipulating the environment in which the RL agent learns and manipulating the rewards that drive the agent's learning. Additionally, in collaborative agent scenarios, adversarial learning can be launched through other agents, making it difficult to manipulate these agents in the environment to mislead the target agent. As a result, this remains a challenging problem in adversarial reinforcement learning research.

**Protecting IP of RL models.** When an RL agent is published, even without releasing the details of the agent, attackers could perform imitation learning to copy the model and even obtain a new agent stronger than the published agent.

**Availability of learning data.** Reinforcement learning (RL) can sometimes require bootstrap data to facilitate training, yet such data may not always be readily available. As such, a significant research effort is devoted to finding effective ways to obtain meaningful bootstrap data for RL.

**Reward function design for RL driven security application.** Creating a reward function for gaming agents is typically a straightforward process, but the same cannot be said for security applications. As such, the primary research question revolves around discovering effective approaches to constructing reward functions for a range of security-related use cases.

**RL model poisoning.** The training process of the RL model may be vulnerable to attacks from adversaries, who can introduce misleading trajectories to the agent, causing a decline in model performance. The research question at hand is how to effectively detect and mitigate the presence of malicious inputs during the model training procedure.

**Auditing and compliance of reinforcement learning systems.** For privacy and compliance requirements, it is essential to determine the dataset on which a model has been trained, tested against. However, for reinforcement learning scenarios, the models are being continuously trained and updated. Therefore, it is hard to keep track of the provenance of the data, apply security, privacy, and security policies. Can there be hybrid reinforcement learning - one trained on encrypted data or differentially private data, and then it learns on plaintext data in a continuous manner. In all these scenarios and vice versa, how can one assess data leakage, privacy leakage and compliance assurance risks, and how to apply such policies on the models. How can auditing be carried out in this regard? Accountability of a decision by reinforcement learning depends on the data - so if an autonomous car using this system leads to a crash, who is accountable for that crash - the real-time data that it has learnt on, the algorithm, the data or algorithm supply chain, the policies, the reward functions. It is an open problem.

**Technical and Societal Impacts**

Reinforcement learning provides exciting advancements in self-driving vehicles, automated supply chain management, personalized learning/training systems, and more. Securing RL will greatly improve the security and reliability of these existing applications. Since RL can be used in multifaceted applications, improvements in security of RL would have significant technical impacts. Furthermore, improvements with RL security can broadly help in making ML secure.

The field of AI in general has advanced rapidly in the past decade, with many technical breakthroughs and rapid adoption of these technologies into society. The advancement of reinforcement learning creates unique opportunities for improving security and safeguarding digital assets. There is concern that reinforcement learning may be used to produce new forms of attacks, which may exploit personal data, deepen extremism and polarization in society, and threaten safety and well-being.

**Cross-Discipline and Outreach Opportunities**

RL can contribute to many application domains. For example, in healthcare, RL can be used for automated medical diagnosis, and even drug discovery design and development. However, confidentiality, integrity, and confidence in such applications can be undermined by the misuse or exploited vulnerability of RL. As another instance, the gaming industry can benefit from improved user experiences by integrating more secure and robust RL into their game design. Enabling RL applications (e.g., auto drive, robot control) and security requires cross-disciplinary research including robotics, control theory, mechanical engineering, electrical engineering, etc.

RL also opens new opportunities for effective pedagogy and the learning process. RL empowered education applications which can enable better personalized learning paths with desired outcomes.

Reinforcement learning has been used in systems that are widely used by society and are increasingly integrated into other applications, e.g., ChatGPT and its integration into search engines. Improved understanding on the security of reinforcement learning can make these critical applications more robust and trustworthy. Adversaries will attempt to attack these RL-based systems given their widespread deployment. It is important for the research community to stay ahead of malicious parties, anticipate potential attacks, and strengthen these systems.

RL can benefit data science and programs that gather data (such as IMR). For many application domains, lack of training data prevents effective development and deployment of powerful Machine Learning techniques. Reinforcement learning has the potential to generate labeled data that can be used for training other models. Understanding the security of reinforcement learning regarding initial bootstrapping data, training environment, and reward functions, is critical to make this a reality.

RL benefits from security experts, since studying the security implications of reinforcement learning can help to prevent potential vulnerabilities and threats that may exist in RL-assisted platforms. This will help to develop robust and reliable protocols with safer data selection mechanisms for training in such critical applications.

RL can benefit from game theory and economics: Game-theoretic reinforcement learning using economic models can be a game-changer: reinforcement learning systems can use game-theory for developing new attacks or defending against attacks and protect data, systems, users from economic-driven attack losses (need not be financial but based on economic models).

For audit and compliance purposes, cross-discipline work is needed for legal and policy development. For fair and ethical reinforcement learning, societal studies as well as legal, lawmakers may be involved for the same.

Security aspects for reinforcement learning models for autonomous driving with sensor fusion may be different than security aspects for financial domains such as stock markets, or healthcare, or education, or security itself.

### 3.3.3 Trustworthy & Security of AI ML

**Area Description**

The importance of trustworthiness in machine learning cannot be overstated. As machine learning becomes increasingly pervasive in critical decision-making processes across domains (e.g., healthcare, finance, and autonomous systems), it is crucial to ensure that these algorithms are reliable, transparent, and accountable. Trustworthiness in machine learning encompasses multiple dimensions, such as fairness, robustness, interpretability, and privacy. Fairness ensures that machine learning systems do not perpetuate bias or discriminate against certain individuals or groups. Robustness guarantees that algorithms can withstand adversarial attacks and unexpected input variations. Interpretability enables humans to understand and validate the decisions made by machine learning models, promoting transparency and accountability. Finally, privacy safeguards sensitive data from being compromised during the learning and inference processes. Prioritizing trustworthiness in machine learning can foster public confidence, promote ethical deployment, and maximize the societal benefits of this transformative technology.

The area of human and AI trustworthy teaming refers to the development and deployment of AI/ML systems in a way that fosters collaboration and partnership between humans and machines, with a focus on ensuring that the resulting systems are trustworthy, safe, reliable, privacy-preserving, and ethical. This area is by necessity a cross-disciplinary effort spanning formal models, statistics, empirical machine learning, and cyber-security.

**Technical Efforts**

**Collaboration and Usability:** Research effort needs to increase understanding of the application domain/content and the role of AI/ML vs. the role of humans to achieve better AI/ML-human teaming. For example, if an AI/ML system is to work on behalf of a human, it needs to have an input validation module to ignore what the application/human would ignore and an output scoping module to produce output that the application/human would have understood and performed. If an AI/ML system is to augment or help a human's work, then it needs to let the human know when it needs human input or when/why it cannot help, or when/why its prediction has low confidence.

More needs to be known about how humans understand the outputs of AI, given the flaws of AI. It is key to be able to predict the circumstances of when and how humans should/will trust AI to provide trustworthy information, assess its correctness, know when to trust it, etc. Relatedly, more research is needed on how AI systems can self-assess their own ability to be credible/trustworthy and convey that to users in a usable way.

Better and more robust mechanisms are needed for task assignment, namely, identifying tasks where humans are better vs. where the ML/AI system is better. ML may be better at certain parts, while humans may be better at others. Determining who/what/when to assign tasks to, is an open research problem.

How to design AI/recommender systems, so they do not recommend harmful things is needed. Part of this question deals with understanding how to design safer AI systems that do not give preference to harmful outputs.

**Data quality.** Creating/curating more representative training sets to train models is critical. One technical approach to this problem is to investigate data-centric or content-centric approaches that sign and authenticate data directly. Reliably knowing the sources of data can provide confidence in the data, and perhaps more importantly, allow blocking or exclusion of known bad/malicious sources.

**Transparency and Interpretability.** There are many questions regarding transparency and interpretability. For example, what are the user needs for explainable AI systems? How to design specific explanation methods for users with different levels of expertise? Can general explanation methods be customized to specific user needs? In addition, alternative ways to establish trust aside from explainability need to be explored. For example, some complex ML models may not be explainable. When such models are deployed in an application/human-interaction context, can additional data/behaviors be used to augment/cross-validate the output of the ML models? One possible direction is transparent training data - if the inputs are auditable, maybe it is okay for the model to be a black box, while still providing trust in the overall outputs.

**Risks brought by human-AI teaming.** The risks of feedback on models need to be explored. For example, if a user/adversary understands how feedback/reinforcement-learning works in the models, he/she can provide specific training data or test cases to cause the models to bias towards his/her input. Such data poisoning can be subtle because the adversary can be a "normal" user, or the model is relying on an open environment such as the Internet for training/validation data.

Another research direction is to understand motivations for why people would poison systems and create user interfaces that would avoid such. This will require creating threat models for AI systems, an empirical understanding of motivations for attacking AI systems, and why predatory AI systems are deployed.

**Policy and regulations for human-AI teaming.** As human-AI teaming becomes more integrated into users' daily lives, policy and regulations for human-AI teaming are increasingly important. Although there have been some efforts to regulate AI, many open questions remain. For example, is the responsibility for the outputs of human-AI teams determined? Are those who use the system held responsible? How about those who create the system? How about those who create the data sets?

**Trust and safety issues that arise due to (malicious) ML-human interfaces**. Human-AI teaming may occur in scenarios where the "user" is unaware they are interacting with the AI/ML system. The combination of convincing dialogue generation systems and human-appearing audio/video will make fraud significantly more difficult to detect and defend against. A wide variety of new technical mechanisms will be necessary to defend this ecosystem. For instance, determining the provenance of multimedia (i.e., being able to tie generated content to a specific model or even a specific instance of a model) will be extremely important. Methods to both explicitly and implicitly watermark such output will be important in this effort. Similarly, techniques that prove the liveness or actual existence of an event (e.g., a political speech and its unmodified contents, an interview with a journalist) will also be important tools in the societal consumption of information.

**Supply chain issues with ML models.** There may be interesting research problems emanating out of having trust in the ML systems. For example, while software supply chains have been an active area of study, fruitful research in understanding whether inclusion of ML models makes more difficult or less useful deployment of existing techniques can be imagined (i.e., because model parameters can be attested to easily, but attesting to how these were generated is material to the trustworthy guarantee).

**Decision-making systems**. There are a number of potential research questions in the context of ML-assisted decision-making systems. This is an intentionally broad framing, reflecting the wide swathe of places where ML-assistance may arise. Traditional ongoing topics such as algorithmic fairness (bail suggestion algorithms, facial recognition systems, recommender systems), and others require more research. At the same time, new contexts will arise where an ML system is directly being used to inform human decisions: as ML is integrates into more technological tools even seemingly banal systems as office applications, email, scheduling, filling out forms, etc., are all going to be transformed by the automation abilities of ML. The potential productivity gains and improvements from these operations issues need not be overlooked.

**Human-robot interaction.** While they fail under the decision-making systems above, the physical safety requirements of semi-autonomous vehicles, other forms of robotics, etc. make this area different from non-physical world decision-making systems. This area is well-recognized as a critical area of study.

### Technical and Societal Impacts

AI/ML teaming is critical for the future of computing and society. Successful efforts in the space of AI/ML-human teams will result in AI systems that are trustworthy, verifiable/certifiable, and safe for humans to team with. As more AI systems are deployed in new tasks and processes, human interface and interaction with these systems will pose new opportunities, challenges, and threats.

Success in this area will improve humans' efficiency and safety and inclusion, and fairness in society. Indicators of success in this area include increased efficiency (in terms of time, money, energy, environmental impact) of AI-human teams, humans' ability to make good decisions about when and how much authority to give to the outputs from AI, AI's ability to amplify appropriate human expertise, and human-AI teams' ability to do work that neither humans nor AI alone could do. How the team impacts the human agency and status within society is a crucial component of measuring success. Replacing humans in teams without providing other productive ways for those humans to contribute to society is an area of further consideration and research.

There are many new challenges in this domain: e.g., security of human-AI teaming, privacy, and ethical risks in human-AI teaming. If the scientific community fails to make progress, the human-AI teaming will make bad decisions in critical domains, which introduces risks in fairness, health, safety, environmental protection, and many other critical domains. For example, research needs to focus on ways to ensure equitable access to AI systems/agents, or there is a risk that humans who do not have access to AI become second-class citizens, relegated to interacting with AI systems instead of humans and having limited access to social resources.

If the scientific community fails to make progress in this space, the issues prevalent in AI systems today have the potential to influence, endanger, and cause mistrust in the humans that interact with them. New methods to enforce users to be informed when they are interacting with an AI system, and for verifying/certifying that AI system will be critical in preventing mass spread of misinformation via bad actors using human-seeming AI systems.

**Cross-Discipline and Outreach Opportunities**

There are multiple opportunities for productive collaboration with areas outside the traditional computer science (CS) boundary. For example, research on human factors in technology has a long history of cross-disciplinary collaboration, drawing from the computer science discipline of human-computer interaction (HCI) but also from psychology and industrial engineering. Continuing and enriching these collaborations is crucial to progress in human-AI teaming research. In addition, the study of law and policy can inform legal frameworks that define boundaries for human-AI teaming and shed light on liability and copyright issues. The fields of ethics and philosophy have long dealt with moral questions that take center stage when designing human-AI teams. Collaboration outside traditional CS boundaries is crucial to developing trustworthy human-AI.

Further, bootstrapping and establishing trust in human-AI teaming would benefit from a broad, cross-disciplinary perspective. Engaging social scientists, psychologists, and decision scientists is of clear importance to model and improve how AI-generated information is presented to humans in a team setting, especially in socially consequential contexts like distribution of welfare resources, prediction of recidivism, and clinical diagnosis. Helping human partners understand how this information was generated, the level of confidence in that information, as well as the counterfactuals will be of particular importance in these contexts. Engaging marketing experts to help understand how to appropriately convey the advantages and limitations of AI systems will also be of interest.

Human-computer interaction researchers will be helpful to engage in designing interfaces that enhance user agency and transparency. For example, as human-AI teaming becomes more mainstream, it will be of increasing importance to provide users with interfaces that helps them understand when they are engaging with AI-generated information as well as controls that can help increase or reduce that engagement on demand. Concepts of seamful design may be particularly useful in these contexts — instead of hiding the AI from people, how can clear touchpoints be provided to allow users to audit and control the AI with which they interact?

Since human-AI teaming will soon be omnipresent and cross-cutting across many problem domains, it will also be important to engage context and domain-specific expertise. For example, for clinical diagnostics, it will be important to engage clinicians and medical professionals. There are several outreach opportunities that such multi-stakeholder engagements might entail as well. For example, in cases where human-AI teaming may be used to help social workers make decisions about resource allocation for the unhoused, researchers and practitioners should engage the communities directly impacted.

Finally, it would be pertinent to engage public policy and legal scholars to help disentangle the policy and legal implications of human-AI teaming for high-stakes decision making. For example, as autonomous vehicles become increasingly autonomous, there will be many questions about accountability and blame when bad decisions are made. In these contexts, it will be helpful to understand policy and legal perspectives to see if these perspectives can be directly integrated into the AI decision-making process as well as to produce outputs that would be useful to reliably attribute blame after failure occurs.

### 3.3.4   Robust Machine Learning

**Area Description**

Conventional ML studies focus on improving accuracy and efficiency of ML. Security is a big challenge for deploying ML for real-world applications such as intrusion detection, autonomous systems, identity detection/verification and precision healthcare. An adversary can compromise the confidentiality and integrity of ML via tampering with its training and/or deployment phases. It is critical to develop theory, algorithms, and tools to quantify and improve robustness of different ML algorithms and systems, including but not limited to, supervised learning, federated learning, self-supervised learning, recommender systems, large, pre-trained models, etc.

The community should rigorously define *Robustness of AI/ML systems* in a security context. That is, not being manipulated or tempered by an adversary for all layers of data collection, ML algorithm/architecture needed for training, and predictions.

**Technical Efforts**

**Realistic datasets for security applications.** The absence of realistic security datasets has seriously hindered the advancement of AI/ML research for security domains. Particularly, different from other domains, security datasets would pose the following challenges. First, realistic datasets for security applications often contain privacy or IP sensitive data fields which can raise serious legal and regulatory concerns. Second, data collected from the wild at large scale are often unlabeled or incorrectly labeled and mostly security datasets are largely imbalanced. It is difficult to generate or obtain timely and realistic datasets for adversarial/malicious activities.

**Observability of system generated logs and traces**. ML robustness depends on quality datasets as well as their data sources. Research should consider how system and network appliances at different layers generate data to be input for ML-based security analysis. The scope of the effort would include computing devices would include, but not limited to, desktop computers, servers, network monitors, IoT devices, CPS systems etc.

**Provably robust ML**. One unique challenge of securing ML is that attackers are adaptive. Provably robust ML aims to defend against advanced, adaptive attacks via providing mathematically sound guarantees against all attacks under some constraints. However, existing provably robust ML techniques assume threat models that are often unrealistic in real-world applications. It is critical to develop new mathematical theories, algorithms, and tools to analyze and build ML techniques that have provable robustness guarantees under realistic threat models in real-world applications.

**System and hardware-level assurance for robust ML.** The community also needs to develop solutions to protect AI/ML systems for their hardware and system level composition and architecture. As the computer architecture community advances the technology with deployments with trusted computing. Robust hardware isolation and protection methods are needed for ML model deployments – another aspect of robustness. More research on applying trusted computing frameworks (TEEs) to models is needed. Attestation techniques should also be applied to data – training and inference.

**Scalable ML systems.** Considering the processing capacity of ML systems is crucial for ensuring their robustness, as we suggest. In security research, it is important to address concerns about applying ML models at scale without disregarding important artifacts from the influx of streaming data input, especially from fine-grained data points. Some of the specific aspects to consider include larger data sets, streamed data, multimodal data, and real-time data.

**ML systems for heterogenous devices.** Scaling ML models to different types of devices is an essential aspect of applying them in a realistic context. However, this process involves various tradeoffs that must be considered for different security applications, making it a topic for research. The application of large ML algorithms to constrained devices should be explored. In this regard, there should be consideration of various aspects of ML system operations, including but not limited to performance, resource consumption, and model update operations.

**Extended threat model.** New threat models are required for different ML applications across various security domains (e.g., temporal dimensions, domain-specific problem space considerations). As the research community sees growing adoption of ML systems in previously unexplored social and economic areas, where fine-grained data collection and enhanced model capacities are expected, renewing the traditional threat model is suggested.

**Future ML-based security and its robustness considerations.** Prospecting the coming decade, ML-based security systems should not only consider statistical data analysis and prediction but should work on automated response to data for particular security objectives. This would help ML to graduate to AI within the security space. The following directions are suggested. First, binding ML predictions to automated actions with security objectives increases the requirements on robust ML. Also, ML-based security should consider adversarial uses of ML.

### Technical and Societal Impacts

The technical and societal impacts of addressing security, robustness, and trust in ML/AI systems are far-reaching. By prioritizing these aspects, we can promote public trust, foster responsible technology deployment, and unlock the potential of ML/AI for positive societal transformation.

*Security and Robustness of ML*: The security and robustness of ML systems have significant impacts on various societal domains. With the widespread application of ML and data science across industries, ensuring the integrity and reliability of these systems is crucial. By addressing security vulnerabilities and enhancing robustness, the potential risks of data breaches, adversarial attacks, and unintended biases can be mitigated, promoting trust in the technology.

*Public Trust in Technology*: Public trust in technology is a vital aspect that affects multiple areas, including public health, democratic institutions, data analysis, and ML-based systems. Building and maintaining trust in these domains is essential for the successful adoption and acceptance of technology-driven solutions. By prioritizing security, transparency, fairness, and accountability in ML systems, public trust can be fostered, enabling the responsible and ethical deployment of technology for societal benefit.

*Robustness in Different Execution Contexts*: Ensuring the robustness of ML/AI systems in various execution contexts, such as Internet of Things (IoT) and Cyber-Physical Systems (CPS), is crucial for accelerating the adoption of ML/AI-based systems in diverse scenarios. Robustness considerations address the challenges associated with resource constraints, varying environmental conditions, and potential adversarial settings. By developing techniques and methodologies to enhance the robustness of ML/AI systems in different execution contexts, the potential for real-world deployment and widespread impact can be significantly expanded.

*Accelerating Technological Adoption*: By addressing security, robustness, and trustworthiness concerns in ML/AI systems, the adoption of these technologies can be accelerated across industries and sectors. ML/AI-based systems have the potential to drive innovation, improve decision-making processes, and enhance efficiency. However, ensuring their security and robustness is essential to gain the confidence of organizations and individuals, enabling them to fully leverage the benefits of ML/AI technology while minimizing potential risks.

## Cross-Discipline and Outreach Opportunities

Fostering cross-disciplinary collaboration and outreach among the ML, information security, statistics, and formal methods communities can result in valuable synergies, facilitating advancements in research, education, and practical applications in these fields. Exemplars of this work include:

*Realistic and Trustworthy Datasets*: Collaboration between the ML, information security, statistics, and formal methods communities can lead to the development of realistic and trustworthy datasets. These datasets would be invaluable for advancing academic and educational communities, enabling researchers and educators to conduct various research activities and educational initiatives with reliable and representative data.

*Interdisciplinary Collaboration:* The intersection of ML, information security, statistics, and formal methods presents an opportunity for interdisciplinary collaboration. Researchers and experts from these communities can come together to exchange knowledge, methodologies, and best practices. This collaboration can foster innovation, address challenges, and advance the understanding and application of techniques related to data analysis, information security, statistical modeling, and formal verification.

*Educational Initiatives:* Cross-disciplinary outreach and educational programs can be established to promote the exchange of ideas and knowledge among the ML, information security, statistics, and formal methods communities. Workshops, seminars, and training sessions can be organized to facilitate learning, skill development, and knowledge sharing. By fostering collaboration and interdisciplinary learning, these initiatives can nurture a diverse and well-rounded pool of researchers and practitioners in the field.

*Practical Applications and Impact:* Collaborative efforts between these communities can lead to the development of practical solutions and techniques that have a real-world impact. By combining expertise in ML, information security, statistics, and formal methods, researchers can address challenges related to data privacy, cybersecurity, algorithmic fairness, and trustworthy AI. These advancements can benefit industries, organizations, and society at large by promoting the responsible and ethical use of data and AI technologies.

### 3.3.5 Sage Generative AI

**Area Description**

Generative AI (GAI) has recently captured public attention by purveying freely available, easily usable services that display a (sometimes surprising) level of human-like responsiveness. The increased ubiquity of this emerging technology raises new security-relevant opportunities and security risks that demand new research. Security risks include GAI information privacy (e.g., information leakage from one GAI-user to another), GAI misuse (e.g., for misinformation or deception), GAI-assisted adversarial attacks (e.g., leveraging GAI to embed secret commands into audio, or backdoors into code), and malicious GAI mis training to mislead and deceive. Security opportunities include improvements to usable and explainable security (e.g., more informative security warning messages), automatic detection of misinformation or plagiarism, and automation of tedious, security-enhancing tasks, such as software vulnerability, attack, and risk summarization for better informing human users.

**Technical Efforts**

**Plagiarism Detection.** Plagiarism is one of the top concerns raised by the public regarding GAI misuse. Future research should address this concern by innovating new plagiarism detection algorithms and methodologies that estimate the probability that a given document or other artifact was generated by various forms of GAI, or that can identify GAI-authored fragments within a larger document. Such research should leverage and extend prior work in authorship detection and similarity classification to accommodate GAI authors and similarities.

**Sociotechnical Systems Fraud.** Generative AI models pose a security and trustworthiness challenge for sociotechnical systems that seek human-generated submissions for contests, such as publication venues, grant opportunities, and fine arts competitions. Guarding against AI-generated content is important to respect the intent of these contests–i.e., to reward human effort, and to avoid overwhelming referees. Accordingly, a task (with many other applications) is to automatically detect GAI content with the assumption that it is aberrative in each context.

A specific instance of this problem is online survey deception and misuse: With powerful large language models like InstructGPT, it becomes much easier to create bots that can automatically complete online surveys with answers that look highly credible. There may be business developed around this idea to cheat in online crowdsourcing platforms like MTurk and Prolific, and eventually pollute the data for empirical research and data for training future machine learning models. Research is needed to investigate the impact of GAI on research areas like social science and AI, and to design new research methodologies that can overcome this potential threat to data integrity.

**GAI's Impact on Safety-critical Human-in-the-Loop Systems.** With the fast advancement of GAI, human-in-the-loop may become more important than ever. Do humans perform security-relevant tasks more accurately and efficiently by reviewing and correcting GAI-provided solutions, or does that strategy backfire by misleading humans into missing subtle GAI-produced vulnerabilities that a human author would not have generated at all? The extent to which humans can provide effective interpretation and safety guidance is an unknown area that should therefore be explored. Logical layered approaches involving GAI and humans might be an interesting direction to investigate.

**GAI Trust Models.** People using GAIs to augment their work is a growing trend, however, the content generated by GAIs can contain inaccurate or harmful information. This could have a severely negative effect when GAIs are used to build real-world products, such as inserting code with security vulnerabilities and generating privacy policies that fail to comply with privacy laws and inaccurately represent the apps' behaviors. Accordingly, significant research is required to design new human-AI collaboration paradigms to help people avoid these issues and hold people accountable when taking advantage of the productivity gain.

Generative AI models have begun to supplant search engines in how people find answers to questions and gather information about topics. This poses related challenges related to trustworthiness. One challenge includes accurately characterizing the limitations of the GAI's output, to report how confident the user should be in each answer's correctness, what kinds of errors could be present, and limitations in scope or applicability. Another challenge is the possibility of self-sustaining loops in information generation: a common answer that a GAI gives to a question may become the commonly accepted answer, regardless of whether it is correct. Techniques must be developed to identify and assign trustworthiness metrics to GAI content. This will require some criteria for establishing trustworthiness, as well as techniques for ensuring that the system-generated content meets those metrics.

**Adversarial Mistraining and Misuse.** Malicious user input poses a security risk for GAI models. For example, ChatGPT has been shown to be persuadable in natural language (i.e., via user input) to violate content generation rules that OpenAI developers instructed it to follow. This kind of vulnerability, expressed in natural language, is novel for computing systems and poses an opportunity at the intersection of natural language processing and security research.

The quality of GAI output has not been systematically studied. There are many unknown aspects in the output that may misguide users. To ensure the quality of the GAI output and safe use of it, a spectrum of assistive tools should be designed to evaluate different aspects of GAI output in terms of appropriate scope, ethical compliance, content accuracy, etc.

**Explainable Security.** There has been a big knowledge barrier issue of security and privacy both for end users and practitioners (e.g., app developers). The potential issue is that current resources of security and privacy advice are out-of-context, scattered, and consist of dense text full of jargon. Future research should explore the potential of using LLMs to analyze these resources at scale and convert them to a form that can be more easily understood, to make better use of the collective wisdom of the crowd to inform users of how to better protect privacy and security or help developers build more privacy-preserving apps.

**GAI-Assisted Secure Software Engineering.** GAI is already being used in many sectors for faster code development and prototyping, but the security implications and opportunities for GAI-enhanced software development cycles is not well understood. Future research should investigate and quantify whether adding GAI-assisted phases to code development workflows introduces more or fewer software vulnerabilities, evaluate the severity and impact of GAI-introduced vulnerabilities, and develop best practices for mitigating the resulting risk. Technologies that can automatically secure or detect insecurities in AI-generated code should be studied.

Leveraging GAI to address these security risks also offers promise. Future research should consider GAI-automation of tedious software security tasks for which humans are error-prone or have difficulty scaling their activities to large workloads. Such tasks include textual summarization of software vulnerability reports and patches, explanatory code comment insertion for better code readability and maintainability, and human-AI collaboration strategies for vulnerability detection in large codebases.

**Technical and Societal Impacts**

The rising influence and effectiveness of GAI suggests that it will become increasingly intertwined with security-sensitive human activities in the future. Success in these research areas is therefore expected to have a significant impact on achieving safe and ethical use of GAIs for our society. The impacts are twofold. First, there are many potential dangers due to the proliferation of GAIs, which include but are not limited to the privacy concerns of sharing data with a centralized GAI server and/or model, the easier creation and dissemination of misinformation, the threat to the data integrity of future social science and AI research, aiding in plagiarism, and introducing more security bugs to software. Future research must characterize these dangers and formulate effective mitigation methods. Second, the research should also explore how to make good use of GAIs, such as building tools with GAIs to help users obfuscate their information to preserve privacy and using GAIs as a design prototyping tools for apps that involve sensitive user data to foresee the potential privacy risks.

**Cross-Discipline and Outreach Opportunities**

One challenge at the intersection of trustworthiness, ethics, and human-centered design is how to communicate the distribution of labor between GAI and humans when both are involved in creating an artifact. This kind of human-machine teaming is likely to become more common as GAI lends itself to increasingly powerful productivity tools. Communicating the work distribution effectively is important to avoid misconceptions about credit for effort and liability for adverse outcomes.

GAI is already having a significant, immediate impact on education, and this impact is likely to increase. On the one hand, there exist benefits of GAI in the educational domain related to its elevation of student interest in AI and the mathematics that drive it, and its capability go beyond static answers to provide interactive, explanatory dialogues for humans seeking knowledge. On the other hand, GAI is also being abused in the educational domain. For example, it has provided students with new cheating capabilities and raised new honesty threats for instructors. Defining proper usage of GAI for educational purposes is challenging but critical for the next generation of educators and students at all levels.

Policy and law are additional cross-disciplinary areas for GAI trustworthiness and accountability. Who is at fault when errors or provocative content from a GAI model leads to adverse outcomes, such as financial loss or death? Relevant parties include entities (people and organizations) that created the training data, entities that trained the model, entities that packaged the model into a usable product, entities that used the model to create the problematic output, entities that audited (or failed to audit) the output, and entities that consumed the output. Determining ethical liability may involve collaboration with philosophers, and determining legal liability may involve collaboration with legal experts.

### 3.3.6 ML/AI for Security

**Area Description**

The area of AI-enabled security and privacy encompasses research and challenges concerning the use of AI and ML technology to improve security and privacy outcomes. AI/ML for security and privacy can substantially increase coverage and monitoring capabilities, as well as take on proactive roles in supporting secure and privacy-preserving systems design and operation.

**Technical Efforts**

**Attack detection and forensics.** ML techniques are traditionally used to detect attacks — differentiate attackers from legitimate users. In the foreseeable future, it is predicted there will be a need for even more sophisticated ML models for attack detection.

**Creating and sharing data relevant for security tasks**. A novel challenge is to find secure ways to share data about security incidents, attack traffic, authentication requests that can be used by researchers to develop and test new AI/ML models. Research progress is hindered due to the lack of availability of such datasets. Methods to securely share such datasets while preserving the privacy of users is an area of need. These datasets can be used by AI/ML models to predict "security weather" (e.g., vulnerabilities being exploited now, vulnerabilities likely to be exploited soon, areas likely to be impacted, etc.) and make precautionary recommendations.

**Risk assessment.** Currently, AI is already being used extensively for "Tier 1" risk detection – finding real-world developments that warrant further investigation by analysts. However, future research could enhance the ability of AI to assist with "Tier 2" risk assessment or carry out that further investigation. AI could help in two ways. First, AI could identify the specific areas of risk posed by the event that are relevant for the "client" for whom the risk assessment is being performed (e.g., government, NGO, company, individual, etc.). Second, the AI could do the initial gathering of additional information in the identified relevant areas. Consider, for example, an earthquake in Country X. AI could identify that for company A, the additional information that is necessary for follow up investigation concerns the stability of the currency of company A, the extent of physical damage from the earthquake, the number of refugees generated by the earthquake, and the functioning of transportation into and out of the impacted area of the country. Then the AI could proceed to scan news reports, social media, statements by local authorities, etc. to provide initial assessments in all these areas, LLM generative chat bots (e.g., Chat GPT) could then compile that information into a report. This could both substantially cut down the amount of human time necessary to carry out the Tier 2 assessment and/or allow the human analyst to target their effort more thoroughly in other areas.

**Code security analysis.** New ML techniques could help automate several security vulnerability analysis steps and help developers write secure code. Research should also investigate how existing ML models (e.g., GitHub Co-pilot) help (or not help) developers in writing secure and privacy-respecting code. How to communicate these ML model outputs to developers, so that they can understand and take steps to mitigate the security vulnerabilities?

**Automating compliance verification.** AI systems can substantially automate compliance verification tasks. This may include the analysis of privacy policies to check them for compliance with regulatory and legal requirements, compliance of a system's data practices with its privacy policy or security obligations. Challenges include analysis of vague textual statements, necessitating interpretation of policy text.

**Privacy and security transparency generation.** While compliance verification looks at disjointly produced artifacts (laws, policy documents, code) to assess their consistency, AI-based approaches could also be used to automatically generate human- and machine-readable reports regarding data practices and security measures used by an application or system. Such reports could then be kept up to date when code changes and could be based on code annotations.

**Personalized security and privacy assistants.** AI agents can help improve privacy and security outcomes both for individuals and for organizations. Such agents can take over certain privacy- or security-related configuration tasks or provide recommendations. AI agents may also help individuals recognize when dealing with scams or phishing and social engineering attacks, as well as provide support with security/privacy hygiene.

**Detecting AI-generated contents.** As generative models become prevalent, it becomes necessary in some contexts to detect the presence of AI-generated content (e.g., deep fakes). Need to develop AI/ML models to do so.

**AI/ML Model Adaptivity.** AI/ML models developed for security and privacy applications need to adapt to changing conditions/landscape (e.g., adversary adaptation, underlying usage pattern changes). It is important to develop AI/ML models that continue to learn/adapt over time without having to retrain them from scratch repeatedly.

**AI/ML models for mis/dis information detection.** Need for ML/AI models to either detect or assist in detection of mis/dis information. A particular challenge here is that a clear ground truth is rarely available, thus potentially requiring human involvement in training and guiding AI approaches.[1]

**Model robustness (explainability, reliability, blind spot detection).** It is very important for ML models to be used in security-critical contexts to explain their decisions, be reliable under attack and unknown security contexts, and aware of their blind spots. A novel research direction is how new ML techniques that provide these properties. How does research create techniques to measure these properties of ML models in use and provide assurance. Additionally, ways are needed to measure the harm these models can cause to certain at-risk populations, or certain groups of people, should they fail?

**Technical and Societal Impacts**

AI-enabled security and privacy approaches can facilitate safer and more trustworthy ICT systems by substantially reducing the effort required to build and operate secure and privacy-respecting systems. For example, these technologies can lead to more robust policies and compliance. It can also lead to more robust public and organizational policy and compliance verification/enforcement around security and privacy by enabling large-scale analysis.

In another vein, *AI*/ML models, if not properly developed, not only undermine the security of systems that rely on them, but also could hamper the safety of users and society at large. AI/ML has tremendous potential to help improve the safety and security of the world, but care must be taken to ensure these systems are trained so as not to perpetuate biases or contain blind spots. Research is needed to ensure AI/ML models are safe to be used in practice.

**Cross-Discipline and Outreach Opportunities**

There is a great opportunity for collaborators from criminal justice, law, business, political science, social sciences, and criminology, as well as policymakers, to work together in this area. New policies can be built that can be enforced using AI/ML techniques and identify policy violations. AI/ML models can help users identify malicious user behavior in the social media sites.

Content moderation around hate speech and misinformation on social media platforms has become a significant area of interest in the fields of political science and communications. The challenges associated with effectively detecting and addressing misinformation go beyond the technical aspects of identifying false or misleading content. They are deeply intertwined with broader societal questions about the sources, dissemination, and reception of misinformation. Understanding the dynamics of misinformation requires studying not only the technological tools for detection but also the social and psychological factors that influence its production, sharing, and belief. By examining these complex interactions, researchers in political science and communications can contribute to a deeper understanding of the underlying mechanisms and develop strategies to combat the negative impacts of misinformation on society.

## 3.4    Applications Security

As computation becomes increasingly integrated into our personal and professional lives, the significance of the applications and how data interacts grows exponentially. However, this surge in importance is accompanied by a parallel rise in the efforts of adversaries seeking to exploit these applications, compromise our privacy, extract sensitive data, and launch malicious attacks against users and infrastructure. Considering these challenges, the research vision outlined here aims to address the pressing need for privacy-preserving and secure applications across different domains. By focusing on diverse classes of applications, this research seeks to develop innovative solutions that safeguard user privacy, enhance security measures, and fortify the foundations of digital systems against adversarial threats.

### 3.4.1   Blockchain Applications

**Area Description**

Blockchain technology can be characterized as distributed, authenticated data structures that provide data provenance and are maintained through decentralized consensus. Such data structures can support applications where trust cannot be placed into a single trusted entity, but instead needs to be distributed across multiple parties. For example, trustworthy science, where the provenance of scientific data sets, analysis, and final recommendations are recorded on a blockchain and can later be replicated and verified by other scientists and the public, helping build public trust in the results of this research. Other examples include such far ranging areas as self-sovereign identities and authentication, health data interconnects, decentralized finance, and decentralized humanitarian aid and disaster relief. The applications can enable both new apps and improve the security and usability of existing apps, both of which will have strong societal benefits.

Critically, projects funded in this area should cover both traditional blockchain technology as well as other blockchain-adjacent technologies that provide authenticated data structures with provenance but lack some features of existing blockchains. This broad definition is essential to allow research into blockchain-like technologies, which may provide the security benefits of a full blockchain system, without unnecessary energy, computational, or storage overheads. Notably, this broad definition has been adopted by industry.

Research in this area should include the identification of applications that are well suited to blockchain and blockchain-adjacent technologies. It should also include the implementation, evaluation, and deployment of apps (i.e., software packages) that implement these applications. Finally, it must include research into the support structure necessary for those applications and apps to be successfully used in the real-world. This includes research into key management and the cryptographic wallets required by most blockchain systems. It also includes research into formal verification of both the consensus algorithms and software backing a blockchain system as well as the smart contracts created by users. Importantly, there is a need for user studies, both short- and long-term, that demonstrate that developed apps will be used correctly by end-users (i.e., non-crypto experts) and meet their objectives under real-world usage.

**Technical Efforts**

**Usable security on blockchain apps.** Some blockchain apps require users to manage hundreds of keys (or even long private keys). They are obviously low on usability. Low usability of these security measures will certainly prevent users from using them (or from using them with the intended secure measures). Hence usable security can be very relevant for blockchain apps.

**User privacy and access control.** Conventional blockchains are fully public: every participant of a blockchain has full visibility into every activity on the chain. Unfortunately, this goes against data privacy (e.g., health records of patients are all put on a blockchain, anyone having access to the chain will be able to access patients' health records) or permission management (e.g., some participants of a blockchain should not be able to access certain information) and may cause privacy infringement issues. Improperly designed permission management on blockchains hamper visibility and transparency, which makes blockchain applications less secure. Research should focus on the integration of blockchain and on-chain permission management.

**Responsible use of blockchains; Social and economic implications of blockchain applications.** Money laundry, Ponzi Scam, Phishing DApp, etc. are prevalent in blockchain applications. As has been well documented in the public domain, some uses of blockchain have had negative side-effects. For example, the energy cost of Bitcoin mining has had indirect effects on the environment. Here identifying critical benign use cases of Blockchain, as well as enabling cost-benefit analysis, will benefit the development of blockchain.

**Resolving conflict between user freedom and reasonable regulation.** Tensions arise between traditional governing bodies (e.g., central banks, governments, or even large companies) that attempt to protect their users or citizens from potential harms caused by new, unregulated blockchain application technologies and users of blockchains who want the absolute freedom to use these technologies. Research is needed to answer the question of how these two parties can work together without sacrificing the distributed consensus and transparency nature of blockchains, while ensuring large governing bodies can still protect their users and citizens, especially to those that are more vulnerable to new technologies.

**Software security and hardware security.** The fundamental assumption in blockchain applications, "code is law," is broken when people realize that (a) code can have vulnerabilities, and (b) code still needs to run on interpreters, operating systems, and of course, hardware. There is huge room for research to apply traditional and novel software security techniques on ensuring the security of software that implements, runs, and supports blockchain apps. Such efforts may work in concert with other areas described in this vision document.

**Fundamental trust model of blockchain.** The security of blockchain apps and applications heavily depend on all parties involved in the design, development, and deployment of the blockchain (and its related software). Research is needed to establish the trust model for the supply chain of a blockchain, identify all involved parties and individuals, and use such information to quantify the trustworthiness (and then the security) of an implementation.

Applications that use blockchain technology will rely on the latter for some of the trust, privacy, and security requirements of the application. The current understanding of how to articulate these requirements for a multi-party application, allocate portions of it to blockchain and select the appropriate blockchain is lacking.

**Technical and Societal Impacts**

Successful technical efforts will benefit users and society in several ways. First, work systematizing the distinct properties of blockchains and their impacts on downstream applications should help concretely define the use-cases and limitations of blockchain technologies, helping stakeholders design when blockchains are and are not appropriate for specific application areas. Work on human-centered security in blockchain applications should help end-users understand and mitigate emergent risks in the blockchain application space — for example, automatically generated nutrition labels for smart contract security, creating user-facing smart contract wallets that helps users recover access to their wallets should they lose their private key. Research on responsible and ethical blockchain applications will help the community understand how to develop blockchain applications that are environmentally conscious and consistent with relevant policy and regulations. For example, beyond consensus, one area of research could explore new ways to encode policy and regulatory rules in terms of what transactions are valid. Currently, rules to prevent double-spends exist — but one can envision systems that ensure that stolen funds, or funds earmarked for unethical causes as defined by regulators can be seized. Working on systematizing and clarifying trust models in blockchain applications can help users better understand what level of privacy and security they can expect when interacting with smart contracts so that they can make more informed decisions. Finally, successful technical efforts can help the scientific community obtain clarity in terms of the likely downstream societal impacts of the blockchain applications they are developing, thereby avoiding negative side effects.

Failure to engage in these concerted efforts could result in significant risks. First, blockchain technologies may be developed in an ad hoc and haphazard manner that fails to consider ethical and policy implications. For example, today, there is an adversarial relationship between regulators trying to grapple with how emergent blockchain applications fit into extant regulation and individual developers and users who are trying to broadly explore blockchain applications. Without the efforts outlined above, there is risk that this adversarial relationship will result in many well-meaning developers and users running afoul of regulation and/or the stifling of legitimate technical advancement.

## Cross-Discipline and Outreach Opportunities

Advancements in blockchain application research have the potential to significantly impact public and societal perception of security with respect to their financial transactions, information privacy, information provenance, models of trust, identity self-sovereignty, and awareness of potential abuses of the technology by criminals. Cross disciplinary research in security and privacy, economics, and social sciences will aid practitioners in making sound and strategic decisions concerning what are societally benefiting uses of blockchain technologies.

Cross-disciplinary research should seek to bridge security concerns at multiple layers of blockchain workflows, such as by flowing cryptographic assurances through to software design and implementation verification, and finally to end-users in the form of understandable guarantees. Blockchain usability is an important concern in this space, such as the ability for users to remember and/or store keys or other credentials for blockchains, possible vulnerabilities in blockchain-manipulating smart contracts, and public understanding of exactly which data integrity and privacy guarantees are supported by blockchain implementations.

Lastly, blockchain has unexpectedly resulted in major environmental impacts. Energy, climate, and environmental concerns are an emerging cross-disciplinary area of study as blockchain technologies become deployed on a mass scale. Computer, economic and climate research is needed to understand which blockchain deployments constitute acceptable tradeoffs.

### 3.4.2   Supply Chain and Intellectual Property

**Area Description**

Supply chain security research focuses on assuring the overall cybersecurity (with a particular emphasis on semantic integrity) of a system that contains components that have been designed, sourced, and/or procured from third parties. Intellectual property security concerns itself primarily with ensuring the confidentiality and attributable use of components that are developed and used by different parties. The components (and the overall system) can be comprised of—for example—data, documentation, software, and hardware.

**Technical Efforts**

**Provenance tracking.** There is a need for methodologies, tools, and technologies to provide transparency into the components (used generically: software, hardware etc.) used to build a system to engender trust in the supply chain and in the produced system. These approaches for transparency should be able to cross "boundaries" — e.g., between development groups, between organizations and maybe even between geographic boundaries. Here the community must develop tools and techniques to allow specifying requirements for suppliers (supply chain) and ensuring that such requirements are followed.

**Software bill of materials.** To the extent they are currently used, Software Bill of Materials (SBOM) contains a list of software components included in a system. While necessary, this is usually insufficient for downstream entities to determine the trustworthiness of such components, their potential impact to the cybersecurity posture of the whole system, and the need for potential mitigations to any risks or threats potentially introduced by a third-party component. Research in providing greater knowledge and transparency on several aspects of the development lifecycle of the components that can be included in the SBOM. Enhancing the SBOM such that it describes software provenance in the presence of modification (e.g., a modified open source library), software design attributes (e.g., the use of sandboxing or memory-safe languages), development practices (e.g., testing), developer attributes, and other relevant information that can allow system composers, administrators, and end users to determine the potential risk and the need to employ additional security mechanisms to address said risk. Research is needed in developing tools and methodologies for generating, updating, verifying (fully or partially), and consuming these enriched SBOMs, and technical risk mitigation techniques that can take advantage of such information. A particularly rich setting to conduct this research is open-source software, both due to its widespread use (on its own and as part of commercial/closed-source software systems) and the visibility it offers researchers in many aspects of the software development lifecycle. The knowledge thereby derived will be more broadly applicable to any type of software.

**Artifact authentication and verification.** There are many verification methods that have been developed, each with its own cost for ease of deployment, ranging from tracking and artifact authentication (e.g., code signing) to full-scale verification. Research is needed to ensure an understanding of the allocation of effort needed to meet the assurance requirements of the supply chain for some artifact.

**Trust management.** Architecture-level approaches to building trustworthy or high assurance systems that reason about components (or providers) of varying degrees of trustworthiness (including development of appropriate validation / testing procedures focused on those components) is a useful direction. These approaches may rely on the evidence or proof discussed in "transparency" and build process requirements and.

**Traceability.** New mechanisms are needed to ensure effective traceability in hardware design and software development point of view. There are some existing solutions in hardware design and software development separately. It is unclear how to ensure traceability when considering both hardware and software integrated as a system down the road to the supply chain. It is important to study the component-level IP and the system-level IP and understand the tradeoff between applying the traceability and maintaining the system performance.

**Intellectual property protection.** Maintaining both trustworthy and verifiable proprietary software while protecting intellectual property presents a challenging trade-off. To verify software, its internal details must be exposed, which can compromise intellectual property. The challenge lies in finding ways to protect intellectual property while still ensuring trustworthiness. One solution that is actively being researched is confidential computing, which allows developers to upload their code and artifacts to protected hardware enclaves, without worrying about code leaks. However, there is still a need to verify that the software performs as intended. Developing techniques to verify such enclave-protected software would be extremely valuable to address this tension.

## Technical and Societal Impacts

The technical and societal impact of work in this area may be substantial. Building better trust amongst various elements in a supply chain is a natural outgrowth of the assurance techniques developed. In a broader setting, supply chain assurance is crucial to building confidence in critical systems such as electrical grids. Supply chain security in general is an important question and applicable to other supply chains (other than software and hardware), such as food, manufacturing, and others. Developed solutions can inform market-based approaches for sourcing multiple components meeting the same specification.

## Cross-Discipline and Outreach Opportunities

Work in this area will be relevant to computer science, economics, business (e.g., supply chain management), human interaction, and risk management. The approaches to supply chain security impacts users as well as maintainers of composite systems. Industry at large can benefit from academic research as this area is still in its infancy, and the problems are hard. The economics of supply chain security should be of interest to people from allied disciplines. Economic incentive mechanisms for downstream developers providing assurance are worthy of investigation.

### 3.4.3   Social Media Security Applications

**Area Description**

Social media applications and systems are a crucial (and often controversial) means of communication in society today. There are broadly three types of social media: social media involving human beings in the loop, social media applications involving non-living entities (e.g., bots), and social media hybrid applications involving both living and nonliving entities. All social media platforms and applications have some influence on society in some manner. Recently research has uncovered how these platforms have enormous impact on society---particularly in the economic, political, and psychological realms. Research in social media should explore aspects of security, privacy, regulatory compliance and compliance with other policies, ethics and fairness and legal requirements.

**Technical Efforts**

**Information integrity**. Misinformation and disinformation, posted and propagated by either bots or genuine users, pose big threats to the information integrity on social media. This is an especially big challenge, given the progress in large AI models – such as ChatGPT and DaLL-E–that can be used to automatically generate realistic looking texts and images on a large scale. One critical research direction is to develop techniques and tools to prevent and detect such fake information.

**Privacy loss measurement and nudge on social media**. Efforts should explore new metrics and ways of designing and deploying a system to measure user and aggregate privacy. Users publish a large amount of data–such as text, image, video, and audio-on social media. An adversary can leverage ML techniques to infer users' private data–such as sexual orientation, political view, gender, location, age, etc.--using their seemingly innocent data publicly available on social media. One research direction is to develop algorithms and tools to measure a user's privacy leakage as he/she publishes data on social media. For instance, when a user likes a webpage on Facebook, a "privacy assistant" informs the user of the quantifiable privacy implications or qualitative examples of potential privacy harms, so the user can make more informed decisions on sharing.

**Cryptographic techniques.** New cryptographic techniques should be developed to secure the content shared on social media platforms. First, the images or messages can only be visible to the intended group of audience. The cryptographic techniques ensure that the content will not be leaked to arbitrary persons. A high privacy-preserving scenario is that the social media platforms only access the encrypted content, and only the users will be able to see the clear-format content. A research challenge is how to operate on encrypted data. Second, it is important to develop cryptographic techniques that prove the ownership or authorship of the shared content on social media. In addition, if a user prefers to take down the shared content, the cryptographic technique serves as an approach to attribute the authorship. This provides a plausible way to withdraw content from social media.

**Decentralized/federated networks**. Recent years have seen a rise of federated social networks such as Mastodon. Since information can be shared across instances yet there is no centralized company running these networks, content moderation will become a challenge. On the other hand, they also provide promises to counter institutional surveillance. More research is needed to investigate the security and privacy implications of these new technologies.

**Detection of evasive scammers on social media**. Scammers are active and they are developing more and more advanced evasive methods to avoid detection. It is critical to design new techniques to detect and defend against such evasive scammers in current and future social media. This is particularly true with respect in light of recent discoveries of large-scale, state sponsored dissemination of misinformation and propaganda.

**Accessibility.** There are new challenges to traditional accessibility barriers (e.g., low vision, hearing impaired). But there are also accessibility issues related to privacy preferences. Those people who have strong privacy preferences are often excluded from social media, and the benefits of social media. Enabling participation in social media networks for people who do not want to violate their own privacy or friends, family, organizations privacy is also an area for accessibility investigations.

**Intersection of emerging technologies such as generative AI with social media.** Research should explore how to help identify and present risks to end users - and society - of technologies that enable the generation of misinformation at scale. This also brings about questions of the nuance between generated false information and generated nearly true information.

**Creating a Taxonomy of Threats.** There is also a need to systematize a taxonomy of pertinent security and privacy threats in the context of social media to facilitate the exploration and creation of pertinent defenses. For example, many threats might be institutional — e.g., the surveillance of individuals by private companies and nation states; others might be interpersonal — e.g., the selective disclosure / hiding of information from individuals in one's social networks. Still others might be related to bystander privacy, or the compromising of an individual's privacy by someone or something else.

**Technical and Societal Impacts**

The largest effect of social media security will ultimately be on the end users. Social media can cover and connect a massive amount of people, and thus information can spread quickly. Therefore, the research community must be especially vigilant at preventing misinformation. Misinformation can not only cause societal disruption but can threaten safety.

Challenges and potential harms can include authentication and verification in novel user interfaces, analyzing social media data, organic data, large scale evaluations. Additionally, challenges and potential harms include issues related to society harmony vs. polarization, network and graph analytics and AI/ML security will be impacted. Other stakeholders such as developers, business owners, regulators, etc., will be impacted by how social media evolves over the next few decades.

As the threat landscape expands, the assumption that real and synthetic media can be fundamentally differentiated is challenged with the latest developments in VR, metaverse, and AI synthesis technologies. In the not too far future, synthesized media will become ubiquitous, and the question of real/fake will become less relevant. Research is needed to determine the correct question to study, to ensure user identities are not abused.

**Cross-Discipline and Outreach Opportunities**

Given that social media is ubiquitous, one can foresee numerous opportunities for cross-disciplinary outreach. With respect to information integrity threats, it would be useful to incorporate perspectives from psychological modeling (e.g., neuroscientists and psychologists) and social psychology to understand how individuals make sense of what is fake and real with respect to content that they encounter via social media applications. It will be particularly pertinent to capture these perspectives as social media evolves and entails new threats — e.g., as deep fake and generative AI technologies get better and facilitate the construction and dissemination of content via these applications.

It will also be relevant to incorporate socioeconomic perspectives to explore novel methods of data ownership and adversarial interoperability. For example, today, private centralized social media companies own the data contributed by individual users and can use this data unfettered. Even if users are unhappy with these platforms, it is non-trivial for users to delete this data and for users to port this data to different, preferred platforms. New models of data ownership that will facilitate adversarial interoperability across social media platforms should be needed - e.g., models where users "loan" data to social media applications but can rescind access and distribute to other platforms as preferred? Perspectives from the social, behavioral, and economic sciences can help generate and assess the downstream impacts of these alternative data ownership models. There is also a need for participatory and co-design perspectives in the development of social media applications: some applications disproportionately harm some populations; therefore, it is worth capturing diverse stakeholder perspectives when designing these applications to better model these disproportionate harms and benefits.

There is also a need to incorporate perspectives from legal and public policy scholarship. It is important to consider enforceable models of governance for social media. For example, today content moderation rules are largely privatized. Research investigating what it looks like to incorporate end-user perspectives in these rules and how could such a system can be implemented in a manner that is, itself, secure. As content moderation rules evolve in the coming decades, there will also be a need to translate regulatory requirements into technical practice.

### 3.4.4   Internet of Things (IoT) / Cyber-Physical Systems (CPS) Security

**Area Description**

IoT/CPS systems operate in resource-constrained environments, including all necessary elements to run computing devices such as network connections and electronic power. Further, IoT/CPS systems involve both cyber and physical elements that constitute traditional sensing-computing-actuation cycles. It is interesting to note that what constitutes an IoT/CPS system is challenging. For example, as general-purpose computing platforms a smart phone is considered an IoT device, but because it has sensors and moves in the physical world, it shares characteristics with IoT/CPS. Outlined below, it is useful to identify characteristics common to IoT/CPS systems.

IoT devices interact with the physical world, sensing various data, and controlling actions that impact the physical world. IoT devices are often purpose-built (such as a dedicated camera, stove, or lightbulb); alternatively, general purpose devices (like smartphones) are used for IoT purposes by dedicating them to a particular task. IoT devices often have limited resources, and often lack a visual interface. Unlike personal devices (laptops, phones), IoT devices are often shared among multiple users. For example, smart home devices are shared in the family and might collect information from guests. Therefore, there are many challenges to ensuring the security and privacy of IoT devices, from physical security/safety, formal methods, usable security and privacy to privacy-preserving machine learning.

IoT/CPS devices are of increasing importance because IoT/CPS deployment is becoming ubiquitous. Devices may be easily overlooked, meaning users may not be aware of devices they encounter in public. The devices may also be customized, so they may not be aware of the capabilities of the devices they encounter. These devices have unique security and privacy risks due to their interactions with the physical environment, including the privacy risks posed by data collected by sensors, and the physical security risks by the control the devices have over their environment. Moreover, interactions between IoT devices (and between IoT devices and the online systems that support them) can obscure which data is collected and how it is used.

**Technical Efforts**

**Interoperability of IoT with legacy device network.** The fragmentation of IoT ecosystems, from manufacturing processes to software authoring and maintenance, is a serious and inherent challenge in securing ever-growing IoT/CPS networks. In the coming decades more users and organizations will install devices to connect physical domain components to cyber digital infrastructures. As a result, the challenge of safeguarding legacy IoT/CPS systems and maintaining interoperability with newly installed systems is expected to worsen. Although existing standards and recommended procedures have been established to address this challenge, these efforts have been largely inadequate. Innovative research proposals that automate the process and provide better security guarantees will assist in advancement in this area.

**Applying defense techniques against physical attacks in IoT and CPS.** Due to the ubiquitous deployment of IoT devices, it is easier for the adversary to physically access IoT devices making them vulnerable to various threats including physical attacks, offline-attacks and side-channel analysis. Thus, the next-generation development of IoT/CPS should address these new attack surfaces. Some defense mechanisms should be developed to counter such potential attacks.

**How to guarantee authenticity of data input?** An open problem in this field is the ability to guarantee that data sent to an IoT server is indeed authentic and indeed collected by the authentic IoT sensor, or that sensor data collected by a CPS system that dictates the next action is legitimate. Theoretically, this can be accomplished using cryptographic proof systems, but that requires large computational overhead. In practice, IoT devices can be physically accessible by attackers, and thus some privacy information (e.g., proving key, signature key) can be compromised via physical attacks or side-channel analysis. This could prove to be fatal as it could result in the collapse of the entire CPS system.

**Side-channel resiliency on physical layer.** The recent advancements in Machine Learning (ML) techniques have enabled adversaries to leverage data analysis to design new attacks against IoT/CPS systems. With the vast number of data points that IoT/CPS systems generate, adversaries can put efforts into finding previously unseen sources of side-channel attacks. Therefore, it is crucial to reconsider and revisit hardware-level attack vectors against IoT/CPS devices. The research community needs to develop a rigorous process to provide safety and security guarantees for IoT/CPS devices, taking into consideration newly emerging attack vectors.

**Secure development, patching, authentication, and device management.** Market forces related to the development lifecycle of IoT devices cause cost efficiency and functionality to be prioritized over security and privacy issues. There are few to no incentives for vendors to maintain their devices throughout a potentially very short product life span. Additionally, IoT devices are resources constrained. Consequently, patching is more difficult than for general purpose devices.

The management of "shared," multi-user devices is problematic. Authentication and secure pairing of devices with no screen or keyboard is challenging. Devices owned by couples that separate cannot be easily (or not at all) reset to a single owner use. Also, second-hand devices may embed undetectable malware or backdoors. Furthermore, the "shared" use of IoT devices increases risks to privacy. For instance, rental cars may leak the contact information of previous users who previously connected the car to their personal/wearable devices. Also, devices placed in hotel/Airbnb pose serious privacy issues.

**Privacy of sensed data, incidental users, standardized consent/notices.** Privacy is a challenging issue for IoT devices, with additional difficulties beyond those for more traditional computing devices like laptops, tables, and mobile phones. For IoT devices that have only a limited user interface, it is often unclear how to notify users that the devices are collecting data about them, or how they can provide or decline consent for that data collection. Consumer electronics devices, like smart doorbells and smart speakers, collect data from the intentional users (i.e., those who set up the device or use it on a regular basis) as well as incidental users who may be unaware that the devices exist or their capabilities. In public settings IoT devices like Bluetooth beacons may be owned by companies or other organizations that collect data for commercial/surveillance purposes.

Accordingly, challenges exist for creating notification standards and formats (i.e., to tell users that data is being collected about them) and for communicating information about data notification in ways that people understand and interact with meaningfully. Beyond notice, providing and respecting data collection choices is a challenge. For some IoT devices, it may be unclear technically or conceptually how to collect data from some individuals while excluding others.

IoT devices currently struggle to provide strong privacy guarantees of collected data due to the computational complexity of state-of-the-art privacy-preserving computation techniques (e.g., Fully Homomorphic Encryption). In the next 10 years IoT devices will have the ability to support customized chips built for certain special-purpose PPC. Note that there are currently research efforts, such as the DARPA DPRIVE project, to accelerate privacy-preserving operations. It is important to explore how the developed technologies can be integrated into emerging IoT.

**Risks to physical security, network data authentication.** The risk of physical harm exists when IoT/CPS devices are compromised. These risks could be from loss of control of the device and an attacker instructing the device to cause physical harm. The risk also is in terms of how to remediate a compromise when the device is in operation (i.e., automobile or airplane) and it is not safe to reboot or take it offline. This brings the challenge of how to design IoT/CPS devices that are difficult to instruct to cause harm. It also brings up the issue of how to limit the attacker when remediation is not immediately possible.

It is important that inputs to sensors be authentic and that malicious inputs can be detected. This is challenging for analog inputs where proof systems are not feasible. These malicious inputs could cause the device to physically harm people even if it is built to not allow this to occur. The attack could be creating fake inputs using analog or digital methods.

**Technical and Societal Impacts**

IoT/CPS systems will present increasing physical safety and privacy issues. Mitigating these will potentially have an enormous positive impact on society. There is a question of trust in IoT devices that, if addressed, will increase safe adoption of modern technologies, and society and individuals will reap the benefits of these technological advancements.

There are promising new applications where IoT and CPS technologies will play an increasingly important role. These applications bring new and unique security and privacy challenges that require research. Wearable computing uses miniature, body-borne computers or sensory devices worn on, over, under or integrated within clothing. They have limited ability to interact physically with humans. Smart devices are increasingly used in everyday life and may impact people in ways that were not envisioned before. Medical devices, including both monitoring devices and surgical robots, will be more prevalent. So will smart vehicles with V2V and Vehicle to Infrastructure communications. Such information is used in safety-critical decisions. Smart city technologies use sensors to detect air quality, traffic, gunfire, and other. Smart speakers will get smarter. Instead of waiting for special wake words, they may be able to monitor the environment and provide reminders and conversations in a more organic way, creating new privacy and security challenges.

**Cross-Discipline and Outreach Opportunities**

Research into the security of IoT/CPS intersects with many other fields. A challenge is how to quantify and communicate security properties such as risks, rights (particularly notice and choice about data sharing and privacy), and options (how to manage devices) of IoT/CPS systems with users. That challenge can benefit from input from education and communication research, as well as human-centered design research. In addition, there is a role for legal and government input on regulation and the role of disclaimers (notice and choice) on what IoT/CPS systems should do.

Enhancing the security and privacy of IoT will enable safer and more reliable deployment of IoT devices in user-critical disciplines such as medical and healthcare that require high-precision and robustness of IoT devices. Other disciplines such as IoT and agriculture also get much benefit from secure and trustworthy IoT ecosystems to keep track of data provenance and operations.

### 3.4.5   Virtual/Augmented Reality Systems Security

**Area Description**

The rapid advancement and increasing popularity of virtual reality (VR) and augmented reality (AR) technologies have opened new possibilities and experiences in various fields, from gaming and entertainment to education and healthcare. As these immersive technologies continue to evolve and integrate into our daily lives, ensuring the security and trustworthiness of VR and AR environments becomes paramount.

The security community must recognize the need for extensive research and innovation to address them effectively, thereby laying the groundwork for making VR and AR trustworthy. Given the immersive nature of VR and AR experiences, the security of these environments extends beyond traditional concerns. The protection of user privacy, the prevention of unauthorized access and manipulation, and the detection and mitigation of malicious activities all become crucial aspects to consider. Security mechanisms must safeguard personal data, prevent identity theft, and defend against cyberattacks targeting VR and AR systems.

**Technical Efforts**

**Threat models that account for the intersection of virtual and physical worlds.** As virtual and physical worlds merge and blend together, it will become crucial to develop threat models that consider attacks that exist at or cross virtual-physical boundaries. To what extent might virtual environments enable attacks that have effects and consequences in the physical world? Virtual worlds might have physical components or connections to the physical world. In mixed reality, IoT devices may be overlaid with virtual functions, etc. Manipulation of human perception in VR environments could cause physical harm to VR users (e.g., have users walking into the wrong place). Attacks on VR/AR hardware and equipment may lead to physical manifestations (controllers vibrating, blasting sound, flashing lights, etc.). One clear threat model that exists in virtual space but does not exist in normal computing models is that attackers might be able to induce physical harm on participants. If able, an attacker can control a VR headset and use information of the physical environment (which the headset has) to guide and manipulate the user into harm (e.g., having them trip over a chair or table). The question of how the emotional reactions of users can be manipulated to induce fear is an interesting, multi-disciplinary problem to attack. This also ties into social engineering attempts in VR environments.

**Social engineering attacks.** Grooming and social engineering attacks in virtual environments may involve manipulation of avatars and other aspects in even more convincing ways than in other online interactions. The possibility of numerous social engineering tricks against users is an area in need of further investigation.

**Digital twins to detect attacks.** There have been limited efforts to compare the physical device behavior and its digital twin version to detect attacks. It is not clear whether having a digital twin potentially improves the security of the physical system. In addition, more research may be needed to understand whether the interaction of digital twins and their physical equivalent creates new attack surfaces.

**Privacy threats from increased sensing capabilities.** VR and AR systems are increasingly equipped with sensors to both track the user's movement and their environment. This raises privacy concerns regarding the inference of health conditions, location, sexual orientation, and other aspects of behavior in virtual/augmented environments. Research is needed to investigate signals present in virtual space (i.e., can be observed from other users) that can be used to track, identify, or violate the privacy of the user.

**Biometric authentication.** Biometric authentication is a direction that VR environments have been taking. In this space, biometric input from users can be evaluated, and based on that (e.g., like fingerprinting the browser of a user), the user can be authenticated. Biometric signals include data points such as retina readings, eye movements, and the movement characteristics of a user. While such techniques can be reliable in identifying users, at the same time, they carry the risk that this authentication information can potentially be stolen in a VR environment and used by the attackers. How such authentication information can be securely stored and managed, and how the tracking of user behavior by unauthorized parties can be prevented is an important area of research. Along with increased sensing capabilities that can be used to violate the privacy of users, these increased sensing capabilities can improve authentication. We found similar parallels to the browser fingerprinting problem in web security: browser fingerprinting is used to track users against their wishes; however, it is also a strong component in risk-based authentication for web applications. Therefore, these increased sensing capabilities can provide extra signals that can be used to fingerprint users, which can be used to enable both detections, as well as risk-based authentication in a virtual space.

**User manipulation in AR and VR.** In the VR space, marketing attempts will require ethical studies as they may involve interesting and novel methods of deception and manipulation. In AR, reality could be manipulated with photorealistic overlays. AR annotations could be hacked or be susceptible to spam. Misinformation could be spread, e.g., by manipulating restaurants' health scores or reviews – both by hacking and changing the underlying information or by overlaying information in the real world with photorealistic renderings. There is also a risk of politically driven misinformation in VR environments.

**Virtual property protection.** During Pokémon Go's popularity, there were reports from some location owners that were upset to have Pokémon Go players in their space (such as churches). There was no way for the physical location owners to control/restrict people on their premises' access to the AR environment, or to opt-out of being included in an AR environment. While Pokémon Go did allow physical location owners to be removed from the game, the larger concern remains: who controls AR access to physical spaces? Interestingly. a similar problem exists with drones and controlling access to airspace (which is another area of security that must be addressed).

Digital assets are common in virtual reality. For some VR applications, the digital assets are even tradeable. It is critical to ensure the security of digital assets, e.g., securing the transaction of the digital assets and safeguarding the storage of the digital asset. Also, it might be important to protect the privacy of the users' digital assets.

Here, VR spaces will certainly require novel access control and content policies. While there will be similarities, because of the unique nature of VR with regards to allowing users to have very realistic interactions, how such users will be authenticated, and how the content provided to them will be moderated will need to be researched.

Private communication and anonymization in VR interactions, both for interaction in oppressive regimes but also to prevent surveillance by platforms, and among individual users.

Humans have developed their own intuitions for securing their communication in the physical world. If two parties want to have a private conversation, then they can whisper, even in a shared space, and be reasonably assured that they were not overheard. A key question is how does this intuition hold in a VR setting: while it appears, to the user, to mimic a physical setting, do these physically developed intuitions harm users when applied in a virtual space? For instance, who can be heard talking in a VR system? People in the same room? People close in physical space? What about the VR hosting provider? Research should: (1) address and identify these mismatches between users' conceptions of security in the physical world and (2) create privacy aware communication platforms in the virtual space.

**Virtual interpersonal attacks.** Besides technical security and privacy challenges there are also safety and privacy challenges on the interpersonal level, such as how to prevent and detect extortion attacks, deanonymization attacks, sexual harassment, etc.

Inferring features of users in virtual environments can lead to de-anonymization attacks. For example, eye tracking and gaze data in VR headsets can provide identifying features to an attacker. This risk is amplified for certain users, such as those who are exploring their identities in virtual environments who have not disclosed these identities in physical space. De-anonymization of these users can lead to extortion and potentially physical harm.

In addition, harm can be suffered based on virtual identities. Sexual harassment of avatars has been a demonstrated problem in virtual environments. New research shows that victims of harassment and bullying in virtual environments experience similar psychological effects as they do in the real world. The feelings of the physical avatar as an extension of the user's physical body are known as embodiment, and because of this phenomenon, an attack against a virtual avatar can cause the biological systems of the physical user to react adversely.

## Technical and Societal Impacts

Advances in AR/VR security and privacy research are crucial to enable secure and trustworthy remote applications with societal impact, such as education, remote driving, telehealth, remote surgery, remote therapy, etc. Safety in such environments will ensure that complex virtual-physical systems can be used with less concern. Failure to make scientific progress would prevent the maturing and emergence of many promising applications of AR/VR technologies. Implications of insecure VR/AR systems may also have more far-reaching consequences as large numbers of people could be victims of an attack that causes physical harm.

## Cross-Discipline and Outreach Opportunities

Virtual Systems Security is a very interdisciplinary field. This research area will require the involvement of psychologists, behavioral analysts, law enforcement, and other fields. Progress in this area would allow bypassing censorship (e.g., it would be easier to get a group of people together in "large" virtual spaces in comparison to having them assembled in physical spaces and risk arrest or violence). This area also has potential for progress in education (e.g., remote teaching) and even application spaces such as remote surgeries involving medical personnel.

## 3.5 Cryptography & Security Theory

Cryptography plays a pivotal role in ensuring security and privacy in computational systems. This truism has become increasingly clear as systems are increasingly interconnected. Here cryptography provides the essential means to protect sensitive information, communications, and transactions from unauthorized access, interception, and tampering. Moreover, cryptography forms the foundation for various security mechanisms, such as authentication, integrity verification, and digital signatures, which authenticate the identity of users, verify data integrity, and establish trust in electronic communications. In the realm of privacy, cryptography empowers individuals to maintain confidentiality, control over their personal information, and the ability to communicate and transact privately in an increasingly interconnected world. Overall, cryptography serves as a critical tool for upholding security and privacy, allowing individuals, organizations, and societies to operate in a secure and trustworthy digital environment.

### 3.5.1 Quantum Cryptography

This line of inquiry asks about the interactions between quantum computing and the field of cryptography and security. There are two distinct sub-areas. The first is post-quantum cryptography (a better name would be quantum-resistant cryptography) which aims to study new types of attacks on classical cryptographic systems that are enabled by (near-term or long-term) quantum computers, and new cryptographic algorithms and protocols designed to resist such quantum attacks. Here, the attacker is quantum, but the deployed algorithms, protocols, systems, and devices are classical. The second is quantum cryptography (a better name would be quantum-enabled cryptography) which uses the principles of quantum mechanics to improve existing cryptographic capabilities or realize fundamentally new ones. This area of study also has the potential to inform basic scientific questions in quantum mechanics and quantum gravity.

Work in this area can be divided into two distinct areas. Within the sub-area of quantum-enabled cryptography, several crucial areas of technical effort emerge, each presenting unique challenges and opportunities for research and innovation. These areas encompass the intersection of quantum mechanics, cryptography, and computer science, with the goal of harnessing quantum phenomena to enhance cryptographic techniques and address emerging threats.

Within the sub-area of quantum-resistant cryptography, researchers are engaged in various technical efforts aimed at developing robust cryptographic solutions that can withstand attacks from quantum computers. These research areas encompass different aspects of post-quantum cryptography, addressing the challenges and opportunities presented by the advent of powerful quantum computers.

## Technical Efforts

**New Post-Quantum Cryptographic Algorithms.** This research area focuses on the development of new cryptographic algorithms that are resistant to attacks by quantum computers. These algorithms are designed based on hard problems other than structured lattices or isogenies, which are commonly used in current post-quantum cryptographic schemes. By exploring alternative mathematical problems, researchers aim to provide novel cryptographic primitives that offer quantum-resistant security.

**Classical/Post-Quantum Hybrid Modes of Operation.** This area investigates the integration of classical and post-quantum cryptographic techniques to create hybrid modes of operation for algorithms and protocols. The goal is to combine the security strengths of both classical and post-quantum cryptography, offering a balanced approach that can withstand attacks from both classical and quantum adversaries.

**Cryptographic Agility.** Research in cryptographic agility aims to enhance the flexibility and adaptability of software and cyber-physical systems, making them more capable of reconfiguring to use new or different cryptographic algorithms. Cryptographically agile systems enable rapid reconfiguration in response to the discovery of weaknesses in widely used cryptographic algorithms, ensuring timely deployment of new and stronger cryptographic mechanisms.

**Integration of Post-Quantum Cryptography (PQC) Algorithms into Security Protocols**. This research area focuses on the integration of post-quantum cryptographic algorithms into existing security protocols. Researchers develop tools and techniques for automatically analyzing and assessing the quantum readiness of existing code, facilitating the seamless integration of post-quantum cryptographic mechanisms. By ensuring compatibility and readiness, these efforts contribute to the adoption of post-quantum cryptography in real-world applications.

**Quantum-Enabled Cryptanalysis.** This area explores how quantum mechanical processes can be leveraged to develop new cryptanalytic techniques. Research focuses on identifying cryptographic algorithms that become susceptible to attacks and break when confronted with quantum-enabled cryptanalysis. By understanding the vulnerabilities and limitations of existing cryptographic schemes in a quantum computing era, researchers can contribute to the development of more secure and resilient encryption methods.

**Quantum-Aware Security Proofs.** Security proofs of cryptographic protocols, even those based on post-quantum hardness assumptions, may become compromised when faced with a quantum adversary. Research in this area aims to rebuild the foundations of cryptographic reductions while considering the capabilities and potential threats posed by quantum attackers. By developing quantum-aware security proofs, researchers can establish stronger guarantees for the security of cryptographic protocols in the presence of quantum adversaries.

**Design of New Quantum Cryptographic Capabilities.** This research area focuses on designing novel cryptographic capabilities using quantum information and computation. Examples include the exploration of quantum money, innovative approaches for delegating and revoking cryptographic keys, and the generation of certified randomness using quantum systems. By leveraging the unique properties of quantum mechanics, researchers can pave the way for the development of fundamentally new cryptographic tools and protocols.

**Foundations of Quantum Cryptography.** Quantum mechanical principles offer opportunities to reduce or eliminate unproven hardness assumptions in cryptography. Research in this area involves rebuilding the foundations of computational quantum-enabled cryptography, aiming to fully exploit the power afforded by quantum mechanics. By establishing robust theoretical frameworks, researchers can advance the field of quantum cryptography, opening doors to new cryptographic primitives and more efficient protocols.

**Cryptography as a Lens to Theoretical Physics.** Recent work has demonstrated that computer science and cryptography can contribute to solving long-standing mysteries in theoretical physics, such as the black-hole information paradox. This area explores the robust interaction between cryptography and theoretical physics, where cryptographic techniques are utilized as a lens to gain insights into fundamental physics problems. This interdisciplinary collaboration between the fields can yield mutually beneficial outcomes, pushing the boundaries of both cryptography and theoretical physics.

**Technical and Societal Impacts**

This area represents the new frontier in the ongoing cat-and-mouse game of cryptography and cryptanalysis. There is an ongoing need to create new, stronger cryptosystems to protect information from attack and disclosure. At the same time, there is an ongoing need to invent new cryptanalytic techniques used to break more cryptosystems. The development of quantum-enabled modes of computation impacts both sides. The development of new cryptosystems that take advantage of quantum-mechanical effects, can develop new cryptanalytic techniques that leverage quantum mechanics to improve attack capabilities. Implementing Shor's algorithm for commonly used RSA moduli is a prime example of the latter, but there may also be intermediate results that do not require a fully scaled fault-tolerant general-purpose quantum computer. For example, it may be possible to leverage a noise intermediate-scale quantum (NISQ) device to better attack certain cryptosystems. Such improvements would certainly impact the broader community as it would potentially change timelines for cryptographic algorithm migrations.

**Cross-Discipline and Outreach Opportunities**

Quantum cryptography involves expertise from several different fields, including physics, mathematics, computer science, electrical engineering, etc. Given it is interdisciplinary research, the cross-disciplinary opportunities are also natural.

One direction is on the theoretical side. For example, an interesting topic is to explore the possible interaction between quantum cryptography and the physics of black holes. The interaction between quantum cryptography and mathematicians could be the involvement of expertise from mathematicians studying areas such as high dimensional lattice structures to examine lattice algorithms.

Another direction is on the application side. For example, developing practical quantum key distribution systems requires collaboration between physicists, electrical engineers, systems designers, and cryptographers. Another example is the integration of quantum cryptography into existing communication infrastructure. Efforts are just beginning in the development of security protocols standards organizations (e.g., IETF, OASIS, ETSI) to integrate PQC algorithms into existing security protocols (e.g., TLS 1.3). To transition to quantum-resistant cryptography, much work needs to be completed across protocol standards organizations.

### 3.5.2    Cryptographic Constructions

**Area Description**

Cryptography plays a role within a large swath of digital infrastructure, and constructions and protocols are the backbone of applied cryptography. Problems in this space can be split into deployment challenges and application-driven challenges (how to build cryptographic protocols for specific applications). Both can be served by theoretical cryptography, but also feed into theory by developing new settings, definitions, goals, and protocols.

How to secure data and technological infrastructure more broadly remains a core challenge facing computer security. Cryptography, as a mechanism, can help set the "rules of the road" and is critical in this effort. Deployment of cryptographic solutions has increased over the last decade. Secure point-to-point communications via TLS are now the norm, rather than the exception, due to large-scale efforts to improve privacy, and end-to-end encryption for messaging and other applications is exploding.

Though this is an exciting development for privacy, there is a need to pay attention to and nurture research ensuring cryptographic infrastructure builds in accountability features that help mitigate abuse and other harms. "Blue-skies research" that develops new cryptographic constructions that anticipate the needs of the future, needs to be incentivized.

**Technical Efforts**

**Key Management.** Many, if not most cryptographic algorithms rely on users to manage some number of cryptographic keys. This can be managing a single key but could scale up to needing to manage thousands of keys. Moreover, if novel cryptographic systems were ever deployed widely, failure to have usable and effective key management systems could have disastrous consequences. Such disastrous consequences were seen in Bitcoin, where people lost millions for losing a private key. As crypto-based systems become more widespread, these challenges will grow. While there has been some research into key management, it is very basic, focusing on managing a small number of keys and does not explore how key management works for disparate groups or in different stages of life. Research into this area will be very wide ranging, including:

*Recovery.* What approaches are there for helping users recover lost keys? It will be insufficient to just say access to the key is gone. Similarly, it is insufficient to give all keys to a single trusted third party. Research should explore new schemes for key storage. Human factors should also be considered to ensure users recognize the threat model and can make informed decisions.

*Synchronization.* Users will need to use their keys across a wide range of devices—potentially hundreds in the ever-expanding world of IoT. Research is needed to explore how to safely transition keys between devices, the potential risks of doing so, and how to help users do so correctly.

*Stage-of-life considerations and vulnerable populations.* If cryptographic protocols were widely deployed, it would not be acceptable to use key management approaches that assume a homogenized population. Instead, research will need to be done to address how key management changes as people age. For example, how would parents help children with key management. How would children eventually assume control of their own keys as they leave the house? How do children help aging parents manage keys, especially as those parents encounter memory constraints. Similarly, research should examine other vulnerable populations that may struggle to use some key management approaches. For example, if a system requires visually comparing hash values or images, this could be a challenge for users with visual impairments. This research will likely require exploration of both new system designs and human factors.

*Key management models.* The cryptographic community has often taken a hostile approach towards centralized key management systems. However, if crypto-systems become an everyday part of our lives, there will be systems that rely on centralized distributed/managed keys—for example, government issuance of cryptographic IDs. Research needs to explore the trade-offs between these different models and to build new models that could work on different institutional needs.

*Scalability.* Research is needed into how key management will work at scale. How can users manage hundreds or thousands of keys? How can developers do the same? Does this change if those keys are based on different mathematical constructions? How do users/developers/systems transition between keys—for example, because a company was hacked, and they need to change the cryptographic keys used by hundreds of systems and the key material in potentially millions of client devices. Considering these questions at scale is critical to enabling the mass adoption of novel cryptographic systems.

**New cryptographic theory, constructions, and protocols.** While most suggestions revolve around deepening research into real-world usage of cryptographic protocols, there remains a need to continue designing novel cryptographic theory, constructions, and protocols. This includes continued research into attributed-based encryption, functional encryption, homomorphic encryption, secure multi-party compassion, post-quantum cryptography, and cryptanalysis.

**Real-world constraints.** Too many cryptographic systems are designed to work in a world of frictionless spheres. There is a need to prioritize research that will work under real-world constraints. To support that effort, there is a critical need for research that better identifies those real-world constraints. While specific examples of these constraints are discussed elsewhere in this document (e.g., key management, existence of authenticated channels), there is a broad research need to identify and construct community standards around these constraints. Foundational cryptographic research should prioritize projects focused on real-world constraints, not theoretical constraints or models that cannot be mapped to practical application scenarios or prove insufficient in practical contexts.

**Real-world instantiations.** There is significant room for research into real-world instantiations and implements of cryptographic constructions and protocols. One such area is to examine how it is possible to formally verify that cryptographic implementations match the specification. While this can be done on a smaller scale, there needs to be research that further automates this process so that it can be done at scale. Similarly, there is a need to understand how existing constructions support novel use cases such as signal processing, machine learning, distributed computation, etc. Where there is not a good fit, new constructions and protocols should be created.

**Cryptographic agility.** Cryptographic research seldom looks at how real-world usage contexts will migrate from one cryptographic algorithm to another. This consideration can be a theoretical issue (cryptographic composites), an applied cryptographic consideration (e.g., how does a block chain migrate to more secure crypto), or a systems issue (how does an operating system or file system or networking stack migration to new crypto). Cryptographic migration is also important for many IETF protocols that incorporate cryptographic operations (e.g., PKI, TLS, KMIP, …) Research is needed on cryptographic agility frameworks for software, hardware, network protocols, and so on. For example, recent white house directives have called out migration to PQC but the academic research literature on cryptographic migration is scant.

**Developer-oriented research**. Cryptographic research will only reach its full potential if it can be correctly used by developers. This involves research that will identify design principles for cryptographic APIs. These designs would involve ensuring developers can understand the implications of using cryptographic constructions and protocols, being able to correctly implement them, and being able to verify that their implementation is correct. How to best support this will require collaboration between cryptographers, human-factors researchers, developers, and industry. There is also room for collaboration with NIST who standard interfaces could (e.g., APIs) for using cryptography constructions and protocols, as opposed to just the protocols themselves. Similarly, there is room for research into communicating cryptographic protocols to industry and developers to facilitate transition to practice.

**Algorithm-specific hardware co-design.** Cryptographic protocols and algorithms such as homomorphic encryption and secure multi-party computing are known to be very slow and expensive, which limits their practical usage in the real world. It is important to explore new research to accelerate these cryptographic protocols/algorithms. One promising direction is to have software-hardware co-design, a methodology that has already been proven effective in AI. There will be many challenges in new algorithm-specific hardware co-design for efficient acceleration of secure multi-party computing and homomorphic encryption. If successful, this could bring a breakthrough to the security and privacy of clouds, as well as many other computing paradigms.

**Quantifying risk within cryptographic implementations.** Quantifying (even as a percent of total contribution to security) the impact of a specific addition to the cryptographic architecture of a system is a black art. This may require threat modeling, as well as a system's view of security (e.g.: impacts of other security solutions in the system, or impact of a failure on a resilient system). Developing a general approach towards developing an approach to such quantification would provide value to system designers and help drive towards efficient cryptographic implementations.

 Adjacent areas of research which can contribute to advances in cryptography and vice versa include static analysis, software engineering, formal methods, model checking, computer architecture, and hardware design. Design of new cryptographic constructions which are power efficient, as well as utilizing hardware acceleration to improve cryptographic protocol efficiency can help with the uptake of such technologies. Additionally, the synergy between hardware for trust and cryptography can lead to new research directions. Another area is the use of cryptographic protocols in securing software and hardware systems. For example, software or hardware supply chain, detecting trojans, etc.

**Cryptographic implementation.** The implementation of cryptographic matters is often a factor in selecting one algorithm over another: performance, hardware acceleration, side channel vulnerabilities and protections, resource requirements, and more. Previous research has shown that in many cases, it is not the cryptographic algorithm that is flawed, but its implementation in the real world. Research that analyzes, studies, and proposes novel methods of improving cryptographic library and algorithm implementations would be useful, and is an important problem to attack. Another interesting problem with respect to the implementation of cryptographic algorithms is how to improve their performance and optimization while at the same time, making sure that these implementations are correct.

Here, cryptographic research should address the needs of NIST standardization. This includes work on implementations across target deployment platforms (x86, Arm, RISC-V) and various contexts of use (IoT), quantifying performance, examining hardware acceleration issues, describing side channel attack vectors and mitigation, discerning algorithm parameterization and implications for performance and research footprint and more. NIST standardization must consider a full picture of usage characteristics and in a comparative way.

**Real world applications.** There is a significant gap between cryptographic algorithms and real-world applications where additional problem constraints and assumptions become important. For example, many SMPC application contexts add problem constraints. Research investment should go beyond basic cryptographic algorithms to include specific problem applications and how cryptographic algorithm should be applied in practice. (See the conference: Real World Crypto.)

**Transition to Practice.** Tension between privacy and abuse detection: A variety of techniques have been built to provide strong privacy guarantees (e.g., end-to-end encryption, tools for anonymous communication), but they are at odds with detecting abuse and other forms of misbehavior. Resolving this tension to provide privacy-respecting collaboration and communication tools, while mitigating negative consequences of abuse is a difficult problem which cannot be easily addressed and requires attention. The goal is not to delegate decision making of what constitutes abuse to a single party (e.g., a government) and instead develop mechanisms of abuse detection and reporting which are flexible enough to provide value to users (safe environment) without creating an impression that privacy is violated.

"Footprint" of cryptography: There is a cost of using cryptographic solutions, be it the need to develop cryptographic constructions, implement them, or the increased runtime for execution of cryptographic solutions on a system. One direction for exploration is measuring the "footprint" or cost of cryptographic techniques and decreasing it. Research can be done toward minimizing the cost of cryptographic solutions (which is currently done to some extent), but also toward lowering the cost of development (e.g., reusing some components for different purposes) or minimizing the component of the systems that is realized via cryptographic techniques (an example is the use of federated learning with a small cryptographic component for the exchange of signals after each iteration to provide privacy-preserving training instead of running the entire computation using cryptographic techniques).

Secure collaboration is the problem of enabling multiple data owners to run computations on their data collective, while maintaining the privacy of their individual data sets. Scalable methods of accomplishing this goal will have a potentially large impact in many fields of human endeavor including medicine, healthcare, finance, insurance, and more. While there are a collection of techniques including secure multiparty computation, homomorphic encryption, differential privacy, and secure hardware-based solutions that address facets of this problem, they do not solve the problem and have had difficulty being deployed widely. There needs to be an incentive for the identification of important application areas for secure collaboration, and a robust translation of existing cryptographic methods into usable and performant solutions. Often, such translational research is painstaking, time-consuming and is not adequately incentivized. It is just as important to incentivize "blue-skies research" into designing novel methods for secure collaboration, including lightweight methods and alternative architectures that solve the problem at hand. Finally, it is important to develop solutions that live in the "goldilocks zone", potentially targeting just the right notion of security for a given scenario.

There are many classic problems that will be exacerbated by increasing use of cryptography. One example is side channels. The nature of cryptographic protocols is that security depends on a small amount of secret information. Attacks that recover just a modest amount of information can lead to huge security failures. With cryptography deployed more widely, side-channels will pop up in an increasing variety of contexts that will need to be understood. In the context of HW trojans there is also this disproportionality. Cryptographic constructions do not exist to address such architecture leakage. One direction that needs research is side-channel resistant construction/implementation of cryptographic protocols. Another is mitigating mechanisms such as threshold cryptographic and big-key crypto. Questions for research include how to make these mechanisms easier to deploy and what kinds of security notions do these mechanisms provide?

One new problem area that will benefit from new cryptographic constructions is unconventional networks where security communication is critical, such as low-earth-orbit satellite networks and richer landscape of networks (where network-aware cryptography may shine). Another new problem area is finding the right balance between privacy and accountability: Access to data is compromised when data encryption is in place. For example, end-to-end encryption helps protect users' privacy, but also creates insurmountable barriers for security analysts and law enforcement when they are investigating or combating illegal content. Balancing privacy and accountability under affordable computation power and reasonable timeframes remains a challenge, and is calling for new cryptographic constructions, such as more scalable homomorphic encryption primitives.

Another fruitful and emerging area can be referred to as cryptography for social good. Examples could include harassment reporting tools (e.g., Calypso), secure computation for empirical studies of social ills, encrypted gun registries, and the broad setting of trust and safety accountability features discussed above. Social good projects can attempt to directly make progress on some complex social ill, but often there is a huge gulf between the technical capabilities of even custom designed cryptography and the challenges faced which may not need or be amenable to cryptographic tooling. Doing meaningful work in this space is incredibly challenging, requiring deep engagement with domain experts to understand what the real problems are, becoming educated on nuances sufficiently to avoid well-intentioned but ultimately harmful technological interventions, and the time and resources to test tools in the real world. It also requires the ability to walk away without career harm, should it turn out that cryptographic protocols are not the real problem. A related set of concerns is about usability of cryptographic tooling, a hard challenge area.

**Technical and Societal Impacts**

The impact of data encryption for information confidentiality, integrity, and authentication cannot be overstated. Data encryption is the fundamental building block of our daily lives in the digital world. For instance, it touches almost every aspect of our lives running from healthcare to ecommerce, financial transactions, and so on.

Lack of interactions between theoretical and applied encryption communities is one of the roadblocks behind translating theoretical developments into real-world applications. Bridging the gap between these communities can pave a way to develop new cryptographic protocols and algorithms that can be deployed/implemented on smart devices. This will allow new technology to flourish, improving the lives of citizens and improving information security and privacy challenges.

Usable and effective key management is critical to the adoption of many, if not most cryptographic systems. Currently, key management is not a fundamental need for most users, however, is slowly moving in that direction (e.g., Bitcoin). Unfortunately, problems are being caused by lackluster key management—consider Bitcoin where people have lost millions of dollars for forgetting a key. This problem will get worse as aging and other effects cause people to forget their passwords or other items that protect their cryptographic keys. By investing in research into scalable and usable key management now, the research community will help prepare the world for a future where cryptographic systems become an everyday part of most citizens life. This may not happen for 10– 20 years, but it is necessary to be ready, instead of scrambling after the fact.

**Cross-Discipline and Outreach Opportunities**

Cryptography and cryptographic protocols can be used to solve or augment a solution to a problem in a wide variety of domains. One example is the health domain, where cryptographic solutions can improve patient care, data analysis, contribute to medical research, secure interaction of medical devices, etc. Another domain is social science or cryptography for social good, where technical solutions can contribute to solving societal problems.

A collaboration in the legal space is also important to treat cryptographic techniques as law compliant and enable their use on a larger scale and in new domains. An existing impediment is the tension between standardization efforts and deployment where progress on one requires advances in the other. Another obstacle in the uptake of cryptographic solutions is their complexity which makes users unwilling to adopt them. Funding priorities, broader education, and discussing cryptographic capabilities with the general public can improve accessibility and understanding.

From a computer science perspective, there is a wide range multidisciplinary collaboration this needs develop its full potential. For cryptographic constructions and protocols to gain relevance and adoption in practice to solve real world problems it will be necessary to work with other fields, including software engineering (e.g., for the proper design of APIs), usability (HCI design, considering social science aspects, etc.), hardware design (e.g., to explore options for acceleration), and formal methods. Efforts should include the engagement of standardization bodies (e.g., NIST) which are provided not only with the primitives, constructions, or protocols themselves but also APIs, parameter choices, implementations (including acceleration), verification of implementations, documentation of assumptions, and supporting research on implications and such. Funding opportunities should encourage and support cross-disciplinary collaborations.

### 3.5.3  Blockchain

**Area Description**

Blockchain is a distributed ledger that records all activities (called transactions) that happen in a system. The ledger is maintained by multiple entities and harnessed by cryptographic building blocks (e.g., cryptographic hash, digital signature) to offer immutability and integrity as the main desired properties of the system. Blockchain has significantly impacted society: cryptocurrency (one application area of blockchain) has entered the mainstream with Superbowl ads and celebrity endorsements. Smart contracts (another application area of blockchain) have enabled digital marketplaces in NFTs. Fundamentally, blockchains provide distributed trust in a decentralized and untrusted environment. Blockchain can be used to improve the trustworthiness and reliability of critical cyberinfrastructure that requires integrity, immutability, and non-reputability with provable security guarantees.

However, today's blockchain protocols, platforms need to evolve to provide the kind of trust, economic models, and security they can support for the applications to come.

A criticism of many current blockchain implementations is the impact of current applications on the environment via resource consumption, in terms of computing hardware, computing cycles, and electricity. This consumption interferes with the ability to scale up blockchain technologies and has environmental impacts (e.g., carbon-intensive electricity production contributing to global climate change). Making blockchain efficient and environmentally friendly is a priority toward making it an environmentally and economically sustainable technology.

One aspect of cryptocurrency that is difficult to ignore is that cybercriminals are using cryptocurrency as the main method of extracting money from ransomware victims. The rise in ransomware is correlated with the rise in cryptocurrency, however it is not clear that there is a direct causational relationship. This real-world impact highlights how easily transferable digital assets can help cybercriminals.

**Technical Efforts**

With the current hype about blockchain applications, it can seem that every problem can be solved by applying blockchain. This mountain of potential applications obscures the actual, real-world applications that can be solved by blockchain. One way forward is to articulate the cases where a distributed ledger system (blockchain) solves the security properties of the problem domain. By identifying these areas, this can drive further architectures, protocols, and cryptographic primitives to solve related problems. Without guiding applications, it can be difficult to focus effort on underlying architectures, protocols, and cryptographic primitives.

Consensus mechanisms are today at infancy. They focus on proof-of-work or proof-of-stake as defined by bitcoin or other platforms. However, from a distributed algorithms and protocols perspective, "agreement" or "disagreement" between decentralized parties need to be supported in a robust and resilient manner so that it remains truly decentralized, adaptive multi-chains, and is resilient from attacks such as majority attacks. What are the open problems and what are the hard problems such as total ordering. Front-running is an important problem today in the cryptocurrency world because of the lack of a total order among decentralized transactions. Consensus mechanisms need to evolve to support fairness as well as completeness. A Multi-layer consensus mechanism can be seen in play across several blockchain networks - DeFi, dApps, and enterprise blockchain networks.

Economic models are key to the stability of a public blockchain or even a private blockchain. There is not much formal emphasis on the economic basis of token omics, virtual currencies, token mint and burn and therefore, several networks are unstable. This dynamic requires a new field of "Computational Economics" to be developed where DeFi, consensus, and physical world macro and micro-economics principles and processes can be applied.

A key focus identified for future technical efforts is to broadly consider user usability in the blockchain areas. Here, the "user" is quite broad and encompasses both the end-user, developers of smart contracts, and developers that integrate blockchain into their applications. A key security challenge is to communicate to the end-user the security guarantees that the underlying cryptography provides. For developers of smart contracts, the usability experience is how to program in a blockchain paradigm and reason about this new programming model, to reduce security risks and vulnerabilities. Developers that integrate blockchain into their applications need to understand the security guarantees provided by the blockchain and how this can impact the security and privacy of their own applications.

The main problem with blockchain is the efficiency and scalability, especially when it is equipped with cryptographic tools to enable privacy-preservation. Currently, blockchain utilizes zero-knowledge proofs (ZKPs) to enable public verifiability with privacy. ZKPs put a significant performance overhead to blockchain systems that make them hard to be scalable to large-scale users. Current attempts focus on public (ledger) setting, establish trust within blockchain (via zero-knowledge proofs): Current problems: applicability and efficiency, trust assumption in exchange of efficiency. Cross-chain transactions, cross-chain smart contracts: multi-chain cooperation, applications and token omics will be key.

Decentralized digital assets are already on the rise such as NFTs. However, data, code, virtual avatars, metaverse-specific resources, intelligence such as codified in ML models are going to be shared, used, and accessed as digital assets.

Current blockchain systems have unique models, infrastructure, and components that make them hard to be deployed in large-scale. Standardization is necessary for almost all aspects.

Regulatory requirements: SEC for financial and digital assets needs to be developed. Regulatory requirements such as what kind of decentralized protocols can be supported for future chains.

Applicability and suitability: How to apply/deploy blockchain to a *correct* and suitable context and application.

The blockchain has largely been introduced to the broader community as a public ledger for cryptocurrency transactions. This introduces privacy concerns, but also undersells the actual capabilities of the blockchain. While it is important to motivate applications that can be effectively implemented on the blockchain, it is perhaps more important to define clear robustness and privacy guarantees due to the distributed nature of the system. As privacy is increased, malicious users may find it easier to stage attacks or commit cybercrime without detection, like how plain secure aggregation allows Byzantine clients to commit malicious updates to the central model without detection. Therefore, it is important to balance a tradeoff between privacy and robustness to ensure that user data stays private, but the system is robust to malicious clients.

**Technical and Societal Impacts**

Enhancing trustworthiness and robustness of the blockchain, while maintaining user privacy, will spur the development and adoption of the blockchain by the broader community. This can also enable a new paradigm of private and robust distributed applications not possible without the use of the blockchain as a distributed ledger.

One of the prominent limitations of the blockchains lies in the scalability, especially when the privacy of the application relies on heavy cryptography, such as ZKPs. Addressing this problem, whether it be through more efficient ZKP schemes or other routes for privacy on the blockchain, allows for applications to no longer be limited by lack of computational power, thus enabling distributed applications to support larger user bases.

Failure to address these problems severely limits the use of the blockchain fundamentally and by the broader community. Problems of scale must be addressed before these applications can be integrated into everyday life, but this problem is intertwined with the problems of privacy and robustness. While there is a tradeoff between the solutions to these problems for each application, the field must ensure that core blockchain protocols do not put user privacy at risk.

**Cross-Discipline and Outreach Opportunities**

Developer perspective: Acute public attention and excitement could be interfering with how developers and the public understand blockchain. Particularly relevant to this session is an understanding of which problems are appropriate to be solved with blockchain and which are not, as well as a gray area between those extremes. Developing that knowledge and finding ways to communicate key blockchain concepts to several audiences (developers, businesspeople, lawmakers, the public, etc.) is a cross-cutting challenge for software engineering, user-centered design, public policy, and possibly other fields as well.

To solve difficult blockchain challenges, several different disciples are needed. Cryptographers can create new protocols or cryptography primitives that can provide different security and/or privacy guarantees. This can allow blockchain to be applied to different use-cases. System security expertise is needed to improve assurance of the implementation of blockchain systems. Game theory expertise is necessary to understand the complex interactions that occur between different actors in a blockchain. Distributed systems expertise is necessary to provide theoretical bounds on problems such as front running, as well as economic impact and incentive mechanisms. Regulatory (such as the SEC) and policy expertise is necessary so that regulation and policy can keep up with the fast-changing landscape of blockchain and attempt to keep consumers safe.

### 3.5.4   Formal Methods

**Area Description**

The goal of the formal method is showing systems behave as they should via proof. The scope of formal methods could include verification of software systems, programming languages, cryptography protocols, and hardware (use of formal verification in hardware design process). Formal methods play many roles in the development lifecycle (may be end user or developer).

In another light, there are different ways to define formal methods, a more traditional viewpoint focuses on software, hardware, or protocol analysis using tools like theorem provers and model checkers. A slightly broader framing is useful, one that encapsulates the more traditional view but includes opportunities for newer directions: formal methods include any framework that attempts to rigorously prove properties of digital technologies. This therefore can include hand-written proofs, game theoretic or economic models, and more. The value of formal methods therefore is the principled process by which one seeks to prove properties, which helps increase the trustworthiness of tools. Within cybersecurity research, formal methods are aimed at security or safety properties in the presence of adversaries.

Historically usability has been a huge issue for formal methods. These include aspects of usability for developers/ designers of the systems as well as getting input from end users to be used for specification of functionalities. The tradeoff between the usability of the software and how much it can be formally verified should be investigated.

**Technical Efforts**

**Formal verification of ML and code by ML models.** New formal methods need to be developed to verify machine learning models to ensure that models do no get into unsafe states that digress them from their specification. Effort is needed to define specifications for ML models that can be used with formal methods. Large language models have been shown to be effective in generating syntactically and (somewhat) semantically correct code. New techniques are needed to help these models produce specifications and proofs about the generated code from the natural language descriptions.

**Formal methods for privacy.** There is a significant need for formal methods in privacy, specifically, in the formalization of privacy from a modeling perspective as well as in terms of the verification of the privacy mechanisms or privacy properties. For example, verifying if solutions provide sufficient privacy and if the appropriate parameters have been selected (i.e., epsilon for differential privacy). More generally, how can the contextual and time-evolving privacy constraints be captured and reasoned? Is this possible across hardware-software, physical-virtual boundaries? While differential privacy provides a clean mathematical formulation, in many cases the usability it provides is quite limited. Are there alternative privacy models that can be developed to better match application settings?

**Formal methods for ambiguous environments (CPS, human users, etc.).** An important challenge for formal methods is the specification of correct behavior, particularly in contexts where correctness depends on a larger context about which the system will have incomplete knowledge. In many such cases, correctness can be specified using only conditions that serve as proxies for what is needed and written in limited ways. For example, while a generative language model should avoid reinforcing stereotypes and biases, this might be specified only in terms of certain topics or terms to avoid. As another example, the goal for a vehicle to not cause fatalities might be specified only in terms of top speeds in certain contexts. In neither case is the specification complete, but automated methods for making the specification more complete poses interesting challenges.

**Formally proving the absence of problematic "extra" functionality.** Many security issues arise from unanticipated behaviors of a system. Side channels that leak information are a typical example. Formal methods that can prove the absence of classes of such "extra" functionality that have been problematic in the past might have significant benefits for system security.

**Formal methods for smart devices (e.g., smart door locks, smart thermostats).** These compact systems usually have a clearly defined usage scope. A challenge for formal methods is deriving specification of expected behavior and violations given a set of user-specified security policies which may take various formats such as IFTTT policies, policies written in free text forms, or other custom policy specifications.

**Research on how formal methods may get integrated into production systems, such as software engineering toolchains.** There are opportunities for projects such as user studies focused on developers' ability to apply formal method tools. Importantly this could include research on how to help security/cryptography developers and researchers define formal preconditions and postconditions for formal methods approaches.

**Domain expansion.** Formal methods in computer science have traditionally been applied to very domain specific potential harms within computer science (e.g., can a program scale up? buffer overflow? vulnerability to hacking?). With the expansion of AI, the potential harms can be propagated by computer programs – and especially ML and LLMs – are increasing. To give two examples, consider first the use of ML in self-driving cars. Harms here could include driving off the road and hitting another car. Is there a possible use of formal methods to rule out either of these possibilities? Alternatively, consider chatbots such as ChatGPT. There are many possible harms here that any firm/company using a chatbot would want to be assured would not occur e.g., encouraging a user to harm themself; teaching a user how to build a bomb; spewing hateful diatribes, etc. While a general formal method of assuring against all harms seems beyond the scope of current research, perhaps formal methods to assure that specific harms are not possible would be? This would involve pushing the methods in new directions beyond how they are currently applied.

**Complex formulation of a property and demonstrating that the property is met.** There are examples above that, e.g., talk about extracting properties from a language-based description or formulating what constitutes racist or hateful language in the context of ChatGPT, but this can be taken more broadly. For example, formulating properties for compliance with privacy laws, for meeting sustainability qualifications, etc. The effort will be to prove systems meeting the expected properties, even if the property formulation is not final and may need to be revised over time (which will require revisiting the proofs).

One question that applies generally is the intersection between formal methods and reproducibility/replicability in different situations and use cases. For example, the inherently stochastic nature of the model learning process in machine learning is considered, can formal methods ensure and communicate basic properties?

**Usability of formal methods tools.** One area of formal methods research that has not received enough attention is usability. Subject matter experts seeking to formally verify systems in their domain may need help from formal methods experts because the tools relevant to their field are not usable. Similarly, it might be difficult for developers to adopt formal methods (e.g., software analysis tools, security type systems, etc.) into their workflows. Right now, the key adoption challenges are not always clear. To improve usability of formal methods: important stakeholders (which may include users, developers, and subject matter experts, depending on the domain) and their adoption challenges must be identified. More research is needed to identify best practices for policy specification, modeling systems, and communicating errors (all of which may vary for different stakeholders, formal methods technique, and the properties being verified).

There is a need to investigate the reasons behind the limited adoption of formal methods by developers and understand the challenges associated with their use. Research should focus on identifying the barriers that make formal methods difficult to employ effectively, enabling the development of strategies to overcome these hurdles and encouraging wider adoption of formal methods in software engineering.

Efforts should be directed towards democratizing formal methods, making them accessible and easier to use for a diverse range of stakeholders. This includes not only developers but also subject matter experts (SMEs) and end users. Research should explore innovative approaches and tools that cater to the needs and capabilities of different stakeholders, promoting their involvement in the formal verification process.

It is essential to examine whether the use of formal methods leads to unwarranted overconfidence in the correctness of software systems. Research is needed to understand the limitations of formal methods and develop effective means of communicating these limitations to stakeholders. This will ensure a balanced understanding of the capabilities and potential risks associated with formal methods, enabling informed decision-making during software development.

Research should focus on developing user-friendly mathematical verification tools that are accessible to key stakeholders in specific domains. Whether it is non-technical users or subject matter experts, their input and expertise should be incorporated into the verification process.

Investigating methods for capturing and incorporating user input within the mathematical verification framework will contribute to the development of more inclusive and effective tools for formal methods verification.

**Concrete notions of guarantees (e.g., privacy guarantees of differential privacy).** Areas such as privacy could benefit from the concrete guarantees that formal methods present. Currently, differential privacy is a popular means of providing guarantees because of privacy bounds that can be represented mathematically. However, differential privacy does not aid in reasoning about other privacy concerns, such as how data is collected and used by third parties. Thus, a research challenge is to determine whether formal methods can be leveraged to provide strong guarantees of these other aspects of privacy to assure the safety of users.

**Compositional/incremental verification (after one small change needs to be reverified).** Formal verification continues to be a substantial challenge for system designers. While formal verification tools have scaled in the size of programs that they can effectively guarantee, real-world systems remain difficult to verify because of their complexity; as an example, developing proofs of the seL4 microkernel required 20 person-years of effort. Such issues are magnified when updates to verified systems are required, as these can potentially require substantial amounts of re-verification. Research is required to make incremental verification feasible. In addition, the composition of verified components could allow for larger-scale verification efforts; cloud computing could aid these efforts.

**Technical and Societal Impacts**

Success in this area has an impact on security, privacy, and safety of individuals. One can define harm broadly, encompassing physical safety (e.g., self-driving cars, harm from insecure devices in cyber physical systems), violations of security and privacy, psychological abuse during interaction with AI systems, etc. Here research in formal methods is critically important in making sure the system functions as it should. There are notable opportunities and challenges. For example, there are options to use natural language interfaces for policy specification and formal methods that will result in real world protections being realized.

Formal methods have had some significant impacts in practice, being used for processor design, aviation system safety, and increasingly for analysis of distributed systems. Should formal methods be developed for new application areas, the expectation/hope is that a similar impact will be achievable in terms of promoting good outcomes for people. The formalization of privacy can have significant impact in terms of the deployment and use of these models and techniques beyond computer science, for applications in health, social sciences, and others.

**Cross-Discipline and Outreach Opportunities**

While traditionally formal methods have been useful within computer science and engineering, much of the future focus may be on exploring their use on a broader set of technology problems that will be critical to solve. For example, understanding how to express safety properties for ML or cyberphysical systems might be able to help provide assurance harm can be avoided, such as hate, harassment, racism, or physical accidents, even in adversarial environments. Research in formal methods can be applicable and benefit from working with legal and policy studies.

The primary communities identified are HCI, human factors, usability, etc. There is a need to make formal methods more usable for stakeholders such as developer teams and SMEs. Experts in formal methods acknowledge that usability is an issue. Likely there is an opportunity to interact with mathematicians. Domain experts in cyberphysical systems, aviation, etc. would also be relevant collaborators. There may also be educational opportunities such as incorporating formal methods training into Computer Science curricula. Formal verification of machine learning is also within scope.

### 3.5.5   Theoretical Cryptography

**Area Description**

Theoretical cryptography is broadly defined as the science of *identifying* capabilities withstanding adversarial behavior, *defining* the security requirements rigorously, and *designing* mechanisms to achieve the said capabilities, using tools from mathematics, theoretical computer science and more. Theoretical cryptography has historically produced game-changing notions such as public-key encryption, digital signatures, zero-knowledge proofs, differential privacy, and secure multiparty computation. Thus, a very important endeavor in this field is to identify the right questions and to come up with the right definitions for the question at hand: an exemplary question, is that of private machine learning. Theoretical cryptography often relies on computationally hard problems to achieve its goals: for example, there are several mathematical ways to construct private-key encryption systems, but many fewer for public-key encryption, and if one imposes the constraint of (presumed) quantum-resistance, the size of this list drops to one. Thus, an equally important endeavor in theoretical cryptography is to develop and expand a robust toolkit of computationally intractable problems that comes from mathematics and perhaps further afield from physics, chemistry, biology and more.

A fundamental nature of research in this field is that the most impactful research is of a "blue skies" nature, where the true impact is hard to predict and is felt long after the problem is formulated and solved in principle. It is important to ensure that fundamental research continues to be pursued, even if the exact nature of the applications is not immediately clear; long term thinking is often needed and should be encouraged by NSF when soliciting and evaluating theoretical cryptography proposals. The initial design of cryptographic mechanisms is typically a proof-of-concept that achieves the stated security goals but is not necessarily performant or usable enough for practical deployment. Thus, theoretical cryptography is tightly connected to the field of applied cryptography that both helps identify problems of practical importance as well as translate theoretical solutions to ones that are performant and usable. Robust interaction between these two fields should be nurtured, encouraged, and adequately supported.

**Technical Efforts**

**Addressing open problems in cryptography.** Theoretical cryptography is a field with abundant open problems. A primary open problem is to formulate new mathematical trust models that provide security that is more robust to failures of various types. Research should explore designs that cryptographic schemes function usefully even when the underlying trust assumptions break down. In general, formulating new and useful mathematical models of security is a core goal of theoretical cryptography. A critical example where this is needed is models for securing hardware against side-channel attacks. While theoretical cryptography has begun this effort with models like wire-probing attacks and more generally, leakage-resilient cryptography, there is significant scope for developing new mathematical models that are relevant to practical attacks.

**Sources of hardness.** Another primary open research area explores identifying cryptographically useful sources of computational hardness: for example, despite decades of research, there are only a small handful of mathematical structures that can be used to construct public-key encryption. Important work explores how the community can reason about hardness in general, demonstrate hardness, and show that these sources are resilient against efficient attacks.

**Unstructured hardness.** Research is needed to explore the potential of unstructured hardness, which does not rely on underlying mathematical structures, and its applicability in building public-key encryption or other cryptographic goals. Research needs to investigate gaining a comprehensive understanding of the power of unstructured hardness and leveraging it to develop secure and efficient public-key encryption schemes or other cryptographic primitives?

**Program obfuscation.** Investigation is required to explore the construction of secure program obfuscation schemes using alternative sources of hardness. For instance, research needs to investigate achieving the original vision of Diffie and Hellman by obfuscating secret-key encryption algorithms to create public-key encryption schemes. One possible approach is to invent a secure program obfuscation method utilizing unstructured hardness and utilize it to obfuscate the Advanced Encryption Standard (AES) algorithm.

**Theory of cryptanalysis.** Research efforts should aim to develop a universal theory of cryptanalysis, increasing confidence in hardness assumptions. By establishing a comprehensive framework for cryptanalysis, the understanding of the computational hardness underlying cryptographic primitives can be enhanced. This would contribute to the development of stronger security guarantees and more robust cryptographic systems.

**Algebraic structures and hardness.** There is a need to identify and analyze algebraic structures that can be effectively utilized in cryptography and reason their computational hardness. Research should focus on exploring new algebraic structures that can serve as foundations for cryptographic primitives. By investigating their computational complexity and security properties, researchers can expand the toolkit of cryptographic techniques and enhance the ability to reason about the hardness assumptions they rely on.

**Cryptographic primitives.** Research is needed to explore the construction of cryptographic primitives not solely based on specific hardness assumptions but rather from large families of hardness assumptions. By broadening the scope of potential hardness assumptions, cryptographic schemes can be developed that exhibit stronger security properties and resilience against attacks. Investigating the construction and analysis of cryptographic primitives from diverse families of hardness assumptions will significantly contribute to the advancement of cryptographic theory and practice.

**Post-quantum cryptography.** Another open question about theoretical crypto is how to build up theoretical foundations for post-quantum crypto. It is critical to understand how different mathematical foundations, computational models and threat models affect the theoretical foundations of post-quantum theoretical crypto. Research should explore the gap between theoretical foundations of traditional cryptography and post-quantum crypto and address the new challenges to build up theoretical foundations for post-quantum crypto.

**Time sensitive cryptography.** Yet another open and challenging issue is that of temporality. Even in cases where data is largely encrypted over its lifetime, the concept of its lifetime is not well-defined. Reasoning in this space could bring new tools and guarantees, such as data that reveals itself after a specific time (i.e., with guarantees beyond current time-lock puzzles) and data that provably degrades or destroys itself after a certain time. There have been high-level attempts to address this problem (e.g., the Vanish paper), but failures in threat modeling and assumptions about the protocols upon which this system was built ultimately resulted in their defeat. Foundational primitives with specific guarantees would make formal reasoning about the lifecycle of data possible. More generally, developing hitherto unseen cryptographic primitives such as vanishing data could make use of new physical properties such as the (apparent) hardness of problems not only from mathematics, but also physics, chemistry, and biology (such as the hardness of maintaining RNA molecules without decay).

In many application settings, encryption tools need to interact with the real systems (e.g., database systems) and somehow what information people should be able to extract from encrypted data needs to be defined. Furthermore, such access may need to be audited later for governance reasons. Understanding such access control and governance requirements in developing novel theoretical encryption techniques could be useful.

Bridging practical cryptography to formal guarantees. Finally, there is a large chasm between fundamental cryptographic primitives used in practice, such as block ciphers, and theoretical constructions thereof, called pseudorandom functions. While AES is widely used in practice, it comes with essentially no theoretical security guarantees. Research must bridge this gap using new mathematical tools. Here an attainable goal would be to show that AES (or other block ciphers) resists a large class of practically relevant attacks including differential and linear cryptanalysis and algebraic attacks.

**Technical and Societal Impacts**

Cryptography such as asymmetric and public key cryptography have become crucial foundations of the information age underlying all transactions. When successful, theoretical cryptography breakthroughs spawn new fields of research and practice for secure and trustworthy computing applications and directions. Advancements in theoretical cryptography will provide foundations for future systems, such as ensuring secure post-quantum computing.

One reliable indicator of success in theoretical cryptography is that a new domain of research and/or application takes off. Secure multiparty computation and differential privacy are two recent examples of ideas that began in theoretical cryptography and then matured into fields with clear applications and application-oriented optimization goals.

Threats to theoretical cryptography mostly come from outside the field. For instance, quantum computing poses threats to existing public key encryption protocols. Staying ahead of the advances that threaten secure information exchange requires advances in theoretical cryptography.

**Cross-Discipline and Outreach Opportunities**

Theoretical cryptography has the potential for positive engagement with current topics in mathematics and automated theorem proving (e.g., Schotze's work on foundations of mathematics and LEAN, and advances in higher-order dependent type theory and calculus of inductive constructions). Efforts to formalize problem formulation in hardware design are also likely to lead to cross-disciplinary work. One challenge for outreach is identifying the right stage in theoretical cryptography research to turn toward application – too early stifles innovation and too late limits the potential impact. Another challenge stems from the various methods of demonstrating knowledge in different domains. For instance, for some fields like theoretical cryptography proofs are the currency of progress, but in other fields systems embodying successful applications are required to recognize progress. There is significant scope for a feedback loop with other fields providing a rich set of goals and directions for theoretical cryptographers to explore.

There are interesting research questions to explore when it comes to the impact of cryptographic techniques on the environment. To move to a world where most data is encrypted in some form - to guarantee properties, what would be the implication in terms of energy consumption and impact on the environment? What is realistic here and to what extent can algorithms provide strong security guarantees while minimizing energy consumption?

Further, while theoretical cryptography has traditionally ignored the human element, there is a need to understand how humans are expected to interact with these techniques, whether assumptions made about what humans can do are considered. The community should consider whether important properties such as "fairness", "democratization", "decentralized controls" be considered in cryptography, as well design cryptograph algorithms that provide these properties.

While advances in theoretical cryptography have had significant societal impacts, it is sometimes difficult to predict those impacts. There may be opportunities for cross-disciplinary engagement with, e.g., social scientists, ethicists, economists, public policy scholars, and HCI scholars to work with theoretical cryptographers to translate how theoretical advancements might have both negative and positive downstream impacts, and to incorporate those perspectives into the goals of the theoretical work itself. It is worth engaging with policy and legal scholars to understand ways regulation and policy requirements can co-exist and be incorporated into advances in theoretical cryptography.

It is also important to create educational opportunities to broaden representation and knowledge of theoretical cryptography. High-level knowledge of the goals of and advances in theoretical cryptography across a broad population might help translate and accelerate those advances into societally meaningful outcomes.

## 3.6    Privacy

The study of privacy within SaTC explores personal, group and systems privacy in the context of technology and data collection and use. Researchers in this field focus on developing techniques, policies, and mechanisms to protect individuals' privacy rights, as well as analyzing the impact of technology on privacy. They examine issues related to data collection, processing, storage, and sharing, and investigate the ways in which privacy can be compromised by inference, eavesdropping, and data breaches. In addition, researchers in this field explore the social, legal, and ethical implications of privacy, as well as the relationship between privacy and other values such as security, free speech, and innovation. Privacy research is crucial in ensuring that individuals' rights are protected in the digital age, and as such, it is an interdisciplinary field that draws on computer science, law, sociology, psychology, and other disciplines.

### 3.6.1 Anonymity

**Area Description**

Researchers and engineers have been working on systems to preserve anonymity for over three decades, yet existing solutions have meaningful weaknesses that leave users vulnerable. Science must continue in this area to develop stronger solutions resilient to current and future adversaries. Before exploring the look of future work, several key questions that illuminate the opportunities for science and impacts should be considered.

Any inquiry into anonymity should start with identifying what are the objects or entities that anonymity applies to. Typically, anonymity relates to people. However, anonymity, as discussed in this group, should also pertain to non-living objects. There is a need for objects that have been given/assigned identity to be anonymous, if needed.

Deanonymizing can also result in different levels of identification. For example, an attack on anonymity of some dataset might identify "pseudonymous person #1", or "some Texan" or "Greg Abbot"—all of these reduce anonymity, but to different levels.

In another light. anonymity is not an "all-or-nothing" property of an entity; an entity may be complete or total anonymity with respect to another entity, or it may have a degree of anonymity (a.k.a. partial anonymity or pseudo-anonymity). For example, anonymity technology may be designed in such a way that every player is anonymous so long as that player does not perform a prohibited action (e.g., try to double-spend a digital coin). Attempting to complete a prohibited action could result in a loss of anonymity, resulting in traceability for the bad actor.

Another form of anonymity that might be desired is pseudo-anonymity, whereby the identity of an entity is hidden but it is possible to link the entity's activities together into a single profile. For example, the Direct Anonymous Attestation cryptographic technology in the Trusted Platform Module v1.2 specification was a form of pseudo-anonymity that protected the identities of attesting TPMs but allowed attestations to multiple websites to be linked to each other (if this option was chosen by the websites).

Further. it is important to recognize that anonymity only makes sense within some defined scope. "Scope" includes what information is shared (and thus subject to deanonymization), and the capabilities of the adversary. While strong results may be possible in narrow, well-defined scopes, anonymity from a global, all-seeing adversary is impossible.

As an example of scope of sharing, compare sharing a dataset, to observing a network from a fixed point, and to observing all data entering and exiting a laptop. A dataset may be prepared ahead of time and scrubbed of identification to expose little information about individuals. Observations from one network point have more information but may only capture part of what an individual sees. All data leaving a laptop provides very detailed information about a single individual. What level of anonymity is achievable varies in each of these domains.

Scope in the capabilities of the adversary matters as well. Continuing with this example, access of dataset through differential privacy mechanisms provides guarantees, but allows only certain types of queries. Access to the raw underlying data provides more information. Access to a network or a laptop represents different levels of invasiveness of an adversary.

**Technical Efforts**

The overarching goal of technical efforts is to identify tradeoffs inherent to tackling anonymity. Here, defining anonymity and the role of anonymity is a multifaceted challenge. Identifying trade-offs needed for anonymity requires clear definitions. It is also important to clearly understand the differences or relationships between privacy vs. anonymity in different contexts. Anonymity of people also depends on anonymity of devices and other non-living entities. Research also needs to address what new challenges new domains and applications such as metaverse bring, and understanding the role of anonymity in large datasets that are shared and used for analytics. Exemplars of research in these areas include:

**Secure MPC and Privacy**: Secure multi-party computation is a powerful tool for achieving anonymity in certain contexts. Secure MPC allows for general computations on private inputs. Many anonymity tasks can be formalized as MPC tasks. An area of research necessity lies in determining when MPC based solutions are feasible for tackling anonymity goals, and if so, what are the performance, privacy, and cost tradeoffs.

**Enabling revocation of anonymity**. There are many serious societal needs that necessitate the revocation of anonymity within deployed systems. There are various challenging questions to answer: what kinds of revocation are technically feasible? How can systems of governance be designed that can authorize such revocation? How can these governance systems be held accountable for abuse of their privileges? How can such systems be audited for discovering abuse, without sacrificing anonymity or with minimal sacrifice of anonymity? A concrete example of where these questions should be investigated is cryptographic ring signatures.

**Definitions of privacy and anonymity**. Various research challenges that need to be explored include anonymity in the context of decentralized identity, controlled (through) anonymity with governance, efficient and scalable cryptographic primitives, anonymous authorization, new technical definitions of anonymity that are alternatives to differential privacy, public verifiability issues around anonymous credentials, issues related to plausible deniability in the context of censorship and surveillance, balancing anonymity with accountability.

**New anonymity network designs.** Anonymity networks such as TOR and other onion-routing services have been shown to be vulnerable to a variety of attacks, many of which are capable of completely deanonymizing users. An additional risk is posed if there is collusion between relay owners (particularly in the case of censorship/anti-surveillance resistance). These issues, however, are fundamental to the currently designed systems. While there could be unexplored defenses to help mitigate these flaws, a promising research direction would be investigating new anonymity network designs that are fundamentally different from current methods. Additionally, efforts toward creating systems that more seamlessly integrate into typical network operations, but provide a greater degree of anonymity, is an open research problem worth exploring.

**Network anonymity and traffic analysis resilience.** Network anonymity helps to protect users' privacy and personal information by making it difficult for third parties to identify the source and destination of online communications. Traffic analysis is a technique used to identify network traffic patterns to infer information about the communication of content. Anonymity and traffic analysis resilience can help to protect against this type of attack, making it more difficult for third parties to monitor and track online communications.

**Holistic anonymity.** Users have various identities and means to ensure that anonymity goes beyond a single device/service. Users will be accessing the internet through various devices, and it is important that anonymity is provided through all devices. E.g., in a smart home, different devices might be able to access the internet, and if any one of them can reveal the user's IP address, that will subvert the anonymity of the user. How can holistic anonymity systems be designed for users who use multiple personal devices as well as share many other devices with other users?

**Design of accountable anonymity and preventing abuse of anonymous networks.** Finding the right balance between privacy/anonymity and abuse is a challenging problem. Such networks are often used for illegal or nefarious purposes. Researchers should measure the usage of such networks to understand their impact on network infrastructure, as well as how these systems are being used by users. A novel research direction is how to develop frameworks for accountable anonymity. Users should be able to maintain a certain level of anonymity while revealing certain quasi-identifiers as required by a service. In certain contexts, the authorities can deanonymize users, should they perform a malicious task. Beyond the technical challenges, there are several difficult legal and policy questions: e.g., what should the accountability framework look like, and who should be the "key escrow"? Similarly, research is needed to understand different levels of privacy/anonymity, so that users can (or are required to) pick one of these levels while using anonymity networks. How can new network protocols be designed, so that they only disclose certain information about the user, while preserving others?

**Censorship / anti-surveillance.** The use of anonymity services can be particularly important for individuals concerned about being tracked or monitored, such as whistleblowers, journalists, activists, and citizens living under oppressive regimes. However, the use of anonymity can also create tension between privacy and security concerns, particularly in the context of government surveillance. While governments may argue that they need to monitor and track individuals to prevent terrorism or other threats to national security, this can lead to the potential abuse of power and undermine civil liberties and individual privacy rights. Therefore, the development of techniques to bypass censorship and surveillance while maintaining anonymity is an important area of research. This research is essential to ensure that individuals can communicate and express themselves freely without fear of persecution and to promote the values of democracy, human rights, and freedom of speech.

**Data anonymization.** With data-hungry machine learning applications, more and more information can be collected and analyzed. This poses new questions about data anonymization.

Techniques such as differential privacy, and k-anonymity have been applied to reduce identifiable information. However, there are still open questions about de-identification risks with advanced attacking techniques. Research needs to investigate 1) how network anonymity works with data anonymity, 2) if users will be more comfortable sharing data in the context of an anonymity network and 3) how to design network anonymity and data anonymity together to achieve a holistic solution for anonymity.

Multimedia content such as images / video (or even background audio) can reveal unintentional information, including information such as location of the action or included bystanders. While data anonymization typically targets discrete data such as identifiers, anonymizing images/video/audio in a meaningful way could be a challenge. Here, the assumption is that the multimedia content still needs to be "usable," i.e., obfuscating faces or identifiable location content may not be desirable for the intended parties consuming the multimedia content.

**Technical and Societal Impacts**

Anonymity is a double-edged sword. Empirical evidence about uses of today's anonymous communication technologies suggests that these technologies are dominated by socially undesirable activities, which gives rise to the accountable anonymity agenda outlined above. That said, the socially desirable (potential and actual) uses of such technologies are numerous. Failure to advance this field of research will leave many of these socially desirable uses underserved.

Anonymity is essential for several applications as well as should not be supported for many other applications. In some cases, the same application requires anonymity to a certain level but not to an absolute level that it is hard to determine some attributes of the entity. In some cases, the same application requires revocation of anonymity to a certain degree. Therefore, for the contradictory requirements, policies, governance and mechanisms need to be supported for anonymity. Anonymity is Inherently dual use. Here, strong anonymity could conflict with other things (e.g., technologies may facilitate crime).

Highest level of anonymity: it can aid whistleblowers but needs to support accountability. It, however, may not have any societal impact nor any accountability. The Bitcoin paper by the pseudonymous author has several implications: who owns those several billion dollars' worth of Bitcoins and how they will be used in future.

On the other hand, pseudonymous, revocable, accountable, auditable anonymity can support several "good for society" applications and impacts. There is a need for some level of anonymity around voting. People may believe they have anonymity but are unaware of fingerprinting, data collection, etc.

How much anonymity is being supported in a cybersocial, cyberphysical world should be measured, monitored and supported. Social trust depends on the level of anonymization, and the amount of accountability. Metaverse adoption as well as adoption of anonymous messaging, anonymous networks are going to be driven by technical progress on "anonymous" vs "accountable-anonymous - social, ethical and legal" principles.

Decentralized computing, blockchain, cryptocurrency and evolution of regulation and compliance around them are driven by and drive the nature of anonymity supported.

Autonomous driving, robots, non-living entities may need to have anonymity to support the "work" they do to help authorized entities who own or rent or use them. Privacy is tied to anonymity at that level as well. In fact, the security and safety of these non-living and living entities are dependent on the level of anonymity. Security attacks are hard to attribute to anonymous actors and are sometimes hard to be carried out against anonymous actors. Similarly, protection of anonymous entities from security attacks is hard. So, anonymity has an impact at the appropriate level for the good of society but beyond a point, it becomes impractical to be adopted.

Similarly, usability of a system may be influenced by the level of anonymity. Anonymous credentials and anonymous computing may make it harder, or easier, depending on the type of application.

**Cross-Discipline and Outreach Opportunities**

Anonymity naturally intersects with civil society, including law, policy, and law enforcement. The technical and civil society/policy communities should agree on the definitions of anonymity, methods to achieve anonymity, as well as tools to trace bad actors, with appropriate legal authority, via forensic analysis of malicious online activities. There are also many outreach opportunities to social science and psychology, such as how anonymity influences people's mindset and online behaviors. Research should emphasize better understanding of the implication of anonymity on society (e.g., harms).

### 3.6.2    Privacy Secure Multiparty Computation

**Area Description**

Secure, multi-party computation (also known as secure computation, multi-party computation (MPC) or privacy-preserving computation) is a subfield of cryptography with the goal of creating methods for parties to jointly compute a function over their inputs while keeping those inputs private. MPC enables distributed trust and prevents a single point of failure/compromise thereby enhancing the robustness, resiliency, and trustworthiness of the cyberinfrastructure. MPC can be found useful in many applications and domains such as healthcare, where multiple entities (e.g., health providers) desire to perform joint data evaluation or analysis while complying with standard data regulations (HIPAA, GDPR).

**Technical Efforts**

**Scalability and Performance.** There are certain improvements in terms of performance and security model in MPC recently.  However, there are several significant scalability challenges that hinder deployment of MPC techniques. First, the amount of the data that needs to be processed currently outpaces the MPC improvement. In some applications, such as those in Machine Learning, data set size increases are outpacing the rate of technical improvement.  Second, the challenge also comes from the complexity of applications. For some simple computations, such as set intersection, there are MPC techniques that scale to a large amount of data. However, current MPC techniques cannot handle the combination of complex computations and large amounts of data. Third, some protocols have challenges when scaling to many parties. Fourth, usage scenarios in practice by and large tend to be very different from the settings typically covered in theory. Specifically, in practice there are dynamic aspects such as arriving and departing parties, as well as the need for repeated execution of the protocol(s).

**Generalizability.** One key problem with MPC designs is their inflexibility. Such systems often rely on certain assumptions, external conditions, and computational goals (e.g., desired output) about the data and the computations that are fixed at the time of design/implementation. When these change, such systems need to be drastically modified or even fully replaced with new implementations. Techniques and design patterns that allow for proportionality between changes in input data and changes in the MPC design are an open problem for research. A subset of the inflexibility challenge is fragility. When the trust assumptions about parties change after the MPC system is designed, the design itself may need to be drastically changed or altogether replaced with a new one. A technical challenge therefore is the design of MPC systems that allow for more flexible handling to changing trust assumptions.

**Usability and Deployability.** MPC is complicated and hard to develop due to its complex cryptographic building blocks composition. Due to its unique model and assumptions, it is a challenge to deploy MPC and integrate it with legacy infrastructure. What is needed are new frameworks and systems that allow complex and performant MPC implementations to be realized in a wide range of environments.

**System/Adversarial Assumptions.** Many MPC protocols have unique systems (e.g., fixed 2,3 parties) and adversarial models (semi-honest, malicious). In practice, the number of parties can be large, many of them may not be available to be online and participate in the joint computation process (e.g., network partition, outage). It is challenging to make MPC adaptive with the flexibility and uncertainty of real-world systems.

**Technical and Societal Impacts**

The development of scalable, generally applicable, and usable frameworks is likely to create real application opportunities in diverse domains such as healthcare and finance. There are however significant deployment challenges. The theoretical work in this area makes multiple different trust assumptions, and the extent to which these different threat models apply to real world situations is unclear. There is also the need to map trust and legislative requirements to be met in documented and verifiable ways.

If successful, MPC would enable focused collaboration among (partially) mutually untrusting entities, which is an extremely common setting. One specific example, is the collection of sensitive census/demographic information or personal financial information needed for certain purposes (e.g., determining the health of the economy) but the contributors may be reluctant to provide it for fear of misuse.

Proper use of MPC may provide incentives or properties that may be of interest or importance to parties in practice. For example, MPC may be used to reduce subjectivity and potential misuse by a single trusted entity, which is significant in real-world applications (e.g., medical applications).

**Cross-Discipline and Outreach Opportunities**

There are domains where MPC techniques are valuable, e.g., healthcare and finance. However, significant efforts are needed to bridge theoretical techniques to practical settings. For example, one opportunity for cross-discipline research is within the legal domain. Deployment of security and privacy techniques are largely driven by compliance. A topic that needs more research is the legal status of MPC. The community and society need answers to several key questions. What is the role of MPC techniques in compliance? Does deployment of MPC techniques satisfy compliance needs? Where does liability lay when things fail?

It is also important to contemplate MPC usability and incentives. In many settings, the stakeholders may have less incentive in deploying complex MPC protocols. Especially, for compliance reasons, the good enough, privacy regulation compliant solution could be preferred to a complex MPC solution. In addition, MPC solutions may be harder to deploy and use from a stakeholder point of view. In some domains such as cryptocurrencies, complex crypto based protocols (e.g., zero knowledge proofs) have been successfully deployed in practice. Therefore, more research on addressing incentive and usability issues in target application areas in collaboration with social scientists could be useful.

### 3.6.3   Statistical Methods

**Area Description**

Our data is collected left and right; it is shared, recombined, and mined for all sorts of different purposes. How can statistical guarantees be provided to people whose data is collected, and how can it be ensured that their data is not used to make undesirable inferences? Can a taxonomy of privacy threats and associated statistical guarantees be identified? What techniques are needed to develop and provide these guarantees? How are these guarantees effectively communicated to lay users, and what sort of control is made available to users when it comes to adjusting/selecting those guarantees under which their data is collected? Is it possible to develop functionality that empowers data subjects to evaluate the likelihood that different types of facts are already known about them, or the likelihood different types of inferences can be made about them by different entities? What types of tools are needed to empower developers to embed statistical guarantees in the software they develop and the data they collect? Is it possible to build developer tools, where developers can select and customize from a taxonomy of guarantees?

Answering these questions is critical to ensuring the successful implementation and deployment of statistical methods to ensure data privacy. For example, there is a fundamental tradeoff between the effectiveness of these approaches in terms of preserving the privacy of individual data subjects and the utility of the perturbed data in terms of allowing analysts and models to make useful inferences. What are best practices for navigating that trade-off? It remains unclear, in addition, when these methods are appropriate and when they are not. Differential privacy, for example, is just one of a broad set of privacy-enhancing techniques, but there remains an opportunity to better understand what threat models' differential privacy protects against, how these threats relate to different application areas and user privacy concerns, and when alternative techniques may be useful complements or replacements for differential privacy.

Developers, today, are the primary touchpoints for how statistical and differential privacy methods are implemented and executed. They make several decisions that have material downstream consequences. For example, developers implicitly or explicitly decide the threats they are aiming to protect data against by employing, and explicitly determine the acceptable trade-off between "privacy" and "utility". Yet not all developers will be experts in these techniques. Supporting these developers, particularly non-experts, with processes, methodologies, and tools to help them make informed decisions with respect to statistical and differential privacy methods is thus of critical importance to ensure the effective use of these technologies to protect data privacy.

In addition, real-world deployments of DP face barriers of identifying appropriate data usage scenarios and privacy requirements. One example is the prevalence of using DP for dataset releasing. This requires research to identify DP misuse to prevent data breaches or information leakages, as well as socio-technical efforts to transfer DP techniques into practice.

Also, the trending generative AI (e.g., ChatGPT) and other variances' data sharing policy and term of use raise concerns around data privacy, data ownership, and privacy-preserving data sharing and usage. New systems that depend on aggregating data about individuals require community efforts to understand the new data privacy risks and come up with new privacy solutions, including guidelines, social-technical privacy frameworks, and new formulation of privacy mechanisms.

**Technical Efforts**

**Differential privacy.** Differential privacy provides statistical-based privacy guarantees; however, this is not always sufficient or suitable in many application scenarios. What, if any, other privacy guarantees need to be provided/supported by privacy-preserving statistical methods? Additionally, the guarantees offered are at the data level, but the model itself is not always kept private. Model perturbation may be necessary for stronger guarantees. More broadly, a threat model for statistical privacy methods is needed, defining privacy needs, malicious actors and underlying assumptions (e.g., knowledge, computational power).

**Data Perturbation or data sanitization or data anonymization.** Data Perturbation or data sanitization or data anonymization represent a conventional methodology for hiding data in the crowd while finding a good balance between data privacy and data utility. Two well-studied data perturbation techniques are k-anonymity and its family of methods (incl. L-diversity, t-closeness) as well as differential privacy and its family of variants.

For k-anonymity related privacy research, one of the key challenges is how to determine the proper k for each sensitive attribute under a large population and how to address the correlation among sensitive attributes to avoid minority partition to suffer from leaks (l-diversity, t-closeness, etc.)

For differential privacy, one of the key challenges is how to determine the privacy budget Epsilon specification for a given data processing or learning function, and how to guide Epsilon selection to ensure better tradeoff between privacy protection and preserving the utility of the function on the input data. However, much of the existing work is exploratory and not foundational.

**Privacy theory**. One important technical effort would be to integrate the state-of-the-art privacy theory and methodologies such as k-anonymity and differential privacy while preserving stronger privacy and better data utility, especially dealing with the skewness from both raw data and from the resulting K-anonymization procedure.

Another challenge is to develop theoretical guarantees for a privacy solution. Important mechanisms include formally defining the privacy threat model, and the data utility a differentially private function should preserve. Local differential privacy, for example, has potential to provide theoretical guarantees in distributed settings, where a single release model does not apply.

This calls for privacy and utility metrics: declaring privacy bankruptcy when the "privacy budget" is used up and still not ensuring the desired privacy protection or not ensuring the expected data utility. One way to handle this problem is to reset the privacy budget to balance privacy and utility.

**Fairness and differential privacy.** Although there is some work at the intersection of fairness and differential privacy, it is essential to understand how differential privacy impacts different stakeholders and devise solutions to eliminate/ minimize the potential negative impacts.

**Matching and translation between theory and application needs**. Foundational research on statistical privacy methods is crucial, at the same time more research is needed on supporting users of such techniques in understanding what type of solutions are available for their specific application and how a specific technique needs to be applied to be effective. More application-driven research would further help to move the problem specifications for statistical privacy methods forward to ensure that fundamental developments are informed by and can be applied to specific areas in need of privacy guarantees.

It is important to have approaches for explanation and comprehension of privacy-preserving statistical methods. This includes both developers and end users. For example, how to empower developers to understand parameters and select the right parameters (e.g., Epsilon in DP) considering the context of the application and the data. Similarly, end users and consumers should be able to gain a reasonable understanding of the protections a certain statistical method provides or does not provide. This is relevant for facilitating trust in an organization's practices.

**Data sharing is critical for statistical methods**. However, a fair number of domains may only share data under privacy regulations. This leads to the synthetic data generation area, especially in healthcare and mobile location-based services. Synthetic data generation is one way to generate less skewed data for applying statistical methods, such as differential privacy. Synthetic data also has the inherent advantage of being "false", therefore risk of exposure in case of leakage or other attacks is greatly reduced. An important aspect with synthetic data generation, however, is to ensure biases are not introduced into the data generation process, and that the data is reflective of the population being modeled.

**Privacy guarantees across layers and contexts**. Similarly, privacy guarantees and statistical methods need to be broadened to account for compounding privacy risks. For instance, one mobile app employing statistical methods to protect the privacy of a user's location while multiple other apps, the mobile platform provider, and the mobile carrier still track an individual's location makes the privacy guarantees offered by that first app less meaningful for the user's actual privacy.

**Auditability of statistical privacy applications**. While statistical methods can provide some metrics regarding certain privacy guarantees, in commercial applications these guarantees are rarely shared or made accessible, yet consumers, regulators and research should be able to assess and audit to what extent their privacy is being protected in practice, rather than having to rely on organizations' promises of what they do. A key challenge here is to develop metrics, processes, and tool support to facilitate auditing of the applications of statistical privacy methods.

**A privacy framework / cookbook.** A very valuable direction would be a privacy framework that would provide a "cookbook" or sorts for informed non-experts and practitioners about the possible approaches for situations involving acquisition, computation and release involving private data. Achieving privacy is not an all-or-nothing, but one that involves making reasonable tradeoffs. In situations that involve one or more data gatherers collecting and using data from data providers, there are many different approaches at play -- zero knowledge proofs, secure hardware, federated/multi-party computation etc., in this set of approaches, it is important to evaluate their costs of adoption, and what combination of them will allow the gatherers and providers to make the right tradeoff regarding privacy. Such a cookbook will be valuable to the research community.

**Technical and Societal Impacts**

Privacy has become an extremely important issue in our society. The field of differential privacy has demonstrated significant growth and success in developing privacy guarantees in the past 20 years. It is remarkable to see big tech firms' efforts on applying differential privacy and promoting the public's awareness of differential privacy. In 2016, Apple devoted several minutes in the keynote of their widely publicized product-release event to explain the idea of differential privacy to a largely non-technical audience. This clearly spoke to how the rapid development of privacy techniques had grown from an academic pursuit to become a quintessential part of the public discourse about privacy.

Looking forward, the scientific community shall consider investing more efforts on guiding users (e.g., developers, data scientists, etc.) to appropriately use data protection techniques such as differential privacy. More importantly, researchers need to effectively communicate the meaning of differential privacy (which aspects of privacy are guaranteed, and which are not), to different stakeholders. Everyday people by and large do not understand what these techniques do with their data and how these techniques protect their privacy. Even experts do not agree on the societal impacts of these techniques, as evidenced by the controversies surrounding the use of differential privacy in the 2020 U.S. Census, the potential negative impacts of differential privacy on the detection of health disparity, the fairness of post-Census resource allocation, or the future of democratic representation.

In addition, it is important to be mindful of potential misuse of differential privacy and other privacy-enhancing technologies. Companies may employ such methods as disguise for transparency and then claim their data is not actually personally identifiable even if the outcomes of such privacy-preserving techniques still impact individuals, e.g., efforts towards more privacy-preserving targeted advertising but those techniques do not address issues of influencing and harming users. Thus, a challenge is to broaden, define, and assess privacy guarantees in a way that takes downstream implications of statistical privacy methods into account. A related challenge is to develop methodologies and techniques to facilitate auditing.

Success in the design and deployment of statistical tools and various forms of privacy-preserving analytics (differential privacy, federate learning, secure multi-party computation, and so forth) would consist in the ability to simultaneously achieve two goals: privacy protection with minimal reduction in data utility. Benefits arising from this success may include an increased willingness by organizations to share, disclose, or allow analysis on datasets previously kept private for security, privacy, and confidentiality reasons. In turn, increased access to otherwise unavailable datasets may aid or foster research across a variety of disciplines.

In this process, some pitfalls should be avoided. Those pitfalls include: an excessive reliance on statistical tools at the cost of investigating or using other viable and effective technological or legal solutions for private data sharing and analysis; the deployment of such tools in scenarios where they are not effective (either because the privacy protection they provide is not sufficient, or because the degradation of data utility they cause is too significant); the co-opting of such tools by organizations for "privacy theater," merely focusing on nominal compliance with abstract requirements without consideration of the concrete privacy outcomes arising from the deployment of these tools.

## Cross-Discipline and Outreach Opportunities

There is a need, as well as an opportunity, to tie technical research in this area (statistical and computer science work in particular) with efforts grounded in social science research. "Human-centered statistical privacy" should aim at striking a balance between empowering users without overwhelming them. Behavioral and communication research could complement HCI and design research in designing mechanisms, interfaces, and educational materials that adequately inform different stakeholders (data subjects, but also end-users of data, as well as stakeholders within the organizations that collect data and need to make decisions about the proper techniques they should apply to datasets to preserve privacy and confidentiality) about the existence, properties, strengths and limitations of various statistical and cryptographic methods that allow the protection of individuals' data in parallel to the extraction of value from data.

Research on these tools should also interface with social science research on bias and fairness, given the well-known risk that certain data privacy techniques (such as differential privacy) can differentially affect marginalized groups.

The application of these tools and techniques differentially can generate costs (for instance, opportunity costs) due to the potential degradation in the utility of the original, raw data (this concept is encapsulated in the notion of a Risk-Utility trade-off curve).[2] Economic research on the way those costs are allocated to different stakeholders (data subjects, data holders, society as a whole) is an important complement to technical research in this area.

To promote the deployment of differential privacy and other statistical methods for privacy, theoretical research should be complemented with more human-centered, application-driven, use-inspired research. This will include investigating how different stakeholders are affected by and should participate in the deployment of differential privacy, as well as exploring novel applications of differential privacy beyond the current ones like adding noise to the collected dataset (e.g., the US Census data). This requires interdisciplinary collaboration across computer science, design, social science, policy, law and various other disciplines.

The study of different stakeholders should include not only the data subjects, the data users (e.g., data scientists), but also the app designers and developers who are oftentimes not privacy experts but need to determine the right use case to incorporate differential privacy and correctly incorporate it. Significant research is required to design better tools and interfaces to communicate the risks and benefits to different stakeholders in an accurate, intuitive, and easily relatable manner, facilitate the negotiation about the privacy budget among different stakeholders, improve the technical literacy about differential privacy and other privacy-enhancing technologies of the public, and support the design and development of apps with differential privacy, which means more research should be encouraged that investigates the problem grounded in concrete applications and acknowledge their contribution to fundamental research.

---

[2] See, e.g., Duncan, George T., et al. "Disclosure limitation methods and information loss for tabular data." *Confidentiality, disclosure and data access: theory and practical applications for statistical agencies* (2001): 135-166; Fienberg, Stephen E., Alessandro Rinaldo, and Xiaolin Yang. "Differential privacy and the risk-utility tradeoff for multi-dimensional contingency tables." *Privacy in Statistical Databases: UNESCO Chair in Data Privacy, International Conference, PSD 2010, Corfu, Greece, September 22-24, 2010. Proceedings*. Springer Berlin Heidelberg, 2010.

If differential privacy is to be used to share data with the broader research community, then the barriers to carrying out research using these data should not be prohibitively high for non-specialists in differential privacy. One cautionary example is the Social Science One Research Initiative.[3] This was a massive, and first of its kind, effort to release data from FB to the broader research community. The first attempt to do so involved a collection of information about links shared on FB (i.e., not even individual level data - the unit of analysis here was the hyperlink) that included numbers of views, shares, likes, etc. of these links disaggregated by different socio-demographic characteristics.[4] The decision to introduce DP had two major unintended consequences. First, it delayed the release of the data by a significant amount of time (likely over a year). Second, as researchers lacked appropriate statistical tools for working with DP data, it limited the research that could be conducted (as well as the pace at which research could proceed). While there may have been idiosyncrasies associated with this project, the larger lesson seems clear: using differentially private data to advance research is going to be dependent on the research community possessing methodological tools for working with such data.

The Social Science One example also illustrates another cross-discipline opportunity: when differentially private data is not an appropriate method for protecting privacy, what other sociotechnical approaches are available? The SSOne scenario highlights a situation when the data users lacked statistical methods or knowledge. "Skill up the data users" is not likely a workable approach to ensure that data users can work with DP data. Methods are needed to work with data users to develop statistical techniques that ensure data (re-)usability, and determining when other methods such as hardware security, network security, social or political access policies, etc. are more appropriate methods for privacy protection than DP and other statistical methods that modify the data shared.

---

[3] https://socialscience.one/blog/unprecedented-facebook-urls-dataset-now-available-research-through-social-science-one

[4] For details on the dataset, see https://solomonmg.github.io/pdf/Facebook_DP_URLs_Dataset.pdf.

How is the need for, and applicability of, DP and related techniques impacted by developments in the area of secure hardware (e.g., more ubiquitous and powerful TEEs) that are increasingly available both in the cloud and on end-devices. Furthermore, progress in FHE and other means of operating on encrypted data has some implications for DP and its relatives. While neither FHE nor secure hardware obviate the need for techniques like DP, how can the latter co-exist with them?

## 3.7    User and Social Issues in Security and Privacy

The field of User and Social Issues in Security and Privacy research is concerned with the study of how people, governments, and institutions interact with security and privacy technology, as well as explore how these technologies can be designed to meet the needs and expectations of users. This field covers a wide range of topics, including human-computer interaction, usability, trust, and risk perception. Researchers in this field strive to understand the social, cultural, and psychological factors that influence users' attitudes and behaviors towards security and privacy, as well as the impact of security and privacy technologies on society. They also explore ways to improve the usability and effectiveness of security and privacy technologies, such as by designing interfaces that are intuitive and easy to use, and by developing educational materials that help users understand the risks and benefits of different security and privacy options. User and Social Issues in Security and Privacy research is crucial in ensuring that security and privacy technologies are accessible, effective, and widely adopted by users, and as such, it is an interdisciplinary field that draws on computer science, psychology, sociology, and other disciplines.

### 3.7.1    Human Centric Perspectives & Studies

**Area Description**

A human-centered perspective encourages researchers, designers, engineers, policymakers, and others to center the needs of people, including people that are using a specific system (a user-centered perspective) as well as the broader society that is affected by individual and collective use of technological systems. For computing systems to be safe, secure, and trustworthy, human-centric perspectives should inform research questions, methods, and communications of findings. Examples of problems and domains within secure and trustworthy computing that are inherently human centered include (but are not limited to) user privacy, cyber hygiene (designing education and systems to support users in being more resistant to hacking, phishing, etc.), organizational cybersecurity, explainable AI, information integrity/cognitive security, cybersecurity for at-risk groups. Human-centered approaches generate solutions that are more usable for individuals and lead to healthier outcomes for individuals, organizations, and society at large.

Further, one interpretation of human-centric perspectives / studies is to build human-empowering technologies that can provide technical mechanisms for users to resist or contradict the efforts of corporations or governments. This research effort would take a complementary approach to the "put everything in the cloud" trend and re-localize data. This would reframe and to some degree resolve many of the privacy challenges that the research community has struggled to address (e.g., new cloud side channels). Although historically users have struggled to understand and properly use existing security solutions, these were technologies that were not designed to be usable, and new technologies can benefit from the learnings of the last decade. Human-centered research requires interdisciplinary teams and receptive communities.

## Technical Efforts

**Human + AI systems**. One area of interest is collaboration between human and AI systems. The integration of human and AI systems can occur in many different spaces, each posing unique challenges. The first area for research should begin with the creation of human/AI teams, or human-in-the-loop systems, wherein humans and AI systems are working together to complete tasks and solve complex problems. However, AI systems are vulnerable in many ways to security threats and can be used by bad actors in ways that harm others. For example, AI systems can be used to generate harmful, misleading content; they can guide users to make maladaptive decisions; and they can be the target of attacks that seek to shape their outputs. Novel configurations of human-AI teams create opportunities for new threats for the humans within those teams and for society at large. Research should attempt to identify and understand these new threats and explore potential protective measures.

**Explainability.** As a second area of research, explainability is an important part of building trustworthy AI systems. However, quantifying explainability and what humans deem to be useful, rather than useless, information is an open problem. Particularly with the upcoming prevalence of human and AI teams, explainability will be key to ensuring these systems and teams function as intended. As a final area, building human-centric AI systems, as systems that are intended to behave and function more similarly to humans, is also an open area of research. These systems will also have new, unique challenges and threats, and bad actors could have significant effects on the trustworthiness of such systems, as well as the trust the general public places in these systems. In systems that are designed to behave as humans do, perhaps the most concerning threat to society will be the inability to distinguish between interacting with such systems or with humans.

**Community Engaged Work**. It is important to understand the security and privacy issues and mitigation strategies of a diverse set of communities including at-risk populations (i.e., journalists, researchers, dissidents, elderly, homeless, etc.). This helps guide the creation of threat models that are aligned with specific communities' issues and identify defenses that are feasible for these communities to deploy which focus on mitigating the harms members of the community are most concerned about. These studies should be done for new technologies and security mechanisms. This will provide information used to determine the need to create new security mechanisms, and if existing ones are harmful to certain communities.

Community-engaged research can be resource-intensive and has a cost to the communities that are studied. It will be valuable to connect lessons learned across different populations. What characteristics of one population (e.g., journalists) can be useful in another population (e.g., gig workers)? Certainly, many communities have important differences that make one-size-fits-all approaches doomed to fail; however, efforts to identify important elements in new communities that have been observed in other previously studied communities would help expand the concept of the "default persona" to more meaningfully and effectively address the needs of all. As such, comparison studies across populations (given multiple different axes) should be facilitated.

**Ethical challenges, strengthening defenses against human-centered vulnerabilities**. It is important to tackle ethical challenges in human-centric security and AI. This also includes responsive design thinking, since humans are central to the design process. Cybersecurity is a dynamic area and various types of attacks can be introduced with advancement in technology. There will be challenges associated with building more robust defenses against human-centric vulnerabilities. For example, for new computing interfaces such as VR/AR, it is important to understand human mental model with these new interfaces, and the security/privacy implications on those who use them.

**Educating users**. It is important to identify the key factors to educate users about security problems and sound defense practices. User-centric studies will provide guidance and principles for cybersecurity education. With different groups of users, categorized by age or security knowledge, the techniques or resources for security-related solutions need to be customized or take into consideration different education strategies. Cybersecurity education must be at the core of every societal entity.

On the other hand, user-centric studies will identify the capacity limits of users, i.e., identifying the capability limit of user education. This can guide the design of the security systems to resolve the security aspects that have high impacts on users and complement existing defenses.

It is important to be cognizant of age and literacy proficiency. Efforts should address the needs of the young (elementary and middle school), consider workplaces, and be aware of cultural and individual differences in attitudes. Areas to consider include decreased user awareness of the risks associated with sharing content online and using mobile devices in public spaces (malicious networks, shoulder surfing, etc.).

**Technical and Societal Impacts**

Successful human-centric security research can have a direct and long-lasting impact on the technical community. Technical solutions driven by a strong human component will lead to the design of safer, more secure, and more effective software and products, whether these are security (or privacy) specific or of general use.

Success in this area could also have significant societal impact. Assisting human-decision making in privacy and security, could create a better "human firewall" (more aware and better equipped users of digital technologies). It could guide industry, policymakers, and other stakeholders in allocating resources to different security and privacy efforts. It could produce better AI tools (and, more broadly, better technology) when machines can learn from humans (biological computation). It could lead to more responsible and ethical design, more responsible coding, and - more broadly - more responsible and encompassing thinking about technology, privacy, and security that interacts with society and affects societal outcomes. Ultimately, the role of human-centric studies in technical security and privacy research is unavoidable if the goal is to practically and meaningfully improve individual and societal welfare: if technical solutions do not account for the human factor, they fail.

**Cross-Discipline and Outreach Opportunities**

This area is intensely interdisciplinary. It connects well with many disciplines such as psychology, sociology, political science, economics, education, law and policy, criminology, medicine, public health, philosophy, human-computer interaction. By bringing together experts from various fields, there is the possibility of new insights and perspectives that would not be possible otherwise. For the success of scientific progress in this area, more funds should be made available to researchers from different disciplines. Schemes should exist to encourage multi-disciplinary collaborations that deeply integrate expertise, to facilitate discussion across disciplinary boundaries, and to build communities. Investment in creating cross-disciplinary communities centered on important security and privacy issues is critical. For example, sponsoring workshops, and travel to attend conferences in other areas. Cross agency calls for proposals (E.g., NIH/NSF) focused on multidisciplinary approaches to specific security and privacy problems will be useful for the development. It is also important for Principal Investigators to bring in researchers from multiple disciplines. In addition, for intensely interdisciplinary research, it is critical to recognize different cultures and give credit for interdisciplinary work. For example, there are different publication venues and styles. It is important for the research community to recognize and value interdisciplinary work, and to develop new ways of evaluating the contributions of researchers from different disciplines. This could include creating new metrics or evaluation criteria that consider the unique challenges and opportunities of interdisciplinary research. Traversing boundaries of different fields and overcoming the challenges of jargon and insider knowledge from different disciplines will be important for the success of this research area.

### 3.7.2 Economics of Security and Privacy

**Area Description**

The breadth of the topic area expands to economic, legal, and societal issues in addition to technical issues in the security and privacy of cyberspace. From a science perspective, the data market, and secure ways of sharing scientific data can directly speed up the cycle of innovation and discovery. From the societal perspective, the importance, and the potential impact of research on this topic area will continue to evolve. For example, one research question is to encourage research into user-friendly security and privacy compliance tools, which will help promote better best practices in businesses and organizations dealing with user data.

The importance of the topic area includes the incentives from businesses and organizations and individuals to invest in cyber economics/insurance/contract for maintaining security and privacy. This is especially challenging when designing incentives to encourage every party to participate rather than only relying on monetary-incentives for creating a small domain-specific, small-scale data market.

Because economic trade-offs and incentives (broadly defined to include both extrinsic and intrinsic incentives, from financial benefits to psychological costs) are pervasive and influence decision making of both individuals and organizations, the consideration of economic factors is a cross-cutting area of investigation that intersects with many other technical areas of research. Examples of areas of research at the overlaps of economics, security, and privacy, include the following: the estimation of cyber security risks and the economic properties of cyber insurance, and the latter ability to help organizations and society allocate security investments more efficiently; the analysis of trade-offs associated with the deployment of security technologies; the empirical analysis of the impact of privacy and security regulations; the investigation of the way privacy preserving analytics (such as differential privacy, federated learning, and so forth) allocate costs and benefits from data usage to different stakeholders (such as data subjects and data holders; for example, privacy preserving analytics can facilitate data sharing for social good; on the other hand, by virtue of obfuscation they may reduce data accuracy, and thus its utility and economic value); the analysis of data markets, data dividends, and other data-propertization schemes – including the assessment of their effectiveness in helping users control and benefit from their personal information; the economic and environmental impact of encryption.

It is important to note that recently released National Cybersecurity Strategy[5] articulates the Federal Government's policy position of shifting liability for insecure software products and services to the product and service providers. Effectively creating this shift necessitates greater understanding of the various aspects of cyber economics, including the role of various incentives and disincentives in effecting better security outcomes.

**Technical Efforts**

---

[5] https://www.whitehouse.gov/wp-content/uploads/2023/03/National-Cybersecurity-Strategy-2023.pdf

**Personal data markets.** The idea of personal data markets is a potentially huge benefit to society at large. To be sure, there is already a thriving implicit data market at play, as well as a thriving B2B data market in which companies trade user data. Personal data markets embody the idea that personal data created (or co-created) by users belongs to them (at least in part), and they should be able to drive its utilization through informed consent and through other incentive mechanisms. Data sharing can potentially be enabled by privacy-preserving technologies such as secure multiparty computation, homomorphic encryption, and differential privacy. This area is rife with exciting questions such as: can design mechanisms to incentivize data sharing be created? How should data be priced? How can truthful reporting of data be incentivized? How can privacy risks be handled when a user's data is correlated to that of others? How can commercial entities buying into data markets be incentivized (when there is an apparent disincentive for them to do so given the status quo)? Research into these questions will require robust interaction between experts in economics, game theory, security, and cryptography, and will require developing new models.

**Costs of security.** User-facing security techniques impose varying degrees of burden, sometimes with unclear benefits. It is important to evaluate and understand the costs of a given user-facing security (or privacy) technology (in terms of, e.g., time→money) and contrast with its benefits to conclude whether it does offer any net benefits. On a similar note, though more broadly speaking, it is also important to evaluate a security/privacy technique with the focus on power consumption. This can help shed light on the overall energy cost (and sustainability) of security.

**Security/Privacy Regulation**. Many of the cybersecurity and privacy technology investments seem to be done due to regulations. For example, new privacy regulations such as GDPR and CCPA have had a major impact because of the sizable penalties that can be imposed on entities that fail to comply. In addition, past game theoretic and economic modeling suggested that in different application domains incentive structures created by the regulations are crucial for incentivizing investments in cybersecurity and data privacy. Therefore, there is a need to build models and experiments considering different possible industry incentives, regulations in the context of current and future cybersecurity tools to evaluate what might produce the best results with respect to desired cybersecurity investment outcomes.

**Economics of privacy and security.** Different privacy enhancing technologies (PETs) such as differential privacy, MPC, HE can facilitate data sharing for social good. On the other hand, by virtue of obfuscation they reduce accuracy or trust in the results (e.g., it is hard to check the quality of the results). Given these concerns, economic incentives could be reconsidered to facilitate the deployment of these tools. For example, government organizations may incentivize the deployment by limiting certain deployment risks and subsidizing costs. Hence, understanding the economic implications of using PETs will be important for future deployment of these PETs.

There is a need to also develop quantitative tools that help industry make better informed decisions when it comes to procuring new security and privacy technologies. Ideally, commercial products would provide quantifiable guarantees or minimally representative estimates of their benefits with these estimates being verifiable in some manner. A general question in this area is what the types of guarantees are, or estimates that can realistically be provided, with the understanding that different technologies will lend themselves to different types of guarantees or estimates.

**Market incentives.** Sound economic incentives need to be based on sound metrics, processes that enable assured development, sensible and enforceable notions of liability, and mature cost risk analysis methods. Without a scientific framework, it is difficult to incentivize good cybersecurity practices and subsequently to make a convincing business case for enhanced cybersecurity mechanisms or processes. The projected benefits must be quantified to demonstrate that they outweigh the costs incurred by the implementation of improved cybersecurity measures. There are no sound metrics to indicate how secure a system is, so one cannot articulate how much more secure it would be with additional investment. There is no scientific basis for cost risk analysis, and business decisions are often based on anecdotes or un-quantified arguments of goodness. Currently, it is difficult to collect the large body of data needed to develop a good statistical understanding of cyberspace without compromising the privacy of individuals or the reputation of companies. The means to identify and re-align cyber economic incentives and to provide a science-based understanding of markets, decision making, and motivators must be investigated. Research is required to:

**Costs of incidents.** Research needs to determine a solid understanding of the economics cost of security incidents and the return on investment of security defenses. Without this understanding, it is challenging for companies and individuals to correctly invest in security. Studies to measure the outcomes of security defenses in terms of how much different mitigation approaches reduce the likelihood and cost of incidents are needed.

**Cyber-relevant reporting and collection.** Reporting and collection of accurate data relevant to cyber economic and insurance/risk decision-making remains a significant obstacle to research progress. Future research must evaluate the impacts of new laws and policies that mandate security incident reporting and document the various societal advantages resulting from making such data available to researchers. Related research must also innovate new methods of collecting such data accurately, and quantifying its reliability, such as in cases where costs and even causes of data breaches are unknown or uncertain to crime victims and law enforcement at the time of reporting.

**Economic-inspired cyber defense.** Every cybercrime is typically for profit. If a criminal's profit can be cut (e.g., increase their attack cost), this could be an effective cyber defense. It is important and interesting to study the economic model of adversaries, e.g., measuring and modeling adversarial relationships and their lifecycles. The community needs new research on developing new economic-inspired defenses against various cyber-attacks/crimes. Research on new game-theoretic models that can apply well into the cyber domains may be an area if need. It is also important to have new human-centered research to study the political/economic intersection of coordinated influence operations.

**Cyber-deception.** Cyber deceptive approaches to security are a rapidly growing area of research and practice, but their economic implications are not yet well understood. Deception technologies include honey potting, honey tokens (e.g., fake document generation), deceptive or fake services, honeywords, and fake software vulnerabilities (e.g., honey-patches). Such defenses often have economically inspired objectives, such as depleting attacker resources or elevating attacker risk. New research is needed to evaluate whether these techniques are effective, and to quantify the expected effectiveness in terms of loss, risk, and information gain (in the context of deception for reconnaissance and counter-reconnaissance). Related research should also investigate public good components of these defenses, such as the potential for deceptive defenses to benefit undefended assets and organizations through a general increase in attacker uncertainty.

**Cyber-warfare and international norms.** Also, although physical warfare and how it can be deterred is well understood, the same does not hold for cyberwarfare. This area needs more research so that the understanding of cyberdeterrence can be better understood (and agreed upon). Including how it can be implemented, and how researchers in different areas can collaborate on this important topic. Cyberwarfare is a strategic competition conducted between adversaries in cyberspace. It allows countries to conduct covert operations on a large scale, cheaply, and anonymously. These latter three attributes are particularly important to understand. On one hand, the attributes of cyberwarfare make it difficult for governments to deter it. On the other hand, governments no longer have a choice but to confront it, so interdisciplinary work that studies cyber ware has become important.

**Stakeholder incentives and costs.** Understanding the economics of stakeholders (e.g., attackers, defenders, users) is key to identifying potentially effective security mitigations especially for-profit motivated attackers. Without this understanding security mitigations often fail because they are expensive to deploy and cheap for attackers to overcome. However, with this understanding it is possible to craft effective strategies that are cheap for defenders to deploy and costly for attackers to evade.

### Technical and Societal Impacts

Studying the economic dimensions of security and privacy would enable the community to take a more holistic view of security and privacy, exploring solutions that combine technical advances with the introduction of new regulatory requirements. This would include developing models that help assess the impact of different options available to society. This includes looking at macro-economic impact (e.g., impact on different economic sectors, impact on employment, etc.) as well as impact on different segments of society, including vulnerable populations (e.g., economically disadvantaged groups). Examples of regulations that could be informed by economic modeling and analysis include requirements for software security and privacy guarantees, regulations on data markets, regulations about requirements to properly inform or even warn users about security and privacy risks they face when they interact with different products, services, and processes, including provisions against dark patterns.

If successful, the impacts in this area would be multi-faceted and across multiple disciplines. For example, research that allows companies and government organizations to determine what is going to be shared, and how, would be very useful. One impact in this area could be with regards to blockchains and smart contracts. The use of these technologies might be in a cyber insurance setting in organizations.

If cyber economics as a research area based on the topics listed above is successful, the costs of cyber-attacks would be better understood. As a result, there would be more informed reasoning with respect to manual and automated defenses. Furthermore, cyber insurance would be better (and more properly) priced, cybersecurity would also be correctly incentivized, and the overall security of the Internet would increase as a result.

**Cross-Discipline and Outreach Opportunities**

Cybersecurity insurance and economics is by nature a cross-disciplined field. It requires collaboration between technologists, economists, and social sciences. For example, technologists and psychologists can guide economists to better quantify risk and perceptions of risk respectively. Additionally, public outreach and education is required to help users understand security/privacy risks and consequences to make educated solutions. Collaboration with political scientists and economists will be valuable for building models of coordinated influence operations online.

Ultimately it is important to remember that the provision of cybersecurity is a challenge that runs through technology, to provide security and human decision making, to build, employ, and react to outputs from cyber security technology; accordingly, it is an area ripe for cross-disciplinary collaboration between computer scientists, engineers, and social scientists.

### 3.6.3 Misinformation & Information Manipulation

**Area Description**

Misinformation and information manipulation is an attractive venue for attackers and malicious entities; by attempting to change the belief state of the target entity (individual or public), adversaries can significantly impact the individual or public's decision-making process. Misinformation has many forms, such as video, audio, memes, etc. Social media is a big enabler of misinformation, and so is mass media. Depending on the type of manipulated information, the process can follow different procedures ranging from traditional spamming campaigns to large scale fake information propagation in social media.

Connection to SaTC: information manipulation can be viewed, in part, as attacks on decision making. From SaTC (security) perspective, it would be useful to think about information manipulation/information integrity research as research on cognitive security.

To achieve a vision of high-integrity information ecosystems, research should focus on the following goals[6]: (a) *Integrity Assessment:* Enhance approaches people use or would like to use to discern between high-integrity and low-integrity information, narratives, and characteristics of information ecosystems. (b) *Harm Mitigation:* Identify strategies that could be used to prevent or minimize avoidable harm caused by information manipulation and assess their effectiveness across populations, cultures, communities, and types of harm. (c) *Resilience:* Identify skills and strategies that make it easier for people to operate in the context of potentially questionable information and enhance communication strategies and tools that enable communities to progress toward and maintain appropriate levels of information integrity within open information ecosystems, even in the presence of uncertainty or active manipulation. (d) *High-Quality Evidence:* Collect rigorous empirical evidence to evaluate strategies and technologies intended to address information integrity challenges; clearly convey high-quality evidence to decision-makers to inform the development of relevant public policy; organization-level communication processes; and decisions about which technologies to adopt, enhance, or retire.

**Technical Efforts**

**Generative AI.** Generative AI as an attack surface and defense technology: Advancements in generative AI (e.g., ChatGPT, DALL-E 2) lower the cost and provide new tools for information manipulation and pollution. Generative AI may be used by adversaries to generate misinformation at scale, such as generating many similar articles designed to look like they come from different perspectives or making bots to create seemingly truthful information that is harder to detect. Hence, more research is needed to design better algorithms to detect information generated by these AI technologies and systematically measure them and their implications in real life.

Additionally, generative AI systems may become the targets of manipulators who seek to influence their outputs. More research is needed to understand the vulnerabilities of generative AI to intentional manipulation, as well as the potential for emergent, resonant effects of the human-AI relationship — i.e., how training data shapes output (and behaviors) which may shape future training data.

---

[6]  https://www.whitehouse.gov/wp-content/uploads/2022/12/Roadmap-Information-Integrity-RD-2022.pdf

These models may also change how users consume information in the next ten years. An example already seen includes chatbot services built on the large-language models (LLM) used as alternative search engines (e.g., the New Bing). An increased understanding of the potential impacts on the information foraging and consumption of everyday users is needed. For example, how do users perceive the answers from a service based on LLM? How do users put trust in different sources, such as the platform (e.g., OpenAI, Microsoft, universities, government agencies) and structured supportive evidence (e.g., links that show the sources of the information)? How do users perceive and interpret results of generative text models, and do they consider them as more or less authoritative than the traditional methods (e.g., Google search)? More research is required to study people's perceptions and trust, design technologies to facilitate people to make decisions, and investigate techniques to improve people's media literacy.

Some new AI technologies may help augment our capabilities of detecting misinformation and online harm on a large scale. The key challenge is in how to do prompt engineering or fine tun the AI models to capture the complicated concepts of manipulation and harm. The power and global aspect should be considered. Technical and human-centric solutions must be designed together.

**Centering provenance in technical and human solutions.** Understanding the origins of information and how it has been processed, modified, or otherwise altered into its current form can be a means of identifying information manipulation. Developing platforms that support *provenance chains* demonstrating the lineage of data, and providing means of querying this lineage, is a potential technical approach to this problem. Significant additional research would thus be required to balance trade-offs for privacy concerns as well as demonstrating the trustworthiness of the platforms and of the provenance itself and ensuring that they are usable and psychologically acceptable. Additional research should then also focus on how such techniques can be incorporated into, or designed to support, information literacy campaigns.

**Measurement techniques / data access.** Measuring the spread, and especially the impact of campaigns is a challenge and will become more challenging as campaigns continue to move away from a small number of popular platforms (e.g., Twitter and Facebook) to private groups, closed door forums, and "alt-tech" platforms. Effort should continue to understand the spread of specific campaigns, including different types of actors and motivations. Industry-academia collaborations should be encouraged to complete such measurements. Effort should also develop models and methods to measure the impact of these campaigns (and interventions to address these campaigns) on society — e.g., on political outcomes, public health, values, etc.

**Platform design: Understanding trajectory of moderation.** Platform moderation can be a way to reduce harm caused by mis/dis-information. Here efforts are needed to investigate how to design platform moderation techniques that are trustworthy and explainable, while also preserving the privacy of users. Understanding how to balance content moderation and censorship of information is essential. Further, works should explore how platforms evaluate and mitigate the potential harm to individuals, society, communities when moderation fails.

**Technology, Law, and Policy.** An area of research needed is understanding how to co-design legal infrastructure (criminal or civil laws) with platform design. As well as ways in which actors can be held accountable, and under what existing or future statutes would be relevant.

**Understanding unintended consequences of the work.** Researchers are working to address the spread of harmful misinformation within and harassment against certain populations, especially within marginalized and vulnerable populations, and to increase privacy. But these efforts can have unintended consequences. For example, increasing awareness of manipulation can lead to increased skepticism, which, if taken too far, can have detrimental effects on trust in information and motivation for democratic participation. It is also important to determine how populations could potentially be placed at risk through efforts to counter misinformation through developing platforms and policies that stifle expression or the ability for individuals to assure their privacy and well-being.

**Behavioral research e.g., on debiasing.** Extant research has considered a number of behavioral intervention strategies aimed at reducing bias in decision making and countering influence or external manipulation of individuals' choices. In the SaTC context, research on debiasing, nudging, asymmetric paternalism – as well as on behavioral interventions designed to counter so-called dark patterns – should complement research on misinformation and how misinformation is used to attack people's decision making (see, above, cognitive security).

**Relationship with other harm categories.** Effort should look for common trends between research on information-related harms with other topics such as hate and harassment. For example, silencing critics online via coordinated harassment campaigns may well be done to maliciously assert control over the online information ecosystem. Opportunities exist to explore how different harms combine in adversarial campaigns, and how mitigation approaches for information-related harms are consonant with mitigating other forms of harms.

**Challenges under the context of AR and VR environments.** It is anticipated that Augmented Reality (AR) and Virtual Reality (VR) environments with rich interaction interfaces are to become widely used and perhaps a significant means of communicating and consuming information. It is expected this richer interaction will offer increased risks, but also opportunities for detecting and countering misinformation/disinformation campaigns. Work in this area should draw from a wide cross-disciplinary spectrum (CS, cognitive psychology, UX design, etc.).

Generally, this space is ripe for the development of meaningful metrics that go beyond simple "eyeball exposure". For example, metrics for the "degree" of misinformation/disinformation inherent in a digital object and/or campaign (e.g., ranging from "objective truth" to "white lie" to "manufactured propagated with whole malicious intent"), as well as metrics for understanding the effectiveness of such manipulation, are particularly important in understanding the nature and scale of problem, as well as evaluating proposed mitigations. This area would have a significant intersection with law, social science, and marketing/advertising (potential starting point for measuring effectiveness).

The use of synthetic avatars driven by AI to enable large-scale, high-interaction, personalized messaging can be used for both misinformation/disinformation and its countering. An example of an open question is the degree to which there exists an "uncanny valley" for such avatars with respect to believability. Effective work in this area would draw from CS, cognitive psychology, and social science.

**Interdisciplinary insights for detection.** Currently, research of misinformation and information manipulation does not sufficiently leverage interdisciplinary insights. Social science and existing marketing research constitute a valuable venue for learning profiles of fabricated information and distinguishing real / fake information. These fields should be leveraged to improve the accuracy and speed to detect misinformation or fake content.

**Future challenges of deepfake detection.** Research efforts on detection of deepfake media (e.g., audios, videos, and images) exist, but these current methodologies are unable to adequately handle future advances in AI and media synthesis tools, particularly incorporating biometric identity traits, e.g., hand movements, head tilt, etc. Future research should focus on deepfake detection to meet such challenges. Digital identifiers demonstrate the possibility to differentiate between synthesized and real media. Research efforts should investigate developing theory and practice for inserting digital identifiers and reliably detecting them. In addition, research effort should focus on developing methods to include information provenance or misinformation source tracking, including origin detection. Sharing it with the user on social media platforms can provide facts about the media. Research efforts focused on integrating natural language processing (NLP) for computer generated text detection, context-aware information, and information provenance should be considered to counter effects of deepfake on misinformation and disinformation campaigns.

**Applications of honeypots/honey-tokens.** Honeypots/honey-tokens can be seen to perform defensive misinformation campaigns and should be further studied in this context. Additionally, such forms of defensive strategies could be leveraged to study information manipulation campaigns conducted by adversaries. For instance, honey-accounts on social media could be used as attractive targets for adversaries. This could in turn enable the collection of data related to disinformation campaigns.

**Network Analysis, Countering Misinformation, and information manipulation.** In order to counter misinformation and information manipulation campaigns it is important to study and understand how people trust information, how beliefs/opinions evolve in a person and in groups based on information, and how misinformation and disinformation spread/propagate in a network.

A better understanding of these aspects is critical to design and deploy counter measures. This presents opportunities (essential) to collaborate with social scientists, psychologists, and marketing/communications experts. As an extension, it is important to understand the effectiveness and limitations of "influencers".

Efforts are needed to study and understand what kind of technical measures might be effective in improving trust in legitimate information while reducing trust in misinformation. Similarly, efforts are needed to understand what interventions and at what stages of a misinformation campaign or belief evolution of groups are likely to be effective. How does individual "trust" in society and "group affinity" influence/intersect with the effectiveness of misinformation campaigns and countermeasures?

**Economics, motivations, and Incentives.** It is important to study and understand economics, motivations, incentives, and opportunities for detecting misinformation. Who should be responsible for detecting or helping prevent misinformation? Is it the content creators? Is it platforms that enable content sharing? Or is it the end consumers? And what combinations of approaches are likely to be effective.

**Technical and Societal Impacts**

Inability to ensure access to secure and trustworthy information presents various harms to individuals, communities, and society at large. For instance, coordinated disinformation campaigns undermine voters' abilities to access and verify candidate positions on issues that impact their voting choices. Information manipulation efforts impede individuals' abilities to access trustworthy information about health care decisions (e.g., vaccine efficacy and risks, nutrition guidance). Information attacks aimed at individuals and communities effectively exclude them from the public sphere and limit their ability to participate in democratic decision-making and to enjoy the benefits of affiliation that digital communities can provide (e.g., social support, belonging) and that help individuals lead physically, mentally, and socially healthy lives.

Success would move the information ecosystem toward a future where individuals and communities are able to access trustworthy information, reap social benefits from affiliation, participate in public conversations without fearing for their personal or community safety, encountering corrupt information, or being manipulated for anti-democratic ends.

Success will require a cohesive design that integrates sociological, psychological, geopolitical, and technical considerations. This will provide an example, or even better, a foundational approach for other SaTC/CISE areas to learn from. For example, the co-design of information provenance (which will entail applied cryptography, networking, NLP, ML, etc.) and user-facing tools can either learn from or benefit other areas where humans interact with computing systems, such as human-robot interaction. Furthermore, given society's growing reliance on AI/ML, assuring the veracity and sources of data will ensure the utility and safety of these models. Success in mitigating the threats of misinformation and information manipulation will also strengthen the research process itself across all disciplines by allowing easy and reliable access to information while simultaneously promoting open discussion about conflicting ideas or points of view.

Success depends on researchers' access to data about information, its origins, its spread, and its changes.[7] Their ability to access this data depends on the data custodians' willingness and ability to share data and the research communities' ability to address the public's privacy concerns. Data custodians can include social media platforms, messaging apps, media outlets, and even individual researchers that have collected or linked data from multiple sources. Data consortia that provide safe, secure, accessible mechanisms for sharing data with researchers and that provide researchers computational and methodological assistance accessing and analyzing data are central to the success of shared efforts in this space. Without access to data or the means to analyze it, there is an increased risk for drawing conclusions that do not generalize or are missing insights from qualified researchers who lack computational and data resources to do their best work. Shared data resources also ensure that goals such as tiered access and privacy protection can be managed more effectively, reduce the risk of data leakage, and limit the negative environmental impacts of (overly) distributed data.

More broadly, democracy and many societal functions depend on the decision-making process of citizens relying on information that is largely generated and exchanged on digital platforms and in a digital form. Lack of adequate measures to detect misinformation and information manipulation campaigns will significantly harm democracy and society.

Developing technical solutions for information provenance, combined with security education, would enable the public to make more informed decisions on what information can be considered as reliable. In turn, this could have a significant impact on mitigating the negative effects of information manipulation on society.

**Cross-Discipline and Outreach Opportunities**

Research on misinformation requires the input and expertise of social scientists. While business school and marketing professionals may have a good understanding of "misinformation," psychologists can help us understand how trust and belief in information evolve. In addition, legislation has yet to catch up with the problem of misinformation and there is a need for regulations from the top. It is crucial to collaborate with legal experts to understand the boundaries of freedom of speech and political speech. Further, educating people to distinguish between misinformation and real news is essential. This may present opportunities for cross-discipline collaboration between computer science and journalism. Overall, it is essential to work together across disciplines to address the complex issue of misinformation.

---

[7]  https://www.brookings.edu/research/how-to-fix-social-media-start-with-independent-research/

Delving deeper, computer science, behavioral and social sciences, economics, epidemiology, linguistics, political science, psychology, sociology, library and information sciences, and other relevant disciplines have developed methods, models, and theories that offer explanations of different aspects of information ecosystems. A challenge for advancing the science of information integrity is to connect relevant methods, models, and theories to capture interdependencies within information ecosystems and explain new conditions as they arise. Transdisciplinary theory is needed to advance scientific understanding and formulate research questions that will lead to effective harm mitigation, evidence-based strategies, and resilient technologies and populations. Additionally, at its essence, the study of misinformation and information manipulation combines both technical considerations and questions about human/social behavior.

The technical questions addressed above largely revolve around questions of detection and propagation. Human interaction with misinformation can be thought of as having four components: production, exposure, sharing, and belief.[8] All of course, are influenced in some way by technical considerations (e.g., GPT chat may make it easier to produce convincing misinformation; social media platforms make it easier to share and be exposed to misinformation) but are also a result of social and behavioral processes. For example, belief in misinformation may be more common when the political bias of a false article aligns with a user's own political preferences. It has also been shown that older people are more likely to share links to false news stories on Facebook than younger people.[9] Questions pertaining to the reasons people share and believe false information should be investigated. Research into these trends and behaviors would benefit from collaborative research across multiple fields of social sciences, including, but not limited to political science, economics, psychology, sociology, etc.

---

[8] For a nice visual of how these relate to each other, see Figure 1 in Van Bavel, Jay J., Elizabeth A. Harris, Philip Pärnamets, Steve Rathje, Kimberly C. Doell, and Joshua A. Tucker. "Political psychology in the digital (mis) information age: A model of news belief and sharing." *Social Issues and Policy Review* 15, no. 1 (2021): 84-113.

[9] Guess, Andrew, Jonathan Nagler, and Joshua Tucker. "Less than you think: Prevalence and predictors of fake news dissemination on Facebook." *Science advances* 5, no. 1 (2019): eaau4586.

Interventions on both the technical and social science sides of these questions require rigorous evaluation to test their impact on desired outcomes. However, it is important to note that technical interventions may have social implications. For example, interventions designed to slow the spread of misinformation on platforms by throttling posts that violate certain guidelines may have the consequence of making people think that fact-checks are politically biased, and therefore making people less likely to seek out and/or trust fact-checks. Thus, a technical intervention that might even "succeed" in its technical goal could end up having a broader impact across the information ecosystem that undermines people's ability to identify the veracity of news in other areas. More generally, it is important to realize that interventions that "succeed" may have different impacts when they are applied "in the wild" in a way which interacts with technical aspects of the way that algorithms deliver news content to users of social media platforms. Both point to the importance of cross-disciplinary research and outreach beyond the purely technical questions of identifying misinformation and adjusting the algorithms that distribute it.

### 3.6.4 Social Implications of Privacy and Security Vulnerabilities

**Area Description**

The scope of social implications of privacy and security are wide and overlapping. It can be described in terms of goals: (1) the need to improve public trust in our technology through better communication, (2) the need to understand and quantify risks, to communicate what is needed, (3) understanding the roles of different stakeholders at reducing and disclosing these risks.

Without effective communication regarding security and privacy, it is difficult to build public trust. Instead of simply telling users how to think, trust is best increased by providing users with all relevant data and letting them make their own informed decisions. As such, there is a critical need to improve software processes and communication techniques to clearly communicate to users the risks and rewards of using any given technology. At the software level, this includes research that allows for the transparent operation of that software. Relatedly, there needs to be a marked improvement in the explainability of systems (AI in particular). Similarly, there needs to be a way for vulnerability disclosure to be incentivized and communicated to users. These properties all work towards creating accountable software. This research should then be paired with risk communication research to identify how to best communicate these topics and possible risks to users. Throughout this process, care needs to be taken to ensure that all user groups are included in this risk communication, including populations who may struggle to understand communication that works for most users—for example, youth, the elderly, those with visual impairments, etc.

There is a broader need to understand the social implications and consequences of new technology (such as AI, virtual/augmented reality, and autonomous vehicle technology) as they are developed, rather than leaving this research to be done reactively. Proactively identifying potential risks can allow the public to make informed decisions about technology adoption. It is also important to acknowledge that privacy and security vulnerabilities can disproportionately affect different people, and different people may face different risks. To mitigate harm, the populations most at-risk to privacy and security risks must also be identified.

Aligning incentives among stakeholders to reduce the impact of privacy and security vulnerabilities is key. Well-crafted and technically informed regulation and liability frameworks can incentivize the stakeholders best equipped to disclose and address vulnerabilities to take actions that benefit society. New technologies will often require technical auditing studies that inform new regulation and liability frameworks based on how the technology functions and the ability of stakeholders to address issues. This will engender trust that critical emerging technologies, such as autonomous vehicles, are reliable and safe.

**Technical Efforts**

The technical efforts associated with studying social implications of privacy and security vulnerability consist of improving transparency, accountability and explainability of new techniques, regulating the vulnerability disclosure/notification processing, and understanding privacy/mental health/censorship issues of VR/AR.

**New Methods to Improve Transparency, Accountability and Regulation for new techniques**. New technology will increase the need for transparency. From a security and privacy perspective, it would be important to understand the context behind information flow and decision making. Technological support for this is central to this, especially from a dynamic, systems-wide perspective. Issues to address here include trust (centralized vs decentralized), granularity of information, and formalization of context.

In addition, to mitigate societal harm, there must be accountability for security and privacy vulnerabilities. There are several key research issues associated with this including how to correctly quantify harm and how to ascribe responsibility across different components, systems, and interactions.

**Explanation of Technologies and Their Vulnerabilities/Limitations.** Society needs to understand the capabilities and vulnerabilities of modern technologies to properly use them. A major challenge is that these techniques are difficult to explain, and their vulnerabilities are difficult to assess and communicate. This can cause users to use technologies in ways that they should not, as well as to avoid technologies when they can benefit from them. Research is needed to develop explanation techniques for modern technologies such as Neural-Network based Artificial Intelligence models. More research is also needed to understand and quantify their known vulnerabilities and unknown risks. The research community must develop techniques to communicate these to end users and help them better use such technologies, such as privacy clinics that can help parents with social network privacy settings.

**Regulating vulnerability disclosure/notification procedure.** Given the critical nature of privacy and security vulnerabilities, it is important to regulate the vulnerability disclosure/notification procedure. First, software companies need to follow a standardized vulnerability disclosure procedure when new vulnerabilities are reported. This is imperative for accountability purposes. Companies particularly need to notify software users who may be potentially impacted by a security or privacy breach. Second, external parties, including individuals or companies, who identify vulnerabilities in each computing program (e.g., apps, a platform, an OS), need to report the vulnerabilities to the responsible vendor using a standardized procedure. This is highly crucial for the following reasons: (1) patching the discovered vulnerabilities, (2) preventing potential hoarding and intentional weaponization for specific (malicious or illegal) purposes, (3) preventing exploitation in the wild by hackers, and (4) commercialization for monetary profit.

Also, given the dynamic nature of privacy and security vulnerabilities, it is important to have an approach to disclosure/notification of vulnerability that captures its dynamicity. Naturally, as the new technologies emerge and are adopted more broadly, new vulnerabilities may be discovered that should be considered in the disclosure/notification process.

**Privacy/mental health/censorship issues of VR/AR.** Virtual reality (VR) and augmented reality (AR) technologies are rapidly developing, offering exciting possibilities for entertainment, education, and even medical applications. However, as with any new technology, there are concerns about its impact on individuals and society, particularly in terms of privacy, mental health, and censorship.

One of the primary concerns with VR/AR is the privacy envision. These technologies can collect vast amounts of personal data, such as eye movements, biometric data, and behavioral patterns, which can be used to track and target individuals. It is crucial to establish strong regulations and guidelines to protect individuals' privacy rights and ensure transparency in the collection and use of personal data. In addition, detection of privacy violations, privacy-preserving data analytics, and usable notifications of better privacy decisions in VR/AR is critical.

In another light, there are concerns that prolonged use of VR/AR could have negative effects on mental health, particularly in children. Research has shown that high screen time can lead to increased levels of anxiety, depression, and other mental health issues. The immersive nature of VR/AR experiences could exacerbate these effects, as users may become overly attached to virtual worlds and neglect their real-world relationships and responsibilities. Therefore, it is important to establish guidelines and tools to enable responsible use of VR/AR, particularly for younger users.

Further, censorship and control of VR/AR content is also an important issue. VR/AR allows creating new content and experiences that are directly shared with users, which could be used to spread propaganda or restrict access to certain types of information. Additionally, VR/AR experiences could be used to manipulate public opinion, particularly in the context of political campaigns. It is important to build tools that can detect potential censorship and misinformation in VR/AR, as well as educating users regarding censorship and misinformation.

AR/VR can also be used to inflict bodily harm, by making users move or make actions in unexpected ways. Security and privacy vulnerabilities can be exploited to cause this.

Next, consider fairness and equity. To understand the implications of privacy and security vulnerabilities properly, it is important that the studies be inclusive. Usability studies should include participants from different groups of population to lead to generalized and equitable results. Perhaps techniques from sociology and anthropology could be used to improve the process. In addition, potential implications of seemingly benign techniques should be considered. For example, when differential privacy is deployed, it could impact different groups in different ways. We should consider potential social implications of privacy and security techniques themselves when used in the real world.

**Technical and Societal Impacts**

Research in this space can improve transparency, accountability and regulation, and broadly impact society. From a technical perspective, natural language processing, machine learning, policy management and usable security/privacy advances would be beneficial.

Research in this space will help mitigate VR/AR privacy/mental health/censorship. Addressing the privacy/mental health/censorship concerns in VR/AR will advance techniques such as automatic policy enforcement, hardware-software co-design for security and privacy, privacy-preserving techniques, usable security, and privacy. Scientific progress in this domain will also benefit the broader community of psychology, sociology, law, and public policy. Failing to make progress may have a negative impact on the mental health and safety of, and transparency of human society.

**Cross-Discipline and Outreach Opportunities**

Work in this area of research will interact with researchers in many other disciplines. For understanding what new societal risks are, research may involve experts in other fields, such as medical doctors (or physical effects of VR), or psychologists (to understand the risks of greater use of technology or use by younger or more vulnerable parts of society), and also domain experts from where technology is used (such as vehicle experts for greater use of technology in and between vehicles). Understanding and communicating societal risk is a important requirement; such communication will benefit from involvement of researchers in education, psychology, and communication. Associated education research questions will be raised and solved, which requires the interdisciplinary research of experts from education and cybersecurity. Finally, to explore roles for regulation of societal risk, government policy experts and economists will be helpful.

### 3.6.5 Human-centered Privacy and Security

**Area Description**

The goal of human-centered privacy and security is to understand privacy and security needs, behaviors, and decision making of diverse groups, to empower people to manage their security and privacy, acknowledging that most/all aspects of our daily lives are intertwined with technology that tracks or records. This area of research intersects with psychology, sociology, law, policy, and many other fields.

Usability is primarily concerned with how easy a technology is to use. Some human-centered security and privacy tasks may be better addressed with new computing solutions rather than better interfaces. For example, some solutions may remove end users from the process of securing systems entirely. However, this does not mean that no users are involved. In this case, developers, IT managers, etc., will still interact with these systems. All computing solutions however must be informed by user needs which can be identified using human-centered methods.

There will be new interfaces that are yet to be imagined that will need to be usable. As novel technologies are developed, researchers will need to adapt to assess the usability of those interfaces. Users of technologies include consumers, developers, policy makers, regulators, etc.

**Technical Efforts**

**Privacy.** One answer to the persistent and increasing complexity of privacy for users is to move privacy management "behind the scenes," in the form of automated assistance that users delegate the implementation of their privacy preferences to. Such delegation will require standard methods for computing applications (e.g., apps, web services, hardware devices, etc.) to expose privacy functionality to privacy assistive technologies. Systematizing these standards so they are extensible and (to the extent possible) future-proof is a technical challenge.

**Understanding privacy/security behavior and decision making.** To support users with better privacy and security interfaces requires a deeper understanding of decision-making processes, behavior patterns, and privacy/security needs of individuals and high-risk populations, and whether and how they translate into protection motivation and actual protective actions. There is increasing evidence for intention-behavior gaps in security and privacy, and it will be a key challenge to understand and overcome those gaps to effectively support users in their privacy and security needs.

**Inferring user privacy needs.** Similarly, assisting users will require the ability to anticipate their privacy preferences: accurately, respectfully, and with the ability to reverse course when mistakes are made. A growing challenge will be to make inferences about privacy preferences cross-domain (e.g., across specific applications or computing modalities). A second challenge will be to anticipate when automated intervention is helpful or wanted, being mindful of the dangers of overreach, which may directly cut against the goals of automated privacy assistance.

Along the same lines, a challenge that needs to be addressed is the automatic extraction of "secure default" for privacy preferences in different computing applications (e.g., smart apps, web apps, IoT specific companion apps, etc.). The automatically inferred preferences need to be automatically synthesized and adopted in the apps. A key challenge is to transition from just literacy and individual support to establishing baseline protections for everyone.

**Automatic evaluation for privacy compliance.** Another challenge that needs to be tackled is evaluating computing apps for compliance with privacy regulations, i.e., developing automatic tools that make sure that developers respect privacy rules. This is challenging to address particularly considering (1) projected customized regulations where different countries and corresponding jurisdictions may require/impose varying privacy regulations, and (2) the closed and highly obfuscated nature of computing apps.

**Education.** It is important to develop training functionality that adapts to the knowledge and context of everyone. Security and privacy are increasingly complex. It is important that people receive training from the earliest age. Training here ranges from awareness of privacy and security threats, literacy to spot and deal with them, and skills to attain privacy and security protections within the context of daily tasks and requirements. There is a need to develop personalized, context-sensitive training functionality that adapts to what each individual user already knows, what they can understand, the specific activities they are likely to engage in, etc. Guidance to people recovering from a data breach, identity theft, etc. could also be improved.

**Awareness.** Several questions arise around raising user awareness. Some important questions that need to be addressed through fundamental research include: How to develop functionality that ensures people are aware of all the relevant security and privacy risks they are exposed to in different contexts? How to ensure that this type of functionality is not overwhelming and that users are only informed about those threats that need to be known, and are not already aware of? What is the best way to effectively convey this information?

**Modeling Expectations.** Security and privacy are in great part about expectations. Not every user has the same expectations or needs when it comes to their security and privacy, including tolerance to different types of risks. How can these models be used to drive interactions with users.

**Usable and useful privacy/security interfaces.** Privacy policies have failed as consumer information tools, similarly security warnings are often ignored. A key challenge will be to better integrate privacy and security information and controls into system's user experiences to provide information and controls that are relevant, understandable, actionable, and aligned with the needs of all/specific user groups. Mobile permissions are a good example here – how should usable and useful privacy/security interfaces like this look like for behavioral advertising of the future, for chatbots, for fully connected smart homes, and many other future technologies?

**Adaptive interactions.** Security and privacy interfaces should adapt to their individual users, considering detailed user models. This includes the types of risks they warn their users about, how they describe these risks to different users, how they motivate users to pay attention, how they remind users about different items they need to be aware of, and more. This also includes keeping track of the different systems an individual user might be interacting with and what information about that user might have been disclosed to different entities or systems.

**Question Answering Functionality.** When faced with situations with which they are not familiar users will increasingly be looking for functionality that can answer their questions and address their concerns. This type of functionality will increasingly be expected to take the form of natural language dialogues and incorporate functionality that can personalize its answers based on models of the user - what the user knows and understands, what risks the user is likely to face in the immediate future, etc.

**Deceptive design in privacy and security.** There is a need to detect deceptive designs (sometimes called dark patterns) in privacy and security interfaces, e.g., interfaces that trick users into sharing more information than they want, privacy/security settings that mislead or confuse people etc.; as well as study how, such deceptive designs affect people's behavior regarding security and privacy. An aspect of this is also to assess the role and responsibility of platforms and third parties (e.g., consent management vendors).

**Reactive interactions and recovery.** Similarly, it will be important to study and develop functionality that can help users recover from situations where their security or privacy is at risk or has been compromised, such as data breaches. This will gain importance as AI and other technologies pervade all parts of people's lives making a "just do not share the data approach" impossible. This area includes challenges around detecting when breaches or security incidents occur, effectively informing affected individuals, advances and guidance on effective remediation approaches and techniques beyond today's "monitor your accounts" approaches, questions of responsibility and penalties after breaches occur – shifting responsibility to companies instead of individuals; as well as transformative advances, e.g., rethinking national ID and identification systems (as well as their privacy implications) to curb the potential for identity theft and the impact of phishing attacks.

**Human interfaces in emerging computer systems.** All the above efforts will require taking into account contexts where users are increasingly likely to interact with AI/ML-driven systems and functionality, IoT systems as well as avatars and other (semi-)autonomous artifacts that are representing them in digital/virtual environments

## Technical and Societal Impacts

All systems involve people. This field of usable security and privacy is about making technology usable and useful to people. This field has advanced significantly in the past decade. For instance, in the context of password management, one significant success was the wide adoption and diffusion of password managers, which reduce users' cognitive burden in password management. In other areas, there have been struggles with email encryption being leapfrogged by broadly available end-to-end encryption in mobile messaging. Moving forward, there is a need for more innovative applications that reduce the amount of time users must actively spend on managing privacy and security, for example in the way password managers facilitate truly usable passwords.

In addition, the research community shall transition from user-centric to human-centric to achieve broader societal impacts. User-centric view positions people as part of the technological platform (and thus their data is also part of products). The human-centric view calls for taking value-sensitive design and inclusive design approaches, by building our fundamental values such as privacy, trust, dignity, respect, safety etc. into our systems. Without human-centric approaches, technologies could reinforce inequality, limit accountability, and infringe on the privacy of individuals. Unreliable and irreproducible AI could misguide global economic policies; ML models trained from biased data could amplify discrimination in the criminal justice system; algorithmic hiring practices could silently reinforce biases and potentially violate the law; privacy incidents and cybersecurity breaches not only erode the trust of consumers, but also expose organizations to legal and financial ramifications.

## Cross-Discipline and Outreach Opportunities

Human-centered security and privacy intersects with all SaTC priority areas as many if not all privacy and security topics involve or impact humans. This includes providing inputs to other areas in terms of what are human privacy and security needs, capabilities, and limitations; and drawing from other areas to study and advance human-related questions in those areas, for instance supporting developers in producing secure and privacy-respecting code and systems.

Like security and privacy topics intersecting with other disciplines that have security and privacy needs due to their processing of sensitive information, human-centered privacy and security is of relevance whenever humans are subject or affected by information practices. Obvious examples where humans are affected include medicine and healthcare, IoT, AI systems, etc.

In terms of interchange with other research areas, human-centered privacy and security requires interchange with, input from, and dissemination to legal scholars (in particular on privacy rights); psychology, sociology, domain experts in any affected field (e.g., physicians in medicine) and communications in terms of understanding, reasoning about, and accounting for human behavior and decision making; research method experts in terms of applying, adapting and advancing qualitative, quantitative and mixed-methods research approaches and methodologies; etc.

As most people interact with systems, the usability and general good design of systems will broadly impact people's everyday experience. They will be impacted by the privacy and security of the systems that they use.

## 3.8    Other Topics

There are numerous areas that are within the scope of SaTC yet are outside technical subareas listed previously. Below are several issues that are cross-cutting and intersecting with the existing areas outlined above.  As such the following areas (e.g., education and ethics) may be viewed as enhancing other research. Indeed, much of the gains in other areas of technology cannot be realized without progress in these fields.

### 3.8.1    Cybersecurity Education

**Area Description**

This area focuses on innovation in cybersecurity education and capacity building to meet the cybersecurity workforce needs of the nation. Without a large and adequately trained workforce society is at risk—as our lives and economy increasingly depend on technology and interconnected systems. Research must explore how to train the next generation of technologies on anticipating, detecting, mitigating, and recovery of future threats.

In a broader light, cybersecurity and privacy threats permeate our everyday life. Children access technology from the earliest age - from smart TVs to smartphones laptops and other everyday IoT devices. As a result, it is imperative that they be provided with some basic training from that earliest age too. More generally, it is estimated that 90% of security breaches have to do with human actions - whether it is someone failing to make the right security decision or someone acting maliciously. In short, there is a need to ensure that everyone has a baseline education in both security and privacy - starting with the basics at a very young age and moving towards somewhat more advanced concepts as people grow older and engage in more specialized activities (e.g., their responsibilities at work). And obviously it is important to also improve training offered to developers and even to educators.

Some examples of challenging research questions that need to be addressed include evaluating the effectiveness of security and privacy training. Prior research teaches us that it is important to motivate people to take the training seriously (e.g., protection motivation theory) by showing them that they are at risk, getting to appreciate how serious these risks are, and getting them to also understand that they lack the necessary knowledge to protect themselves. Cybersecurity training has undergone a major transformation over the past 10-15 years with the emergence of more effective training based on learning science principles. At the same time, a good understanding of how effective cybersecurity training is lacking - it is one thing to measure someone's score on a training game, yet an entirely different thing to see if people were able to apply their training when faced with the actual threat during their daily lives. While some of this data exists, there is a need to develop a more systematic understanding of what effective training is and how much can be accomplished with training - there is only so much training people can tolerate. To make up for these limitations, it will be important to develop just-in-time/in-context training, training where people are told about risks and ways of mitigating risks as they are about to face them.

It is also important to better understand why there is such a huge shortage of cybersecurity professionals and how the field might be able to motivate more people to consider careers in this area.

When it comes to professional education - training the developers, there is a need to develop content that can be embedded in programming and development courses - rather than offering separate courses on secure programming or programming for privacy. Today way too many developers think they know how to write programs, yet they have little or no understanding of secure programming principles.

**Technical Efforts**

**End user cybersecurity/privacy awareness / security hygiene***. There is not enough cybersecurity literacy among end users. There is a dire need to have targeted training programs to educate and make aware end users of cybersecurity threats. For example, basic media literacy programs should be created to train users who are not very familiar with technology. Similarly, a different pedagogical approach should be used to train HS students, who are daily users of technology but are not aware of the cybersecurity risks. SaTC should focus on developing targeted, personalized cybersecurity education materials. Gamification of training materials could also help quicker adoption as well engagement from users from different groups, e.g., High School students. Research needs to develop more attack-based training that shows the consequences of poor cybersecurity hygiene, to motivate users.

**New educational techniques for cybersecurity and privacy.** New techniques for teaching cybersecurity materials are needed. Focus group believes researchers investigate just-in-time or on demand cybersecurity education. Research needs to develop more hands-on training materials, materials that fit new emerging technologies, such as AR/VR, 3D videos, and voice-based interfaces.

Knowledge gaps in cyber skills and awareness among industry and government professionals currently pose severe threats to national and world security. This is in part because cyber training and those levels may be reduced to a "check-the-boxes" obligation that is met with pedagogical methods of limited effectiveness (e.g., non-interactive videos or quizzes with contextually irrelevant questions), or whose effectiveness has not been measured. To address this gap, research is needed that innovates more effective training strategies that leverage higher interactivity, a greater variety of media (e.g., VR/AR, new UIs, just-in-time / on-demand), and content that is customized to individual professions, roles, and employees. This should be supplemented with research that rigorously evaluates the effectiveness of various training strategies for thwarting realistic cyberattacks and mitigating organizational risk.

**Study and create motivations for stakeholders**. One key challenge in promoting cybersecurity and privacy education is that people often lack motivation for learning about these problems. Hence, more research needs to be conducted to help different stakeholders understand the severity of the problems and help create motivation for them. For example, for engineers, teaching them how to do privacy attacks will help them better understand the system vulnerabilities and the consequences of attacks and get more motivated to learn how to mitigate the risks. For lay users, people may get more motivated to learn cybersecurity skills when they feel their cybersecurity and privacy is in danger. Research is needed to crowdsource motivations for cybersecurity and privacy self-education in the wild. Also, research should explore how to establish motivations when training people (especially at a younger age) such as using gamification.

**Measurement**. Given the gaps in our understanding of a lot of critical issues in cybersecurity education, more efforts implementing scientific measurement research is needed to gain a more holistic and realistic understanding of the landscape. For evaluating existing educational techniques and designing new techniques, research needs to develop novel methods to measure the efficacy of the training/teaching methods in the wild and over time. For informing the design of educational techniques for training/teaching developers/programming's and making sure technical research in security and privacy provides the needed tools, research should investigate how so systematically gather and analyze the real-world problems in software engineering about security and privacy and evaluate the gap between the problems in the practice and the problems in the research.

**Curriculum Innovation and Capacity building.** Research efforts need to investigate integrating emerging technologies and topics into cybersecurity education and build capacity around them. Some topics/technologies of interest include AI/ML, post-quantum, misinformation/disinformation, fake information, social media and privacy, digital twins etc. What if a framework or model keeps pace with ever evolving and constantly changing technologies and toolsets (continuing education model?)

**Learning and Education Innovations.** Research efforts need to effectively leverage emerging technologies in teaching/learning methods. Some emerging technologies of interest include AI assisted education and learning; designing and evaluating online cyber education and learning at a larger scale; evaluating effective cyber education and learning strategies (e.g., gamified learning, virtual reality augmented learning etc.)

**Repository and Education Facilities.** Research efforts need to be aimed at creating a curated repository of curriculum developed for others to build on/adapt and use. Research needs to create common facilities and datasets, especially in resource intensive areas (e.g., quantum, critical infrastructure).

**Industry-Academia Gap.** While approaches to innovate education are important, it is important to meet the needs of industry/government and reduce the gap between academia and industry needs. Is the "certification" model working? Are there better alternatives? Is there a need or opportunity for a "bar" or "PE" exam for cybersecurity professionals?

**Cybersecurity FOR ALL.** Cybersecurity awareness and knowledge is no longer just for computer engineers; basic cybersecurity and privacy awareness in all of engineering and in all disciplines is needed; including basic cybersecurity and privacy awareness for all age groups K - "Grey" (credit to Gula Foundation for the term); Designing and evaluating cybersecurity curriculum for all is priority —- K-Grey, general population, outside of computer science and engineering, and beyond engineering. Ways to inculcate "security or adversarial" ("security mindset") thinking into all disciplines and general population.

**Inclusivity.** Questions exist around rethinking common terminologies in Cyber when engaging a broader audience. Research needs to investigate how cyber technologies can be made more accessible to people with no technological background or with special needs?

**Bringing other disciplines into Cybersecurity.** There is a need to include other topics that have a bearing in cybersecurity implementation and practice and should be included in cyber security education. Some topics include psychology, sociology, policy, law, organizational and behavioral concepts, block chain, economics, insurance, etc.

## Technical and Societal Impacts

There is a visible shortage in the cybersecurity workforce. From industry, it is hard to find qualified cybersecurity professionals. Companies invest in on-board training but cannot be compared with formal training from higher education institutions. Without increased capacity, models for engaging broader audience, and security thinking in general education, our society will continue suffer cybersecurity challenges.

**Cross-Discipline and Outreach Opportunities**

A challenge is being or keeping ahead of the technical curve. Outreach opportunities include bringing other disciplines into cybersecurity education and developing non-STEM cybersecurity curriculum. The community must develop innovative ways for the academic community to work closely with industry. Leveraging the best talent and foundation of academic in the new workforce for real life challenges would lead to bridging the gap in skills in cybersecurity. Multiple disciplines can be helpful in developing cybersecurity education. Obvious disciplines include education, social science (e.g., psychology, sociology).

Also note that specific sub-disciplines of computer science such as introductory course professors and professors of computer security can help introduce cybersecurity as a topic that may interest some students as awareness of cybersecurity as a profession still seems needed.

This area is very well suited for broad outreach. It is hard to think of a population that would not benefit from basic cybersecurity education. For example, elementary, middle, high school students, undergrad and graduate students, as well as working professionals, management, C-suite executives, regulators, need cybersecurity education. Each of these populations will need different types of training. For example, elementary school students will need to be taught online safety and technology awareness, high school students will need to develop awareness of cybersecurity as a profession, foundational cyber safety skills, getting diverse groups (women, minority groups) more interested in cybersecurity professions, industry/government professionals need to learn cybersecurity best practices, management, software developers need to learn basic details about computing environments, computer science students need to learn about integration of cybersecurity into all aspects of CS, older adults need to learn about phishing, fraud, scams, antivirus, regular software updating, ads, safe online banking, and faculty need to motivate and train non-security CS faculty to incorporate critical security concepts into their courses.

Another opportunity is in novel tech transfer programs with for example tech colleges that do two-year degrees specifically for training cybersecurity technicians. Support faculty and/or senior graduate students at research universities to do rotations at these schools to help update curriculum, for example.

### 3.8.2 Ethics

### Area Description

The field of ethics in security research can be categorized into three areas: ethics around the process of carrying out research (for example, is data anonymized or does it have private information in it, and if so, is it kept private), ethics around research topics and where research is applied (for example, will the results of research reduce inequity or increase risk to humans), and new technologies that can improve ethics in research; and our goals to support education about ethics.

The *ethics of how technology is applied* (e.g., chatbots interacting with humans, IoT devices taking observations) and the *ethics of producing technology* (e.g., dataset building, research methodologies) are distinct, but both are both important to consider and with scope. There are well-known tradeoffs in applications of security research, like the tension between technologies for surveillance (e.g., for preventing crime), and the right to privacy. These tradeoffs should be treated as ethical questions. Another factor to consider is equity in applications of security research–will these results be usable by everyone, or only certain segments of population. Finally, we recognize that sometimes it is necessary to relax ethics during research (in a controlled, limited way; for example, attacking anonymization to investigate its strength).

Another area that is in scope are *methods and technologies that help carry out more ethical research.* Some examples include the role of Institutional Research Boards (IRBs) to provide ethical oversight. Another example are technologies for anonymization (such as data scrubbing tools) and privacy preservation (such as differential privacy and secure multiparty computation). A third example is the development of common shared datasets that are anonymized but allow researchers to compare results.

A third avenue to consider is *education about ethics during research and use of research results.* Specifically, it is crucial to consider how the ethics in privacy and security should be integrated into educational programs on one side and more generally how to create awareness at large. The next generation to work as cybersecurity professionals must be properly educated in this topic to be well-prepared for developing, implementing, and deploying systems, software, and such in a socially responsible manner. More generally, concepts for creating awareness at large need to be developed. Furthermore, there needs to be an increased awareness of ethics amongst those carrying out research in security and privacy.

## Technical Efforts

**Exploring ethics in security and privacy.** Security and privacy intersect with ethics in a number of ways. One general direction is examining ethical ramifications of security measures and privacy-enhancing technologies. Efforts to increase security, for example, can limit autonomy of users. While privacy is generally perceived to be a positive quality, privacy enhancements (and research toward them) may require access to personal data that poses ethical concerns.

Another avenue of inquiry is in examining the roles of different organizations in promoting ethics in education and research. Relevant entities include NSF, as funding research and promoting consideration of ethics in the proposal review process; ABET, as the organization that accredits computer science and engineering programs at many universities; faculty committees, which create courses and curricula; and individual faculty, who make fine-grained decisions about course content. Exploring carefully what the ethical responsibilities for all people and organizations employed (directly or indirectly) is essential to the future of security and privacy research.

**Socially Responsible Development**. It is important for there to be a standardization of socially responsible guidelines. For instance, even if information is public, should it be easy to access it? (e.g., home address). This is an area to be addressed, as AI chat bots grow in prominence. There should be guardrails (as there are now, but solely because the chat bots have not trained on that data yet) that protect certain information so that a user cannot ask about personal information about someone else.

## Technical and Societal Impacts

The scientific success in ethics research, involving the interaction of ethics with privacy and security, will have crucial impact on three broad categories of audiences: technical science and engineering community at large, in addition to CS, the general public, and the industry in general.

Without ethics awareness, the research community may produce software systems and AI models that aggravate the ethics problems already inherent in our physical society. For example, differential privacy (DP) research may produce a DP guaranteed algorithm that could further worsen the biases induced due to data skewness.

Without ethics awareness, commercial products and services may further aggravate the ethical biases already existing in our social society. For example, people from minority groups may receive fewer effective recommendations. Without ethics awareness, the governments may not be able to promote equitable access to all government policies.

**Cross-Discipline and Outreach Opportunities**

There are many cross-discipline opportunities in security ethics. Depending on the domain, there may be opportunities in healthcare (for the ethics of sharing genetic information, which implicitly shares genetic information about people in the same family tree), or gender studies (for reducing bias in datasets to promote algorithmic fairness). Philosophical techniques may be relevant to help researchers learn how to develop and justify their own ethical frameworks and promote the transition of some of these discussions to public policy. Finally, there is a great opportunity for education–both for researchers and the public.

### 3.8.3   Offensive Security[10]

**Area Description**

Offensive security research refers to exploring novel offensive strategies, attack vectors and techniques in the following fields: networking, architecture, cyber-physical systems, software, and social engineering. Specific examples include software reverse engineering, software vulnerability proof-of-concept (POC) generation and stabilization, cyber-attack modeling and assessment, and social engineering and deception.

Novel exploitation techniques at all layers of our computer and communication systems and the users of such systems, including but not limited to, user, software, operating system, hypervisor, trusted execution environment, micro-architectural, circuit, and the underlying hardware layers.

Offensive software security refers to the art of turning a vulnerability into an exploit that comprises a system. This art, often kept in the realm of industry and government practitioners, should become a science. This area should explore attacks or techniques that teach the community something new about defenses (e.g., their gaps in threat models), or how the complex interplay of software and hardware in computing.

---

[10] It is important to acknowledge that offensive security research is a sensitive topic in the security research community. This section offers a perspective on the importance of offensive security research, comments on how to raise the community's awareness on the importance and significance of offensive security research, and how offensive security research efforts can be proposed and conducted in a responsible manner. In leu of a detailed technical discussion, this section considers that that area might look like and explores the reasoning and ethics of that field.

There are mainly three reasons why offensive security, when proposed and done right, is critical to the cybersecurity community and the general good of the public:

First, offensive research can contribute to enhancing and improving existing defense mechanisms, benefiting the community. A successful example is Google's kCTF program, which rewards whitehat hackers for submitting new exploits that demonstrate the ability to bypass the latest defense systems. Once the exploits are confirmed, the Google research team improves their defenses. The whitehats must then find new exploitation methods to bypass the updated defense. Over the years, the program has identified many new exploitation methods and greatly advanced defense systems.

Second, offensive research can also assist the community in assessing risk. The ability to find bugs has greatly advanced in recent years, resulting in an increasing number of reported bugs to software vendors. However, the resources available for vendors to remediate bugs are limited. Software developers need techniques to assess the risk of bugs. Offensive research can greatly expedite the remediation of bugs and their associated risks.

Third, offensive research can help software security researchers who focus on defenses to gain a better understanding of the attack surface. It is arguable that real, usable, and transitionable defense is only achievable by conducting real offense research. By identifying potential vulnerabilities and attack vectors, offensive research can provide insights that enable researchers to design more effective and comprehensive defense strategies. Understanding the attacker's perspective is crucial to building robust and resilient security systems. Therefore, offensive research can play a significant role in improving the overall security posture of software systems.

### Technical Efforts

This section explores the reasoning and ethical considerations of offensive research.

**Justifications of offensive research.** Offensive research is important for a number of reasons, and directly supporting research that has a more offensive spin rather than a defensive spin will continue to be more important in the future. For example, in mid-2005, many companies were against students acquiring offensive capabilities in class because the general established consensus was that these students would suddenly turn "blackhat" and then engage in illegal behavior. Today, however, teaching students to compromise systems in controlled environments (e.g., exploiting buffer overflows, misusing input for web applications, etc.) has become routine. Our group expects a similar change of mindset for public security research that is more concerned with attacks as its focus.

Offensive security research is helpful to understand what defenses work in practice, and what defenses are not effective. Furthermore, it helps us define the scope of the problem. For example, some attacks may be well-known, but become even a larger problem after some time as the Internet landscape evolves. For example, discrepancy attacks on the web where different web server implementations interact, and result in security problems has become a larger issue today just because there are many more CDNs on the Internet, and many companies are using the CDNs. Clearly, in order to construct better defenses, the community needs to know which attacks relevant, and which ones are work well in practice.

Offensive security is important from an educational point of view. There is a huge demand in industry and government organizations for employees who are well-aware of offensive techniques, and who have offensive skills. If research directly funds projects that focus on offensive security topics such as the scaling up of attacks, these projects will help recruit talent, and educate many students and equip them with the skills they really need outside of academia.

An example of demonstrating the importance of software exploitation techniques is Google's kCTF program. kCTF is an open platform developed by Google that mimics their internal Kubernetes setup, and the challenge is for participants to demonstrate that they can compromise the infrastructure and steal a flag and pays up to $133,337. Because the core security mechanism is containers, many successful exploits are done by exploiting Linux kernel vulnerabilities. Even more interesting, kCTF, unlike many other "bug bounty" programs, will pay for exploits for N-day vulnerabilities (i.e., those vulnerabilities that are already known to the community). The reasoning behind this is that Google wants to learn about exploits and, more specifically, exploit techniques that can bypass modern defenses. Google studies the submitted exploits and exploit techniques, and this helps them to develop better defenses: limiting the attack surface by removing access to unnecessary functionality and by developing better defenses that can prevent entire classes of exploitation technique. This demonstrates that industry sees the value in funding

research into novel exploitation techniques—precisely because they can learn from them and then the entire community benefits from the defenses that are created to prevent exploitation techniques.

**Ethical issues and concerns.** The ultimate goal for offensive security is to understand the attacks (before the attackers do) and build stronger defense. Without deep understanding of the attacks and the vulnerabilities, without seeing the consequences, the defense may be incomplete and less robust. However, it inevitably involves intentionally trying to exploit vulnerabilities in computer systems and software applications in order to identify weaknesses and improve security. It is a must to consider ethics, propose actionable plans to mitigate ethical concerns and minimize potential harms that offensive research may bring to real-world systems, and conduct offensive research in an ethical and responsible manner. Community practice such as informed consent and responsible disclosure must be in place, in addition to IRB approvals. Additionally, proposals that are mainly about offensive research should include a clear and actionable plan of risk assessment. Yet, export control regulations may apply to offensive software security tools and technologies, particularly those that are designed or adapted for military or intelligence purposes.

## Technical and Societal Impacts

Current attacks and defenses in computer security are largely ad hoc (e.g., memory safety: DEP, ASLR; web security: XSS and CSRF defenses). The security community is working towards a more foundational basis for security to understand risk exposure and better defend critical computer assets. Like the theory of cryptography and general inductive/deductive model of science, this foundation for security evolves in a back-and-forth manner. The community should 1) build crypto systems / theories / security defenses, 2) discover ways in which these models/systems break down and fail to represent reality, and then 3) re-build better, more fundamental crypto / scientific theories / security defenses. The current landscape of security has stagnated due to a focus on defenses and a corresponding dearth of offensive security research; in other words, the rate of incidental / in-the-wild vulnerability discovery largely dictates the slow pace of security evolution. To accelerate security development and fortify computer systems against future threats, the community should decrease the latency between offensive and defensive security efforts. To that end, it is important to increase research funding for offensive security research to match that of defensive security.

**Cross-Discipline and Outreach Opportunities**

Conducting offensive research will benefit many CS and non-CS areas. Exemplars of opportunities in this area include: (a) *Cybersecurity education.* Offensive security professionals can participate in educational materials and seminars to educate the general public about cybersecurity threats and ways to protect against them. (b) *Connection with training and awareness of general audience.* Offensive security professionals can conduct training sessions and awareness campaigns for employees of organizations to raise awareness about cybersecurity best practices and to promote a security culture within the organization. (c) *Collaboration with law enforcement.* Offensive security professionals can work with law enforcement agencies to help identify and track down cybercriminals. (d) *Connection with social engineering.* On one hand, social engineering techniques can be used to spread and deliver offensive software. Additionally, social engineering techniques can be used to bypass security measures put in place to prevent the installation or execution of offensive software. For example, an attacker may use social engineering to convince an employee to disable security software or to grant them administrative access. (e) *Collaboration with the medical industry.* Offensive security professionals can work with medical professionals and device manufacturers to identify and address security vulnerabilities in medical devices and systems before they are released or produced.

### 3.8.4 Digital Trust and Safety

**Area Description**

Trust and safety are the traditional industry terms for a qualitatively different set of concerns relative to traditional computer security efforts. It includes issues like misinformation (another section is dedicated to this), but also hate, harassment, stalking, cyberbullying, and other types of harm. Recent surveys show that fully 50% of people report experiencing online threats such as stalking, cyberbullying, or account takeover by someone they know. It is clear that industry does not have solutions to the immense problems here, and research needs to build up the academic community to help conceive of future directions.

One convincing viewpoint is that traditional computer security has been framed around protecting technologies – the confidentiality, integrity, and availability of data and computing devices – and digital trust and safety is an expansion to bring into scope as well harms to individuals, groups of people, and society more broadly. Research should center on mitigating technology-based harm to people. Some have argued that this is a critical and needed expansion of the topics that SaTC has mostly focused on in the past, and the computer security and privacy communities should form research in this area in basic and applied research. In part because trust and safety mechanisms are not possible without good computer security and privacy – data breaches and other traditional issues can cause immense harm to people. In some sense this is pushing on the "trustworthy" part of SaTC and thinking of S as expanding security to the broader view of "safety".

**Technical Efforts**

**Research Focused on Target Populations.** After many years of research efforts studying online harms, one consistent finding is that nefarious online behavior (and exposure) often follows a power-law distribution: large proportions of the population are uninvolved/exposed, but small proportions of the populations are heavily involved/impacted/exposed. Many cases studies of representative samples of the population reveal (important!) null effects, but that does not mean that the smaller portions of the population where activity/impact is concentrated are not important. The solution is to invest in **"tail focused" research** – that is, specific research designs to better understand what is happening in the tails of distributions.

This means research focused on population groups, such as at-risk user groups (groups that are disproportionately at risk of attack or will be disproportionately harmed by an attack should it happen) which often includes historically marginalized or vulnerable communities. For example, those latter communities are particularly vulnerable to being silenced or excluded from social and political conversations via online hate and harassment.

At the same time, research needs to study perpetrator populations. The small communities of people that may be engaging in nefarious behaviors online (harassment, coordinated influence operations, online radicalization efforts) the sheer size of the population means that even small proportions of users can have serious online impact and offline consequences. There is also the huge amount of "small scale" abuse, such as interpersonal abuse involving a small number of people (possibly just two).

Studying these populations – both perpetrators and their targets – can be useful for improving trust and safety not only for these important groups, but for users in general as well.

**Addressing harmful algorithms including predictive and generative AI.** Algorithms can be accidentally harmful, for example by perpetuating and amplifying biases in the training data. Algorithms can also be intentionally manipulated to be harmful, for example by purposefully shaping the underlying data of both predictive and generative machine learning models. SaTC should continue to support research that seeks to understand how AI creates and perpetuates harms — as well as solutions for mitigating those harms.

**Understanding and developing a formalized framework of harms.** Research into safe, secure, and trustworthy computing should center potential harms to people, groups, and society at large. To do this work, researchers, and practitioners (from designers to moderators to journalists) need a formal definition of harm and a framework for characterizing different types of harm, across individual, group, and societal levels. This can be challenging, because harm can be highly contextual, and in some cases specific to individuals or communities (e.g., certain high-risk populations). Common vocabularies and a formalization of harm will facilitate knowledge sharing across projects and support the operationalization of harm for detection and prevention. This will require interdisciplinary research by social scientists, software engineers and cybersecurity professionals.

**Focusing on what is exploited for harm rather than potential vulnerabilities**. An open problem is how to focus on what is exploited for harm rather than potential vulnerabilities. This takes the user-centric view to analyze the harms happening to users. The analysis can transit to where the designed affordances are vulnerable.

**Designing reporting mechanisms that are not exploitable.** An open problem is to effectively provide reporting mechanisms for at-risk populations, so that the victims can get extra aid in a timely manner. One design dimension is to have ethical frameworks to guide, which mitigate intentional false reporting. For example, users may report arbitrary content as negative cases. Potential solutions may need authentication and accountability of the reporter. Another design dimension is to make the tools easy to use for the victims, which may need interface design and customization for specific groups of users.

**Understanding the unintended consequences of solutions.** Sometimes, the solutions to one dimension of the problem open new challenges. For example, platform affordances that allow people to block harassers can be exploited by one group to harass or silence another. This form of "indirect censorship" can empower states to get platforms for carry out censorship for them. Similarly, solutions that help one vulnerable group may put another at risk. For example, policies designed to compel platforms to stop human trafficking resulted in reduced safety for sex workers. Researchers should be encouraged to anticipate these unintended consequences.

**Adversarial manipulation of trust and safety tools.** Tools designed to improve trust and safety, such as reporting other users for violations of community rules, can themselves become Research on adversarial manipulation of tools designed to reduce harm (e.g., Saudi government reporting activists as violating Twitter policies to get them kicked off Twitter).

**Investigating security mechanisms from trust and safety viewpoint.** Trust and safety issues surface new threat models that current computer security approaches fail to address. Examples from prior work include authentication failing because abusers know their partner's or family member's passwords and log in from the same location; encryption to provide privacy may make verifiable harassment reporting more difficult; second-factor authentication makes it easier to lock victims out of compromised accounts; and anti-automation approaches prevent coordinated hate campaigns run by large groups of individuals. At the same time trust and safety mechanisms such as reporting are also vectors, themselves, for new attacks, and need to be treated as such (see above on secure reporting mechanisms).

**Shared public data resources.** Many of the research areas identified above require access to platform data that is ostensibly "owned" by large corporations. Securing access to the necessary data is a non-trivial challenge, at best costing a great deal of researcher time and at worst preventing important research from ever being conducted. Thus in order to advance goals of better understanding and supporting trust and safety online, efforts to make data more accessible to researchers ought to be an important part of this broader research portfolio.[11] Many nascent efforts are developing in this regard,[12] but research to determine the best ways to do so safely, securely, and legally could be catalytic, including building an ethical/legal framework to support scraping publicly available data for academic research.

**Technical and Societal Impacts**

---

[11] For a series of white papers on these challenges, see https://securelysharingdata.com/overview.html.
[12] https://informationenvironment.org/about; https://mddatacoop.org/members/.

Addressing these issues will generate new technical directions for the SaTC community. As discussed above, rethinking basic security tooling to help mitigate trust and safety issues is necessary. If not addressed these issues may result in significant harm; hate and harassment may be the biggest ways in which people suffer due to technology.

**Cross-Discipline and Outreach Opportunities**

A thorough understanding of trust, safety, and online will clearly have to involve both technical and social scientific research because this is both a technical and social problem. Social science research generally to understand human behaviors, vulnerabilities, needs, and harm. Legal scholars will be needed to help translate research into policy, as well as to provide frameworks for securing the necessary data access to carry out this work. Political scientists will be needed to help understand the political context in which security measures (or any solutions) will be carried out, and communication and media scholars will provide vital insights into understanding how to structure discourse.

### 3.8.5    Cyber-Physical System Security and Autonomous Security

**Area Description**

Cyber-physical systems (CPS) domains are diverse, including platforms such as autonomous cars, industrial automation, airplanes, smart homes and buildings, and robotics applications. CPS domains are defined over physical infrastructure, emphasizing sensors and real-time controls. Autonomous systems add to this infrastructure a component that reasons (e.g., plans) and reacts to conditions in the environment when executing some mission.

CPS/autonomous security explores threats and countermeasures in these environments, e.g., adversarial tampering, denial of service, and damage to the physical infrastructure. Research is needed to characterize these challenges systematically and comprehensively.

Autonomous security also addresses security and privacy mechanisms that are self-configuring, self-organizing, self-tuning, self-managing, and self-repairing, i.e., operates independently of human configuration and management.

**Technical Efforts**

**Holistic security.** CPS systems, for example an autonomous vehicle, bring together multiple components and subsystems. Security of subcomponents is necessary but not sufficient for securing the CPS system. This is because components are integrated with one another and therefore exhibit complex and subtle dependencies and interactions. There is communication, data flow, and control flow between components. A major research challenge in CPS security is, how to add holistic (orchestrated) security that goes beyond component analysis. Many CPS systems do not have a centralized management layer. They are designed as an interacting set of subsystems that manage the underlying operation of physical infrastructure.

**Human-in-the-loop.** An interesting opportunity to explore is how CPS security systems could provide information and monitoring for human operator consumption. Future research may reconsider the role of humans who are part of the control infrastructure and oversee and manage it—and therein allow for more secure (and a possibly safer) operation. It is often thought that autonomous security replaces humans. Autonomous security can provide monitoring, analytics, and response functions across a complex infrastructure (e.g., smart building or city) that serves human overseers and decision-makers. CPS security could provide the dashboard at which humans sit and the toolset delivering insights – providing advanced monitoring, analytics, and query support for an extensive, complex, multi-layer infrastructure that is beyond human capacity to manage.

**Real-time security.** A key attribute of CPS infrastructures is real-time control loops that are highly sensitive to delay and interruptions. Consider, for example, autonomous vehicles. It is important that the security design manages observability and automated analysis without generating delays in operation. For example, by the time security analysis of a situation is completed, the autonomous vehicle may already have crashed due to adversarial tampering. Or the addition of security monitoring may have caused a crash because it interfered with tight timing loops in vehicular sensing and control. In general, many CPS systems pose complex challenges to security design because of the tradeoffs between accuracy, coverage, and response time. Research is needed for real-time security and safety in CPS. Two potential directions include: (1) automatic security analysis, (2) responses with different feedback loops.

**Automatic security analysis.** The scale and complexity of CPS systems imply the need for more rigorous security analysis. Yet, the scale and complexity of those environments also renders human-only analysis intractable. Consider, for example, automated manufacturing, where dozens or hundreds, or thousands of robots operate. Several interesting (and important) research questions include: How do researchers design autonomous security capable of overseeing complex infrastructure with a massive, distributed, hierarchical attack surface? How do researchers model and detect threats? What does the analysis of attacks look like? Answers to these questions will provide the science on which highly complex, autonomous CPS systems can be managed and secured.

**Realistic testbed.** A major problem in CPS security research is a lack of access to realistic testbed environments. One issue is that companies who might provide access to real environments cannot share data for performance, stability, privacy, legal, or intellectual property reasons. Moreover, many companies are reticent to share data about known attacks or allow analysis for fear of negative PR. All of this would imply the need for investments in testbeds and simulators. For simulators, there are challenges in the realism and comprehensiveness of the simulator. Few existing simulators can manage the detail and complexity of CPS systems in realistic ways, but future investments in this area may facilitate research on security in real-time contexts.

**Insider threats.** Autonomous CPS domains offer many opportunities for insider threats since attack surfaces are large and often hidden by complexity. One of these threats is data poisoning attacks, which can be challenging to detect in real-time or remediate in systems fed by streaming data. Exploration of insider detection (possibly aided by physical access) is an important area.

**Compliance and legal issues.** An interesting challenge in CPS security is compliance and related legal issues. CPS component vendors often do enough to meet regulatory requirements on their interfaces, features, and specifications. They also want to avoid legal responsibility for security failures. But creating the right regulations to push vendors toward more secure solutions is possible. Several questions arise from this line of inquiry. How best to exercise compliance incentives in the right way? What regulations can be created to positively influence complex systems where humans and AI-based elements interact? Research—in concert with legal scholars—may help address these issues.

**Privacy and utility tradeoffs in CPS.** There are often privacy tradeoffs in CPS security. Consider, for example, a smart home in which the observability of IoT device data is needed for monitoring and analysis but causes privacy issues when information is revealed about the occupants. CPS security may also require access to control systems which create tampering or denial of service opportunities that would not xist otherwise. One area of future may be to explore techniques to identify and manage these tradeoffs.

**Self-driving car security.** Note that "autonomous" has different levels of meaning within CPS research. Consider autonomous vehicles: level 1 consists of driver-assist features that merely augment human drivers, while level 5 refers to fully autonomous vehicle control in completely unstructured environments. Security research should consider the spectrum of system autonomy, and how each level presents unique risks, threats, and possible countermeasures.

Note further that there is an interesting interplay between human and autonomous system security and reliability. Monitoring should consider when the autonomous system is not behaving correctly and make corrections. It might be imagined that a human needs to oversee the autonomous system and step in when, for example, the autonomous vehicle is making a control mistake in a complex reactive situation. On the other hand, an autonomous system may feature faster and more comprehensive monitoring and can step in when a human driver is making mistakes or not following the rules of the road. Security research should comprehend this interplay between an autonomous system and a human operator.

**Supply chain risks.** CPS systems are filled with challenges in supply chain risk. In practice, a vehicle is a complex mash-up of vendor-supplied components. Vendors often do not supply specifications needed for their components in a uniform way. This makes CPS security analysis and verification challenging. Any individual component may have supply chain risks, and the interactions between components can likewise have risks.

**Technical and Societal Impacts**

CPS systems are a growing field as nearly all physical infrastructure is becoming digitized. Furthermore, in many environments CPS systems present failure modes that can profoundly impact human health and safety – automotive safety, aircraft safety, health domains, home IoT domains, etc. Here, security research is sorely needed as current CPS solutions often lack adequate security and represent critical vulnerabilities.

**Cross-Discipline and Outreach Opportunities**

CPS is a large conglomeration of vertical domains where physical infrastructure is intertwined with cyber systems. As such, CPS security touches almost every vertical domain and could become a collaboration with associated disciplines – biomedical, manufacturing, transportation, electrical and power systems, aeronautics, automotive, civil engineering, agriculture, etc.

# 4    TRANSITION TO PRACTICE

Note that transition to practice is not a research area, but a means by which the impacts of research can be realized within the public and private sector. Here, commentary from the community relative to transition to practice resulting from interviews and the vision workshop is provided and suggestions for future efforts and policy changes for SaTC identified.

The current practice of tech-transition work at research-funding agencies and across the entire government has been a federal priority since 2015. SaTC has a lot of success stories in transitioning research results to practice. Examples include companies founded by SaTC PIs and open-source software and developer communities based on artifacts out of SaTC PIs' research projects.

Commercialization is one pathway, but there are also other pathways for translation of research to practice. One is translation to policy, e.g., around the use of a specific technology or its application. Another could be the delivery of a service (or technology in service) of a community organization. Another could be educational interventions and the development of curriculum based on SaTC-funded research. In general, it would be helpful if research agencies were to shepherd some of these pathways for inexperienced PIs.

There are positive and negative lessons in both academia and industry. Many transitions failed because transition teams do not abide by development best practices. Academic research prototypes may ignite internal prototypes in industry, which then become what is tech-transferred. Bridges: (1) Interns, as the "bridge" between academia and industry, play an important role and are extremely important in the transition process, (2) research that is more strongly reproducible, replicable, accessible, and generalizable.

An investigation of the current state of transition to practice yielded further recommendations for improving the frequency, reach and quality of transitions.  These include:

- Transition processes should formalize the success metrics, which will help PIs and industry make decisions on what they should do next.
- Transition to practice processes needs guidelines/training to help PI to make code objects and artifacts documented, archived, maintained, so that code objects are usable by industry.

- Research agencies should work with universities that value or give credit to PIs for failed Transition to practice attempts. This might require a culture and policy change with external organizations and may not be fully addressable in the short term.

- A SaTC event focused on highlighting agency-funded research artifacts and attracting potentially interested parties would foster new communities and opportunities. Audiences could be tailored to specific topics (e.g., software reliability), or be made broader to include a wide range of participants (e.g., industry, education, etc.). Research agencies should work with other programs to identify transition partners and customers.

- Research agencies could encourage academic venues to view transition to practice as valued contribution. For example, some conferences could create new tracks of papers at conferences that explore how to build robustness and reproducibility of research artifacts. The focus is on the lessons learned during this process (which are many) but are often viewed as either "just engineering" and "not novel". New tracks would offer an avenue for such work.

- Proposals that include a "prior research results" section can include research Transition to practice output. Perhaps adding "top-5 transitions" to the bio sketch. Large proposals should have (or strongly encouraged to) provide a separate section documenting past transition achievements.

- Some research agencies are viewed by some as being too slow and result in too little money, so it may be beneficial for research agency program managers to have some discretion to rapidly provide small funding to efforts. With quicker turnaround times (e.g., 45 days), such money could be the difference between an early prototype and a project being abandoned.

# 5    CONCLUSIONS

The Secure and Trustworthy Computing (SaTC) program within the National Science Foundation (NSF) stands as a vital catalyst for groundbreaking research in security and privacy, not only in the United States but with far-reaching impacts worldwide. This research vision document represents the culmination of an extensive collaborative effort, incorporating the insights and perspectives of diverse stakeholders encompassing academic and industrial researchers, government agencies, practitioners, policy makers, and the general public. The resulting mosaic of ideas emphasizes the criticality and breadth of work required in these crucial domains.

Within this comprehensive tapestry, several prominent themes emerge, illuminating the opportunities and challenges presented by emerging technologies, particularly artificial intelligence. Additionally, there is a clear call to reconceptualize the Internet, hardware and software architectures, recognizing the evolving landscape of security and privacy. Furthermore, the intricate connections between security, privacy, and society are increasingly recognized as essential focal points for research and innovation.

It is imperative to acknowledge the immense contributions made by countless individuals to shape this collective vision. The depth and richness of this document owe their existence to the dedication and collaboration of many stakeholders. The NSF, along with the security and privacy committee, wholeheartedly applauds these collaborative efforts, recognizing their direct and indirect impacts that will resonate across generations.

As this research vision unfolds, we hope it will serve as a guiding light, inspiring and directing future endeavors in security and privacy. The NSF, in partnership with researchers, industry leaders, policymakers, and the broader community, is committed to fostering the realization of this collective vision, ensuring that the transformative outcomes will shape a more secure and privacy-conscious world. With unwavering dedication, the NSF remains at the forefront of advancing fundamental science in security and privacy, driving innovations that safeguard our present and future societies.

# 6    APPENDIX

**List of Attendees:**

Adam Doupé, Arizona State University
Ahmad-Reza Sadeghi, Technical University Darmstadt
Alessandro Acquisti, Carnegie Mellon University
Alexandra Dmitrienko, University of Wuerzburg
Aman Mohindru, Wipro
Ambareen Siraj, National Science Foundation
Amit Sahai, University of California, Los Angeles
Andrew Pollington, National Science Foundation
Angelos Keromytis, Georgia Tech
Anna   Squicciarini, National Science Foundation
Ashish Kundu, Head of Cybersecurity Research, Cisco Research
Balakrishnan Prabhakaran, University of Texas at Dallas
Bhavani Thuraisingham, University of Texas at Dallas
Blaine  Hoak, University of Wisconsin-Madison
Brian LaMacchia, Farcaster Consulting Group
Chun Sheng Xin, National Science Foundation
Cindy Bethel, National Science Foundation
Cliff Wang, National Science Foundation
Damon McCoy, New York University
Dan Cosley, National Science Foundation
Daniel Takabi, Georgia State University
Daniela Oliveira, National Science Foundation
David Ott, VMware
Dongyan Xu, Purdue University
Engin Kirda, Northeastern University
Farinaz Koushanfar, University of California, San Diego
Florian Schaub, University of Michigan
Gang Qu, National Science Foundation
Gene Tsudik, University of California, Irvine
Giti Javidi, University of South Florida
Guofei Gu, Texas A&M University
Hafiz Malik, University of Michigan-Dearborn
Heng Xu, Kogod School of Business, American University
Jaideep Vaidya, Rutgers University
James Joshi, National Science Foundation
Jeremy Epstein, National Science Foundation
John Heidemann, University of Southern California/Information Sciences Institute
Joshua Tucker, New York University
Kang Shin, University of Michigan
Kangkook Jee, University of Texas at Dallas
Kate Starbird, University of Washington
Kelly Caine, Clemson
Kevin Butler, University of Florida
Kevin Hamlen, University of Texas at Dallas
Li Yang, National Science Foundation

Libby Hemphill, University of Michigan
Ling Liu, Georgia Institute of Technology
Marina Blanton, University at Buffalo
McKenna McCall, Carnegie Mellon University
Melissa Schreier, University of Wisconsin-Madison
Michael Reiter, Duke University
Murat Kantarcioglu, University of Texas at Dallas
Neil Gong, Duke University
Ninghui Li, Purdue University
Nojan Sheybani, University of California, San Diego
Norman Sadeh, Carnegie Mellon University
Patrick Traynor, University of Florida
Patrick McDaniel, University of Wisconsin-Madison
Quang Dao, Carnegie Mellon University
Rahul Chatterjee, University of Wisconsin--Madison
Rakesh Bobba, Oregon State University
Ravi Sandhu, University of Texas at San Antonio
Rob Beverly, National Science Foundation
Roberto Perdisci, University of Georgia
Ruoyu "Fish" Wang, Arizona State University
Scott Ruoti, University of Tennessee, Knoxville
Shomir Wilson, Pennsylvania State University
Sterling Martin, Epic Systems
Susanne Wetzel, Stevens Institute of Technology
Tarun Soni, Northrop Grumman
Thang Hoang, Virginia Tech
Tianshi Li, Carnegie Mellon University
Tom Ristenpart, Cornell Tech
Tomas Vagoun, NITRD/NCO
Venkat Venkatakrishnan, University of Illinois System
Vinod Vaikuntanathan, Massachusetts Institute of Technology
Wenbo Guo, University of California, Berkeley
Wenke Lee, Georgia Institute of Technology
Xiaojing Liao, Indiana University Bloomington
Xinyu Xing, Northwestern University
Yevgeniy Vorobeychik, Washington University in Saint Louis
Yingying Chen, Rutgers University
Yousra Aafer, University of Waterloo
Yuan Tian, UCLA
Yulia Gel, National Science Foundation
Zahra Ghodsi, Purdue University
Zane Ma, Georgia Institute of Technology
Zhiqiang Lin, Ohio State University
Ziming Zhao, University at Buffalo